\documentclass{ws-ijmpd}

\begin{document}

\def\nocropmarks{\vskip5pt\phantom{cropmarks}}
\let\trimmarks\nocropmarks      

\renewcommand\ArtDir{./}

\markboth{R. Ruffini, C.L. Bianco, P. Chardonnet, F. Fraschetti, S.-S. Xue}
{On the structure of the burst and afterglow of Gamma-Ray Bursts I: the radial approximation}

\catchline{}{}{}

\title{On the structure of the burst and afterglow of Gamma-Ray Bursts I: the radial approximation}

\author{\footnotesize REMO RUFFINI, CARLO LUCIANO BIANCO and SHE-SHENG XUE}
\address{ICRA --- International Center for Relativistic Astrophysics and Dipartimento di Fisica,\\ Universit\`a di Roma ``La Sapienza'', Piazzale Aldo Moro 5, I-00185 Roma, Italy.}

\author{\footnotesize PASCAL CHARDONNET}
\address{ICRA --- International Center for Relativistic Astrophysics and Universit\'e de Savoie,\\ LAPTH - LAPP, BP 110, F­74941 Annecy-le-Vieux Cedex, France.}

\author{\footnotesize FEDERICO FRASCHETTI}
\address{ICRA --- International Center for Relativistic Astrophysics and Universit\`a di Trento,\\ Via Sommarive 14, I-38050 Povo (Trento), Italy.}
 
\maketitle

\begin{abstract}
We have recently proposed three paradigms for the theoretical interpretation of gamma-ray bursts (GRBs). (1) The relative space-time transformation (RSTT) paradigm emphasizes how the knowledge of the entire world-line of the source from the moment of gravitational collapse is a necessary condition in order to interpret GRB data.\cite{lett1} (2) The interpretation of the burst structure (IBS) paradigm differentiates in all GRBs between an injector phase and a beam-target phase.\cite{lett2} (3) The GRB-supernova time sequence (GSTS) paradigm introduces the concept of {\em induced supernova explosion} in the supernovae-GRB association.\cite{lett3} In the introduction the RSTT and IBS paradigms are enunciated and illustrated using our theory based on the vacuum polarization process occurring around an electromagnetic black hole (EMBH theory). The results are summarized using figures, diagrams and a complete table with the space-time grid, the fundamental parameters and the corresponding values of the Lorentz gamma factor for GRB~991216 used as a prototype. In the following sections the detailed treatment of the EMBH theory needed to understand the results of the three above letters is presented. We start from the considerations on the dyadosphere formation. We then review the basic hydrodynamic and rate equations, the equations leading to the relative space-time transformations as well as the adopted numerical integration techniques. We then illustrate the five fundamental eras of the EMBH theory: the self acceleration of the $e^+e^-$ pair-electromagnetic plasma (PEM pulse), its interaction with the baryonic remnant of the progenitor star, the further self acceleration of the $e^+e^-$ pair-electromagnetic radiation and baryon plasma (PEMB pulse). We then study the approach of the PEMB pulse to transparency, the emission of the proper GRB (P-GRB) and its relation to the ``short GRBs''. Particular attention is given to the free parameters of the theory and to the values of the thermodynamical quantities at transparency. Finally the three different regimes of the afterglow are described within the fully radiative and radial approximations: the ultrarelativistic, the relativistic and the nonrelativistic regimes. The best fit of the theory leads to an unequivocal identification of the ``long GRBs'' as extended emission occurring at the afterglow peak (E-APE). The relative intensities, the time separation and the hardness ratio of the P-GRB and the E-APE are used as distinctive observational test of the EMBH theory and the excellent agreement between our theoretical predictions and the observations are documented. The afterglow power-law indexes in the EMBH theory are compared and contrasted with the ones in the literature, and no beaming process is found for GRB~991216. Finally, some preliminary results relating the observed time variability of the E-APE to the inhomogeneities in the interstellar medium are presented, as well as some general considerations on the EMBH formation. The issue of the GSTS paradigm will be the object of a forthcoming publication and the relevance of the iron-lines observed in GRB~991216 is shortly reviewed. The general conclusions are then presented based on the three fundamental parameters of the EMBH theory: the dyadosphere energy, the baryonic mass of the remnant, the interstellar medium density. An in depth discussion and comparison of the EMBH theory with alternative theories is presented as well as indications of further developments beyond the radial approximation, which will be the subject of paper II in this series.\cite{rbcfx02e_paperII} Future needs for specific GRB observations are outlined.
\end{abstract}


\section{Introduction}\label{int}

\subsection{The physical and astrophysical background}

Gamma-ray bursts (GRBs) are rapidly fuelling one of the broadest scientific pursuit in the entire field of science, both in the observational and theoretical domains. Following the discovery of GRBs by the Vela satellites,\cite{s75} the observations from the Compton satellite and BATSE\footnote{See http://cossc.gsfc.nasa.gov/batse/} had shown the isotropic distribution of the GRBs strongly suggesting a cosmological nature for their origin. It was still through the data of BATSE that the existence of two families of bursts, the ``short bursts'' and the ``long bursts'' was presented, opening an intense scientific dialogue on their origin still active today, see e.g. Schmidt (2001)\cite{s01} and section \ref{new}.

An enormous momentum was gained in this field by the discovery of the afterglow phenomena by the BeppoSAX satellite and the optical identification of GRBs which have allowed the unequivocal identification of their sources at cosmological distances.\cite{c00} It has become apparent that fluxes of $10^{54}$ erg/s are reached: during the peak emission the energy of a single GRB equals the energy emitted by all the stars of the Universe.\cite{rk01}

From an observational point of view, an unprecedented campaign of observations is at work using the largest deployment of observational techniques from space with the satellites CGRO-BATSE, Beppo-SAX\footnote{See http://www.asdc.asi.it/bepposax/}, Chandra\footnote{See http://chandra.harvard.edu/}, R-XTE\footnote{See http://heasarc.gsfc.nasa.gov/docs/xte/}, XMM-Newton\footnote{See http://xmm.vilspa.esa.es/}, HETE-2\footnote{See http://space.mit.edu/HETE/}, as well as the HST\footnote{See http://www.stsci.edu/}, and from the ground with optical (KECK\footnote{See http://www2.keck.hawaii.edu:3636/}, VLT\footnote{See http://www.eso.org/projects/vlt/}) and radio (VLA\footnote{See http://www.aoc.nrao.edu/vla/html/VLAhome.shtml}) observatories. The further possibility of examining correlations with the detection of ultra high energy cosmic rays, UHECR for short, and in coincidence neutrinos should be reachable in the near future thanks to developments of AUGER\footnote{See http://www.auger.org/} and AMANDA\footnote{See http://amanda.berkeley.edu/amanda/amanda.html} (see also Halzen, 2000\cite{h00}).

\begin{figure}[htbp]
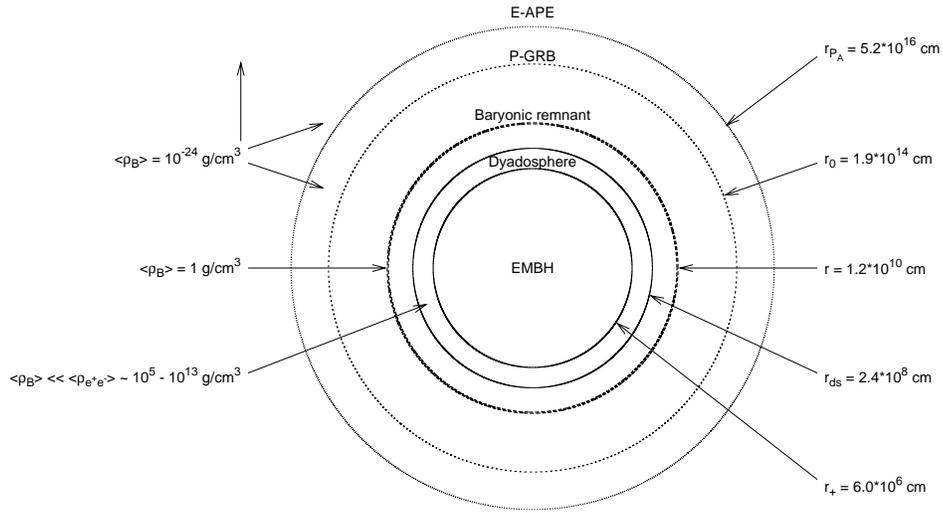

\PSFIG{raggi2}{\hsize}{0}
\caption{Selected events in the EMBH theory are represented. For each one the values of the energy density of the medium and the distances from the EMBH, in the laboratory frame and in logarithmic scale, are given.}
\label{raggi2}
\end{figure}

From a theoretical point of view, GRBs offer comparable opportunities to develop entire new domains in yet untested directions of fundamental science. For the first time within the theory based on the vacuum polarization process occurring in an electromagnetic black hole, the EMBH theory, see Fig.~\ref{raggi2}, the opportunity exists to 
theoretically approach the following fundamental issues:
\begin{enumerate}
\item The extremely relativistic hydrodynamic phenomena of an electron-positron plasma expanding with sharply varying gamma factors in the range $10^2$ to $10^4$ and the analysis of the very high energy collision of such an expanding plasma with baryonic matter reaching intensities $10^{38}$ larger than the ones usually obtained in Earth-based accelerators.
\item The bulk process of vacuum polarisation created by overcritical electromagnetic fields, in the sense of Heisenberg, Euler\cite{he35} and Schwinger\cite{s51}. This longly sought quantum ultrarelativistic effect has not been yet unequivocally observed in heavy ion collision on the Earth.\cite{ga96,la97,la98,ha98} The difficulty of the heavy ion collision experiments appears to be that the overcritical field is reached only for time scales of the order $\hbar/m_pc^2$, which is much shorter than the characteristic time for the $e^+e^-$ pair creation process which is of the order of $\hbar/m_ec^2$, where $m_p$ and $m_e$ are respectively the proton and the electron mass. It is therefore very possible that the first appearance of such an effect occurs in the strong electromagnetic fields developed in astrophysical conditions during the process of gravitational collapse to an EMBH, where no problem of confinement exists.
\item A novel form of energy source: the extractable energy of a black hole. The enormous energies released almost instantly in the observed GRBs, points to the possibility that for the first time we are witnessing the release of the extractable energy of an EMBH, during the process of gravitational collapse itself. We can compute and have the opportunity to study all general relativistic as well as the associated ultrahigh energy quantum phenomena as the horizon of the EMBH is approached and is being formed.
\end{enumerate}

It is clear that in approaching such a vast new field of research, implying previously unobserved relativistic regimes, it is not possible to proceed {\itshape as usual} with an uncritical comparison of observational data to theoretical models within the classical schemes of astronomy and astrophysics. Some insight to the new approach needed can be gained from past experience in the interpretation of relativistic effects in high energy particle physics as well as from the explanation of some observed relativistic effects in the astrophysical domain. Those relativistic regimes, both in physics and astrophysics, are however much less extreme than those encountered in GRBs.

There are three major new features in relativistic systems which have to be properly taken into account:

\begin{figure}[htbp]
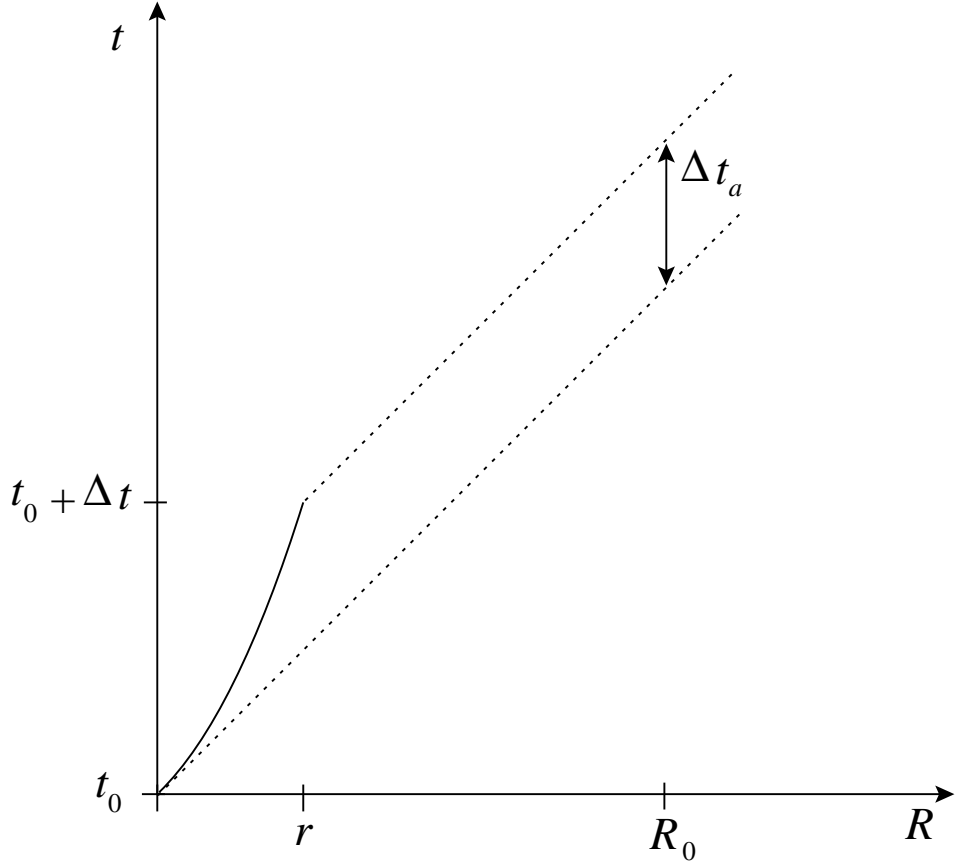

\PSFIG{ttasch}{\hsize}{0}
\caption{This qualitative diagram illustrates the relation between the laboratory time interval $\Delta t$ and the arrival time interval $\Delta t_a$ for a pulse moving with velocity $v$ in the laboratory time (solid line). We have indicated here the case where the motion of the source has a nonzero acceleration. The arrival time is measured using light signals emitted by the pulse (dotted lines). $R_0$ is the distance of the observer from the EMBH, $t_0$ is the laboratory time corresponding to the onset of the gravitational collapse, and $r$ is the radius of the expanding pulse at a time $t=t_0 + \Delta t$.$^1$}
\label{ttasch}
\end{figure}

\begin{enumerate}
\item Practically all data on astronomical and astrophysical systems is acquired by using photon arrival times. It was Einstein\cite{e05} at the very initial steps of special relativity who cautioned about the use of such an arrival time analysis and stated that when dealing with objects in motion proper care should be taken in defining the time synchronization procedure in order to construct the correct space-time coordinate grid (see Fig.~\ref{ttasch}). It is not surprising that as soon as the first relativistic bulk motion effects were observed their interpretations within the classical framework of astrophysics led to the concept of ``superluminal'' motion. These were observations of extragalactic radio sources, with gamma factors\cite{bsm99} $\sim 10$ and of microquasars in our own galaxy with gamma factor\cite{mr99} $\sim 5$. It has been recognized\cite{r66} that no ``superluminal'' motion exists if the prescriptions indicated by Einstein are used in order to establish the correct space-time grid for the astrophysical systems. In the present context of GRBs, where the gamma factor can easily surpass $10^2$, the direct application of classical concepts leads to enormous ``superluminal'' behaviours (see Tab.~\ref{tab1}). An approach based on classical arrival time considerations as sometimes done in the current literature completely subverts the causal relation in the observed astrophysical phenomenon.
\item One of the clear successes of relativistic field theories has been the understanding of the role of four-momentum conservation laws in multiparticle collisions and decays such as in the reaction: $n\rightarrow p+e^-+\bar\nu_e$. From the works of Pauli and Fermi it became clear how in such a process, contrary to the case of classical mechanics, it is impossible to analyze a single term of the decay, the electron or the proton or the neutrino or the neutron, out of the context of the global point of view of the relativistic conservation of the total four momentum of the system. This in turn involves the knowledge of the system during the entire decay process. These rules are routinely used by workers in high energy particle physics and have become part of their cultural background. If we apply these same rules to the case of the relativistic system of a GRB it is clear that it is just impossible to consider a part of the system, e.g. the afterglow, without taking into account the general conservation laws and whole relativistic history of the entire system. The description of the afterglow alone, as has been given at times in the literature, indeed possible within the framework of classical astronomy and astrophysics, is not viable in a relativistic astrophysics context where the space-time grid necessary for the description of the afterglow depends on the entire previous relativistic part of the worldline of the system (see also section \ref{bf}).
\item The lifetime of a process has not an absolute meaning as special and general relativity have shown. It depends both on the inertial reference frame of the laboratory and of the observer and on their relative motion. Such a phenomenon, generally expressed in the ``twin paradox'', has been extensively checked and confirmed to extremely high accuracy as a byproduct of the elementary particle physics (g-2) experiment.\cite{vd77} This situation is much more extreme in GRBs due to the very large (in the range $10^2$--$10^4$) and time varying (on time scales ranging from fractions of seconds to months) gamma factors between the comoving frame and the far away observer (see Fig.~\ref{gamma}). Moreover in the GRB context such an observer is also affected by the cosmological recession velocities of its local Lorentz frame.
\end{enumerate}

\subsection{The Relative Space-Time Transformations: the RSTT paradigm and current scientific literature}

Here are some of the reasons why we have recently presented a basic relative space-time transformation (RSTT) paradigm\cite{lett1} to be applied prior to the interpretation of GRB data.

The first step is the establishment of the governing equations relating:\\
a) The comoving time of the pulse ($\tau$)\\
b) The laboratory time ($t$)\\
c) The arrival time at the detector ($t_a$)\\
d) The arrival time at the detector corrected for cosmological expansion ($t_a^d$)\\
The book-keeping of the four different times and corresponding space variables must be done carefully in order to keep the correct causal relation in the time sequence of the events involved. 

As formulated the RSTT paradigm contains two parts: the first one is a necessary condition, the second one a sufficient condition. The first part reads: ``the necessary condition in order to interpret the GRB data, given in terms of the arrival time at the detector, is the knowledge of the {\em entire} worldline of the source from the gravitational collapse''.

Clearly such an approach is in contrast with articles in the current literature which emphasize either some qualitative description of the sources or some quantitative description of the afterglow era by itself.

In the current literature several attempts have addressed the issue of the sources of GRBs. They include scenarios of binary neutron stars mergers,\cite{elps89,npp92,mr92mnras,mr92apj} black hole~/~white dwarf\cite{fwhd99} and black hole~/~neutron star binaries,\cite{p91,mr97b} hypernovae,\cite{p98} failed supernovae or collapsars,\cite{w93,mw99} supranovae.\cite{vs98,vs99} Only those based on binary neutron stars have reached the stage of a definite model and detailed quantitative estimates have been made. In this case, however, various problems have surfaced: in the general energetics which cannot be greater than $\sim 3\times 10^{52}$ erg, in the explanation of ``long bursts'',\cite{swm00,wmm96} and in the observed location of the GRB sources in star forming regions.\cite{bkd00} In the remaining cases attention was directed to a qualitative analysis of the sources without addressing the overall problem from the source to the observations. The necessary details to formulate the equations of the dynamical evolution of the system are generally missing.

Other models in the literature have addressed the problem of only fitting the data of the afterglow observations by a phenomenological analysis. They are separated into two major classes:
 
The ``internal shock model'', first introduced by Rees \& M\'esz\'aros (1994),\cite{rm94} by far the most popular one, has been developed in many different aspects, e.g. by Paczy\'nski \& Xu (1994),\cite{px94} Sari \& Piran (1997),\cite{sp97} Fenimore (1999)\cite{f99} and Fenimore et al. (1999)\cite{fcrsyn99}. The underlying assumption is that all the variabilities of GRBs in the range $\Delta t\sim 1\, {\rm ms}$ up to the overall duration $T$ of the order of $50\, {\rm s}$ are determined by a yet undetermined ``inner engine''. The difficulties of explaining the long time scale bursts by a single explosive model has evolved into a subclass of approaches assuming an ``inner engine'' with extended activity (see e.g. Piran, 2001,\cite{p01} and references therein).

The ``external shock model'', also introduced by M\'esz\'aros \& Rees (1993),\cite{mr93} is less popular today. It relates the GRB light curves and time variabilities to interactions of a single thin blast wave with clouds in the external medium. The interesting possibility has been recognized within this model, that GRB light curves ``are tomographic images of the density distribution of the medium surrounding the sources of GRBs'' (Dermer \& Mitman, 1999\cite{dm99}) see also Dermer, Chiang \& B\"ottcher (1999),\cite{dcb99} Dermer (2002)\cite{d00} and references therein. In this case, the structure of the burst is assumed not to depend directly on the ``inner engine'' (see e.g. Piran, 2001,\cite{p01} and references therein).

All these works encounter the above mentioned difficulty: they present either a purely qualitative or phenomenological or a piecewise description of the GRB phenomenon. By neglecting the earlier phases, their space-time grid is undefined and as we will explicitly show in the following, results are reached at variance from the ones obtained in a complete and unified description of the GRB phenomenon. We show in the following how such a unified description naturally leads to new characteristic features both in the burst and afterglow of GRBs.

\subsection{The EMBH Theory}

In a series of papers, we have developed the EMBH theory\cite{rukyoto} which has the advantage, despite its simplicity, that all eras following the process of gravitational collapse are described by precise field equations which can then be numerically integrated.

Starting from the vacuum polarization process {\it \`a la} Heisenberg-Euler-Schwinger\cite{he35,s51} in the overcritical field of an EMBH first computed in Damour \& Ruffini (1975),\cite{dr75} we have developed  the dyadosphere concept.\cite{prx98}

The dynamics of the $e^+e^-$-pairs and electromagnetic radiation of the plasma generated in the dyadosphere propagating away from the EMBH in a sharp pulse (PEM pulse) has been studied by the Rome group and validated by the numerical codes developed at Livermore Lab.\cite{rswx99}

The collision of the still optically thick $e^+e^-$-pairs and electromagnetic radiation plasma with the baryonic matter of the remnant of the progenitor star has been again studied by the Rome group and validated by the Livermore Lab codes.\cite{rswx00} The further evolution of the sharp pulse of pairs, electromagnetic radiation and baryons (PEMB pulse) has been followed for increasing values of the gamma factor until the condition of transparency is reached.\cite{brx00}

As this PEMB pulse reaches transparency the proper GRB (P-GRB) is emitted\cite{lett2} and a pulse of accelerated baryonic matter (the ABM pulse) is injected into the interstellar medium (ISM) giving rise to the afterglow.

\subsection{The GRB~991216 as a prototypical source}

Until this stage, the EMBH theory has been done from first principles based on the exact solutions of the Einstein-Maxwell equations implied by the EMBH uniqueness theorem as well as on the quantum description of the vacuum polarization process in overcritical electromagnetic fields. Turning now to the afterglow, the variety of physical situations that can possibly be encountered are very large and far from unique: the description from first principles is just impossible. We have therefore proceeded to properly identify what we consider a prototypical GRB source and to develop a theoretical framework in close correspondence with the observational data.

We present the criteria which have guided us in the selection of the GRB source to be used as a prototype before proceeding to an uncritical comparison with the theory. It is now clear, since the observations of GRB~980425, GRB~991216, GRB~970514 and GRB~980326 that the afterglow phenomena can present, especially in the optical and radio wavelengths, features originating from phenomena spatially and causally distinct from the GRB phenomena. There is the distinct possibility that phenomena related to a supernova can be erroneously attributed to a GRB. This problem has been clearly addressed by the GRB supernova time sequence (GSTS) paradigm in which the time sequence of the events in the GRB supernova phenomena has been outlined.\cite{lett3} This has led to the novel concept of an induced supernova.\cite{lett3} This problem will be addressed in a forthcoming paper.\cite{rbcfx02d_supernova}

\begin{figure}[htbp]
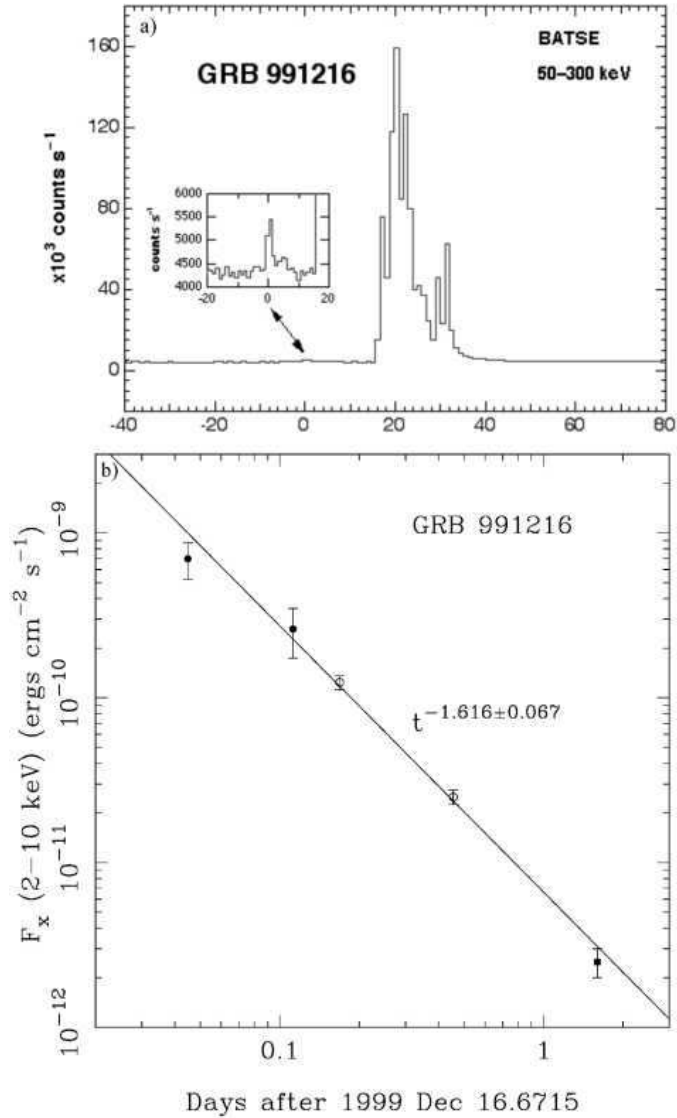

\PSFIG{gb991216}{9cm}{0}
\caption{{\bf a)} The peak emission of GRB~991216 as seen by BATSE (Reproduced from BATSE Rapid Burst Response$^{53}$); {\bf b)} The afterglow emission of GRB~991216 as seen by XTE and Chandra (reproduced from Halpern et al., 2000$^{54}$)}
\label{grb991216}
\end{figure}

In view of these considerations we have selected GRB~991216 as a prototypical case (see Fig.~\ref{grb991216}) for the following reasons:
\begin{enumerate}
\item GRB~991216 is one of the strongest GRBs in X-rays and is also quite general in the sense that it shows relevant cosmological effects. It radiates mainly in X-rays and in $\gamma$-rays and less than 3\% is emitted in the optical and radio bands.\cite{ha00}
\item The excellent data obtained by BATSE on the burst\cite{brbr99} is complemented by the data on the afterglow acquired by Chandra\cite{p00} and RXTE.\cite{cs00} Also superb data have been obtained from spectroscopy of the iron lines.\cite{p00}
\item A value for the slope of the energy emission during the afterglow as a function of time has been obtained: $n=-1.64$\cite{tmmgk99} and $n=-1.616\pm 0.067$.\cite{ha00}
\end{enumerate}

\subsection{The interpretation of the burst structure: the IBS paradigm and the different eras of the EMBH theory}

The comparison of the EMBH theory with the data of the GRB~991216 and its afterglow has naturally led to a new paradigm for the interpretation of the burst structures (IBS paradigm)) of GRBs.\cite{lett2} The IBS paradigm reads: {\itshape ``In GRBs we can distinguish an injector phase and a beam-target phase. The injector phase includes the process of gravitational collapse, the formation of the dyadosphere, as well as Era I (the PEM pulse), Era II (the engulfment of the baryonic matter of the remnant) and Era III (the PEMB pulse). The injector phase terminates with the P-GRB emission. The beam-target phase addresses the interaction of the ABM pulse, namely the beam generated during the injection phase, with the ISM as the target. It gives rise to the E-APE and the decaying part of the afterglow''}. The detailed presentations of these results are the main topic of this article.

\begin{figure}[htbp]
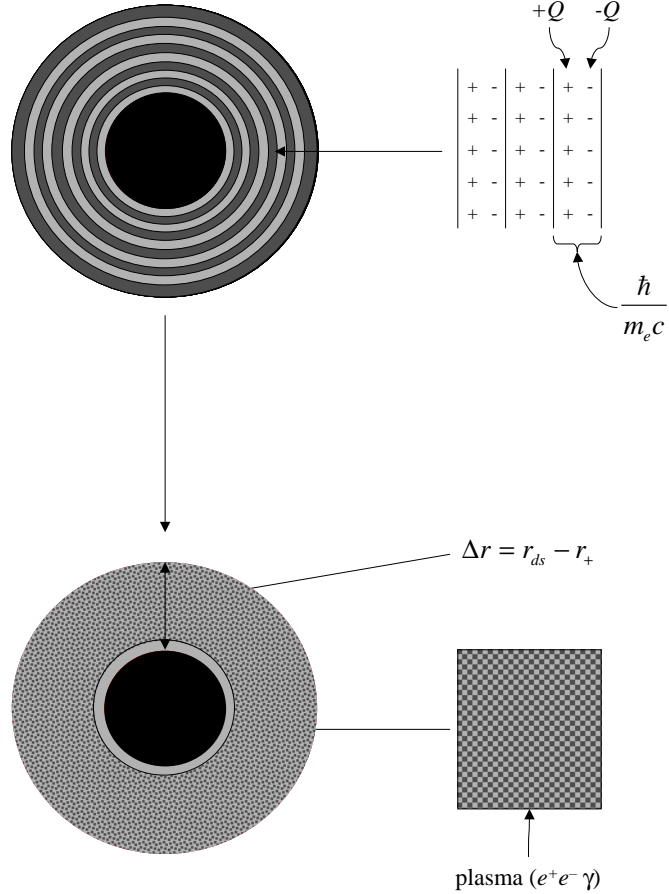

\PSFIG{dyaon}{9cm}{0}
\caption{The dyadosphere of a Reissner-Nordstr\"{o}m black hole can be represented as constituted by a concentric set of shells of capacitors, each one of thickness $\hbar/m_ec$ and producing a number of $e^+e^-$ pairs of the order of $\sim Q/e$ on a time scale of $10^{-21}$ s, where $Q$ is the EMBH charge. The shells extend in a region $\Delta r$, from the horizon $r_{+}$ to the dyadosphere outer radius $r_{\rm ds}$ (see text). The system evolves to a thermalised plasma configuration.}
\label{dyaon}
\end{figure}

We recall that the {\bf injector phase} starts from the moment of gravitational collapse and encompasses the following eras:

{\itshape The Zeroth Era: the formation of the dyadosphere}. In section \ref{dyadosphere} we review the basic scientific results which lie at the basis of the EMBH theory: the black hole uniqueness theorem, the mass formula of an EMBH, the process of vacuum polarization in the field of an EMBH. We also point out how after the discovery of the GRB afterglow the reexamination of these results has led to the novel concept of the dyadosphere of an EMBH. We have investigated this concept in the simplest possible case of an EMBH depending only on two parameters: the mass and charge, corresponding to the Reissner-Nordstr\"{o}m spacetime. We recall the definition of the energy $E_{dya}$ of the dyadosphere as well as the spatial distribution and energetics of the $e^+e^-$ pairs. See Fig.~\ref{dyaon}.

In order to analyse the time evolution of the dyadosphere we give in the three following sections the theoretical background for the needed equations.

In section \ref{hydro_pem} we give the general relativistic equations governing the hydrodynamics and the rate equations for the plasma of $e^+e^-$-pairs. 

In section \ref{arrival_time} we give the governing equations relating the comoving time $\tau$ to the laboratory time $t$ corresponding to an inertial reference frame in which the EMBH is at rest and finally to the time measured at the detector $t_a$ which, to finally get $t_a^d$, must be corrected to take into account the cosmological expansion. 

In section \ref{num_int} we describe the numerical integration of the hydrodynamical equations and the rate equation developed by the Rome and Livermore groups. This entire research program could never have materialized without the fortunate interaction between the complementary computational techniques developed by these two groups. The validation of the results of the Rome group by the fully general relativistic Livermore codes has been essential both from the point of view of the validity of the numerical results and the interpretation of the scientific content of the results.  

{\itshape The Era I: the PEM pulse}. In section \ref{hydro_pem} by the direct comparison of the integrations performed with the Rome and Livermore codes we show that among all possible geometries the $e^+e^-$ plasma moves outward from the EMBH reaching a very unique relativistic configuration: the plasma self-organizes in a sharp pulse which expands in the comoving frame exactly by the amount which compensates for the Lorentz contraction in the laboratory frame. The sharp pulse remains of constant thickness in the laboratory frame and self-propels outwards reaching ultrarelativistic regimes, with gamma factors larger than $10^2$, in a few dyadosphere crossing times. We recall that, in analogy with the electromagnetic (EM) pulse observed in a thermonuclear explosion on the Earth, we have defined this more energetic pulse formed of electron-positron pairs and electromagnetic radiation a pair-electromagnetic-pulse or PEM pulse.

{\itshape The Era II}: We describe the interaction of the PEM pulse with the baryonic remnant of mass $M_B$ left over from the gravitational collapse of the progenitor star. We give the details of the decrease of the gamma factor and the corresponding increase in the internal energy during the collision. The dimensionless parameter $B={M_Bc^2}/{E_{dya}}$ which measures the baryonic mass of the remnant in units of the $E_{dya}$ is introduced. This is the second fundamental free parameter of the EMBH theory.

\begin{figure}[htbp]
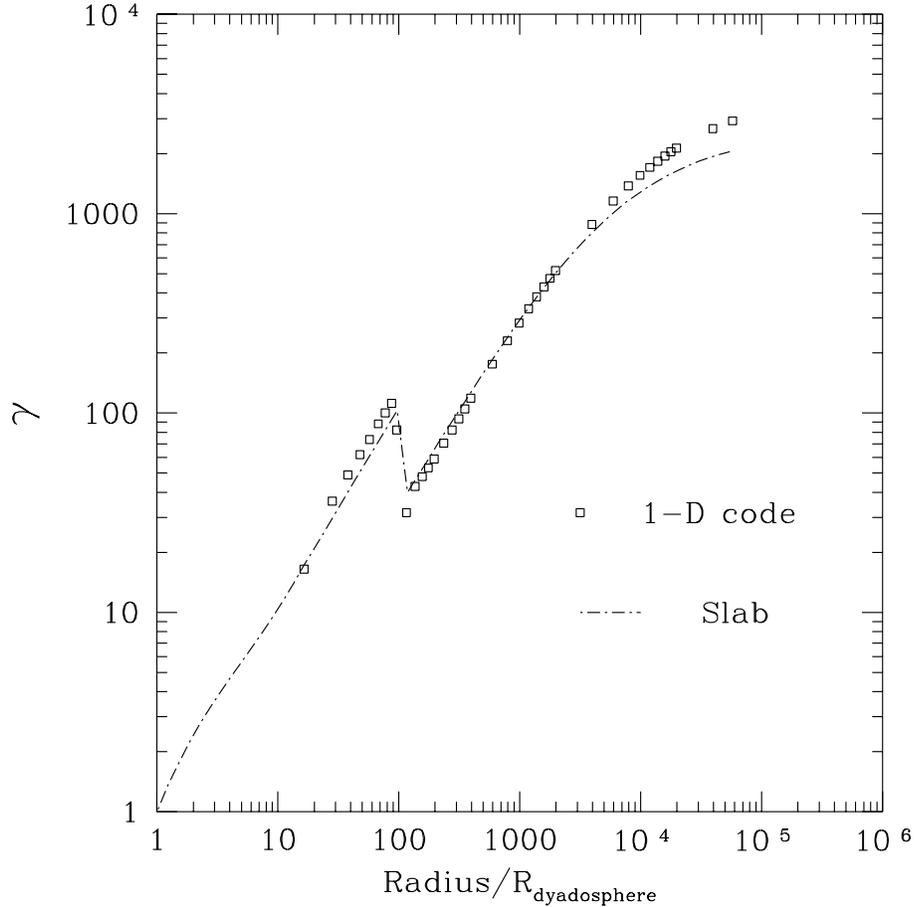

\PSFIG{h2047f7}{\hsize}{0}
\caption{Comparison of gamma factor for the one-dimensional (1-D) hydrodynamic calculations (Livermore code) and slab calculations (Rome code) as a function of the radial coordinate (in units of dyadosphere radius) in the laboratory frame. The calculations show an excellent agreement.}
\label{twocodecompare}
\end{figure}

{\itshape The Era III}: We describe in section~\ref{era3} the further expansion of the $e^+e^-$ plasma, after the engulfment of the baryonic remnant of the progenitor star. By direct comparison of the results of integration obtained with the Rome and the Livermore codes it is shown how the pair-electromagnetic-baryon (PEMB) plasma further expands and self organizes in a sharp pulse of constant length in the laboratory frame (see Fig.~\ref{twocodecompare}).
We have examined the formation of this PEMB pulse in a wide range of values $10^{-8}<B<10^{-2}$ of the parameter $B$, the upper limit corresponding to the limit of validity of the theoretical framework developed.

In section~\ref{fp} it is shown how the effect of baryonic matter of the remnant, expressed by the parameter $B$, is to smear out all the detailed information on the EMBH parameters. The evolution of the PEMB pulse is shown to depend only on $E_{dya}$ and $B$: the PEMB pulse is degenerate in the mass and charge parameters of the EMBH and rather independent of the exact location of the baryonic matter of the remnant.

In section~\ref{at} the relevant thermodynamical quantities of the PEMB pulse, the temperature in the different frames and the $e^+e^-$ pair densities, are given and the approach to the transparency condition is examined. Particular attention is given to the gradual transfer of the energy of the dyadosphere $E_{dya}$ to the kinetic energy of the baryons $E_{Baryons}$ during the optically thick part of the PEMB pulse. 

\begin{figure}[htbp]
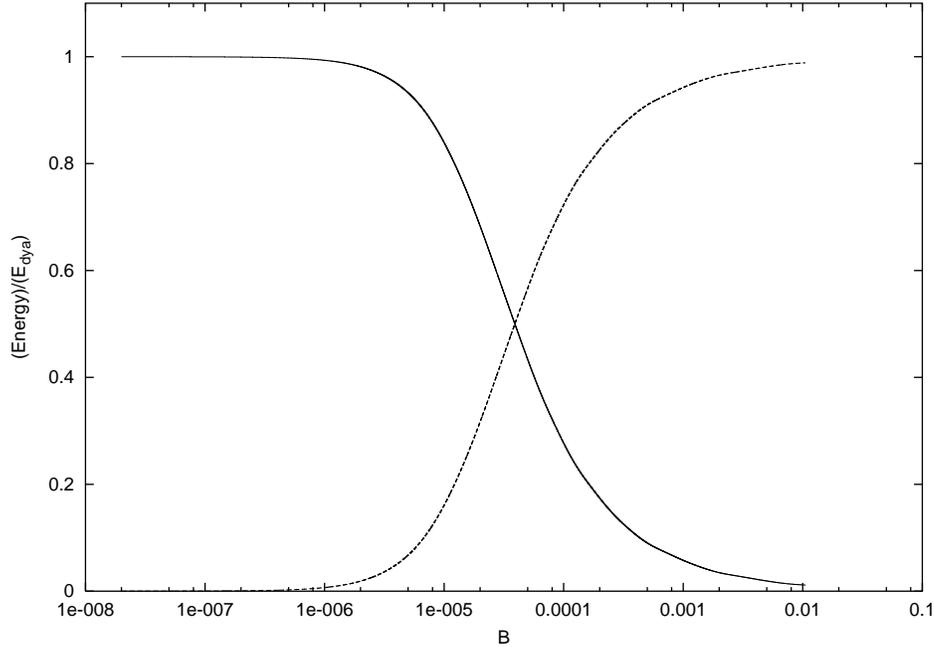

\PSFIG{h2047f11}{\hsize}{0}
\caption{At the transparent point, the energy radiated in the P-GRB (the solid line) and the final kinetic energy $E_{Baryons}$ of baryonic matter (the dashed line) in units of the total energy of the dyadosphere ($E_{dya}$) are plotted as functions of the $B$ parameter.}
\label{fintkin}
\end{figure}

\begin{figure}[htbp]
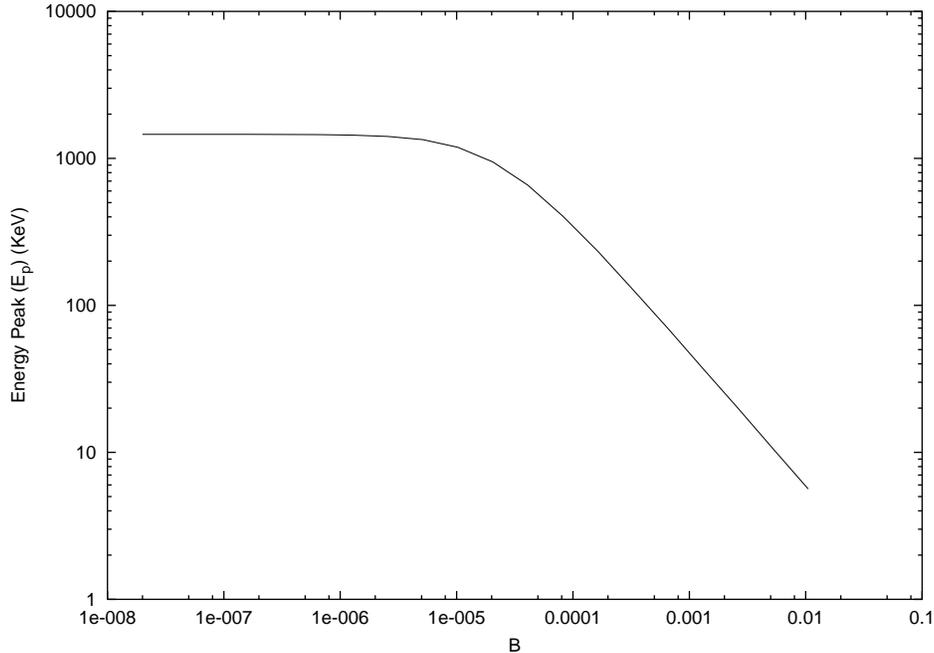

\PSFIG{h2047f10}{\hsize}{0}
\caption{The energy corresponding to the peak of the photon number spectrum in the P-GRB as measured in the laboratory frame is plotted as function of the $B$ parameter.}
\label{energypeak}
\end{figure}

In section~\ref{new}, as the condition of transparency is reached, the injector phase is concluded with the emission of a sharp burst of electromagnetic radiation and an accelerated beam of highly relativistic baryons. We recall that we have respectively defined the radiation burst (the proper GRB or for short P-GRB) and the accelerated-baryonic-matter (ABM) pulse. By computing for a fixed value of the EMBH different PEMB pulses corresponding to selected values of $B$ in the range $\left[10^{-8}\right.$--$\left.10^{-2}\right]$, it has been possible to obtain a crucial universal diagram which is reproduced in Fig.\ref{fintkin}. In the limit of $B\rightarrow 10^{-8}$ or smaller almost all $E_{dya}$ is emitted in the P-GRB and a negligible fraction is emitted in the kinetic energy $E_{Baryons}$ of the baryonic matter and therefore in the afterglow. On the other hand in the limit $B\rightarrow 10^{-2}$ which is also the limit of validity of our theoretical framework, almost all $E_{dya}$ is transferred to $E_{Baryons}$ and gives origin to the afterglow and the intensity of the P-GRB correspondingly decreases. We have identified the limiting case of negligible values of $B$ with the process of emission of the so called ``short bursts''. A complementary result reinforcing such an identification comes from the thermodynamical properties of the P-GRB: the hardness of the spectrum decreases for increasing values of $B$, see Fig.~\ref{energypeak}.

The injector phase is concluded by the emission of the P-GRB and the ABM pulse, as the condition of transparency is reached. 

The {\bf beam-target phase}, in which the accelerated baryonic matter (ABM) generated in the injector phase collides with the ISM, gives origin to the afterglow. Again for simplicity we have adopted a minimum set of assumptions:

\begin{enumerate}
\item The ABM pulse is assumed to collide with a constant homogeneous interstellar medium of number density $n_{\rm ism} \sim 1 {\mathrm cm}^{-3}$. The energy emitted in the collision is assumed to be instantaneously radiated away (fully radiative condition). The description of the collision and emission process is done using spherical symmetry, taking only the radial approximation neglecting all the delayed emission due to off-axis scattered radiation.
\item Special attention is given to numerically compute the power of the afterglow as a function of the arrival time using the correct governing equations for the space-time transformations in line with the RSTT paradigm.
\item Finally some approximate solutions are adopted in order to determine the power law exponents of the afterglow flux and compare and contrast them with the observational results as well as with the alternative results in the literature. 
\end{enumerate}
In this paper we only consider the above mentioned radial approximation and a spherically symmetric distribution in order to concentrate on the role of the correct space-time transformations in the RSTT paradigm and illustrate their impact on the determination of the power law index of the afterglow. This topic has been seriously neglected in the literature. Details of the role of beaming and on the diffusion due to off-axis emission will be studied elsewhere.\cite{rbcfx02a_sub,rbcfx02b_beam}

We can now turn to the two eras of the beam-target phase:

{\itshape The Era IV}: the ultrarelativistic and relativistic regimes in the afterglow. In section~\ref{era4} the hydrodynamic relativistic equations governing the collision of the ABM pulse with the interstellar matter are given in the form of a set of finite difference equations to be numerically integrated. Expressions for the internal energy developed in the collision as well as for the gamma factor are given as a function of the mass of the swept up interstellar material and of the initial conditions. In section~\ref{approximation} the infinitesimal limit of these equations is given as well as analytic power-law expansions in selected regimes.

{\itshape The Era V}: the approach to the nonrelativistic regimes in the afterglow. In section~\ref{era5} it is stressed that this last era often discussed in the current literature can be described by the same equations used for era IV. 

Having established all the governing equations for all the eras of the EMBH theory, we can proceed to compare and contrast the predictions of this theory with the observational data.

\subsection{The Best fit of the EMBH theory to the GRB~991216: the global features of the solution}

\begin{figure}[htbp]
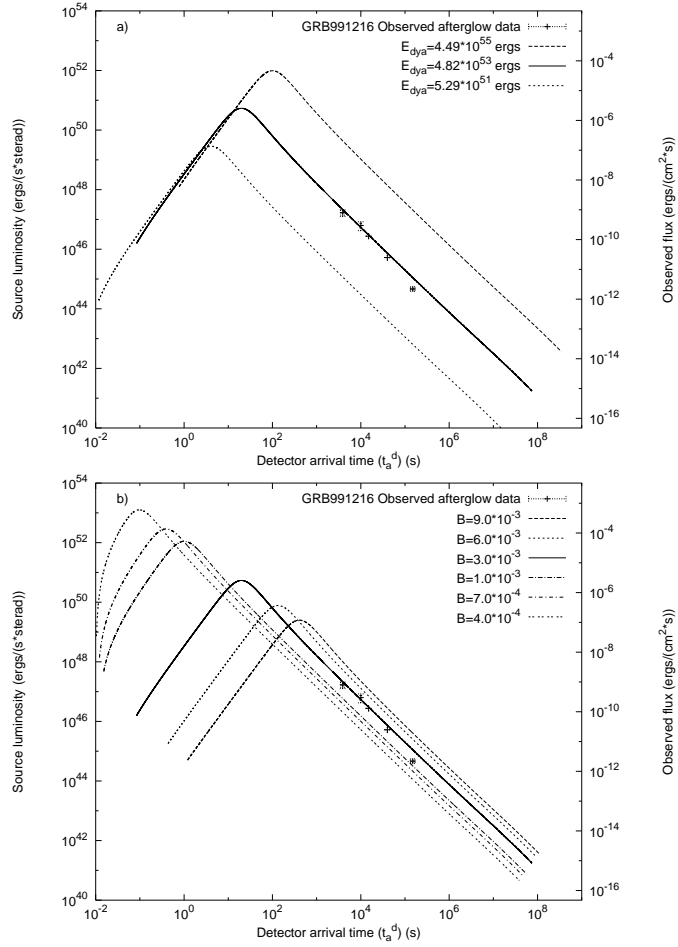

\PSFIG{ii-fig2}{9cm}{0}
\caption{a) Afterglow luminosity computed for an EMBH of $E_{dya}= 5.29\times 10^{51}$ erg, $E_{dya}= 4.83\times 10^{53}$ erg, $E_{dya}= 4.49 \times 10^{55}$ erg and $B=3\times 10^{-3}$. b) for the $E_{dya}= 4.83\times 10^{53}$, we give the afterglow luminosities corresponding respectively to $B=9\times 10^{-3}$, $6\times 10^{-3}$, $3\times 10^{-3}$, $1\times 10^{-3}$, $7\times 10^{-4}$, $4\times 10^{-4}$.}
\label{ii-fig2}
\end{figure}

As expressed in section~\ref{bf}, we have proceeded to the identification of the only two free parameters of the EMBH theory, $E_{dya}$ and $B$, by fitting the observational data from R-XTE and Chandra on the decaying part of the GRB~991216 afterglow. The afterglow appears to have three different parts: in the first part the luminosity increases as a function of the arrival time, it then reaches a maximum and finally monotonically decreases. In Fig.~\ref{ii-fig2}, we show how such a fit is actually made and how changing the two free parameters affects the intensity and the location in time of the peak of the afterglow. The best fit is obtained for $E_{dya}=4.83\times 10^{53}\,erg$ and $B=3\times 10^{-3}$. 

\begin{figure}[htbp]
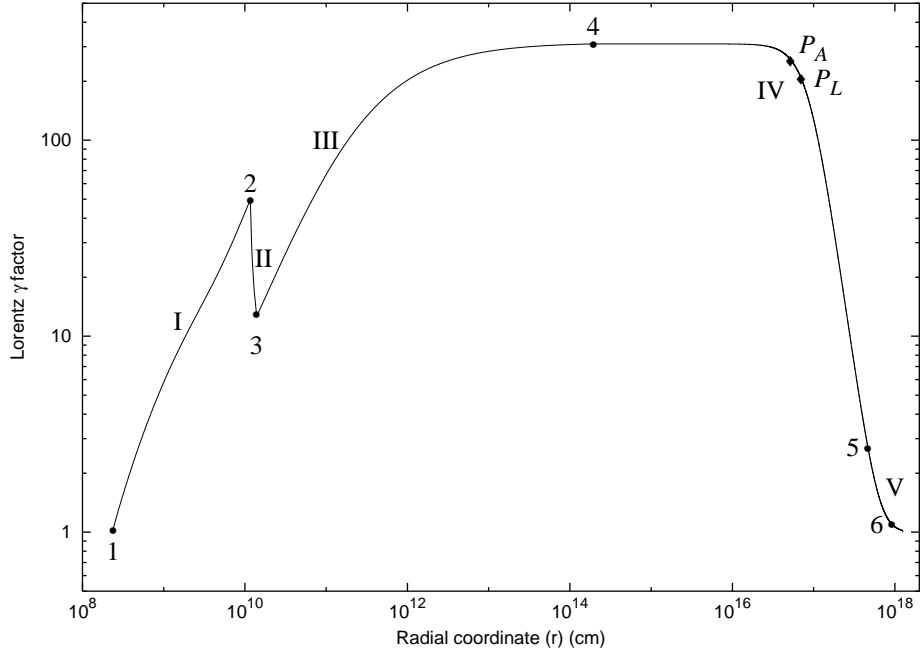

\PSFIG{gamma}{\hsize}{0}
\caption{The theoretically computed gamma factor for the parameter values $E_{dya}=4.83\times 10^{53}$ erg, $B=3\times 10^{-3}$ is given as a function of the radial coordinate in the laboratory frame. The corresponding values in the comoving time, laboratory time and arrival time are given in Tab.~\ref{tab1}. The different eras indicated by roman numerals are illustrated in the text (see sections~\ref{era1},\ref{era2},\ref{era3},\ref{era4},\ref{era5}), while the points 1,2,3,4,5 mark the beginning and end of each of these eras. The points $P_L$ and $P_A$ mark the maximum of the afterglow flux, respectively in emission time and in arrival time$^2$ (see and sections~\ref{era4},\ref{approximation}). The point 6 is the beginning of Phase D in era V (see sections~\ref{era5},\ref{approximation}). At point 4 the transparency condition is reached and the P-GRB is emitted.}
\label{gamma}
\end{figure}

\begin{figure}[htbp]
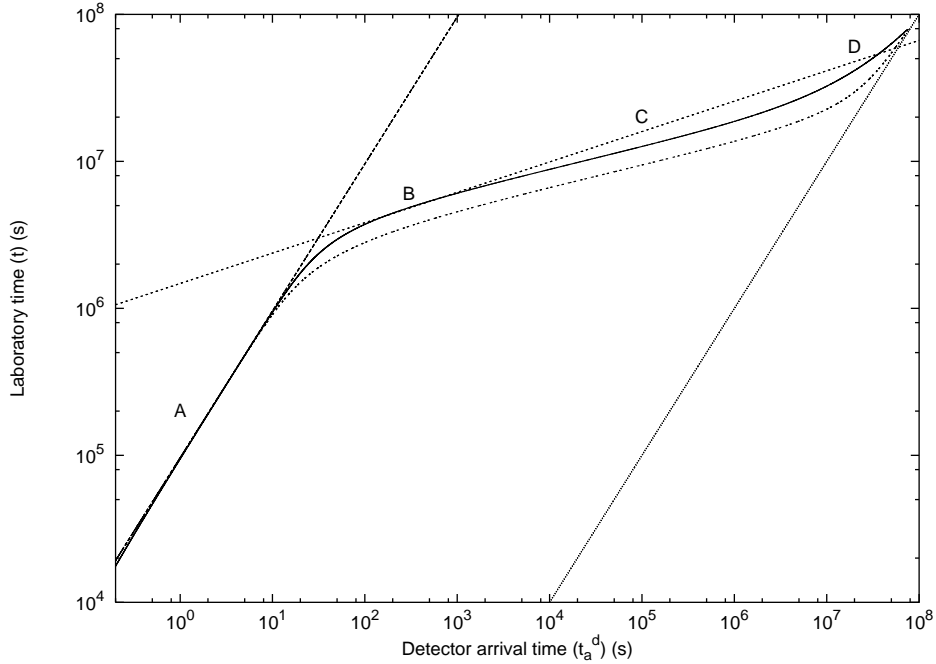

\PSFIG{tvsta}{\hsize}{0}
\caption{Relation between the arrival time ($t_a^d$) measured at the detector and the laboratory time ($t$) measured at the GRB source. The solid curve is computed using the exact formula given in Eq.(\ref{tadef}). The dashed-dotted curve is computed using the approximate formula given in Eq.(\ref{taapp}) and often used in the current literature. We distinguish four different phases. {\bf Phase A}: There is a linear relation between $t$ and $t_a^d$, given by Eq.(\ref{appA}) in the text (dashed line). {\bf Phase B}: There is an ``effective'' power-law relation between $t$ and $t_a^d$, given by Eq.(\ref{appC}) (dotted line). {\bf Phase C}: No analytic formula holds and the relation between $t$ and $t_a^d$ has to be directly computed by the integration of the complete equations of energy and momentum conservation (Eqs.(\ref{heat},\ref{dgamma})). {\bf Phase D}: As the gamma factor approaches $\gamma=1$, the relation between $t$ and $t_a^d$ asymptotically goes to $t=t_a^d$ (light gray line).$^1$}
\label{tvsta}
\end{figure}

Having determined the two free parameters of the theory, we have integrated the governing equations corresponding to these values and then obtained for the first time the complete history of the gamma factor from the moment of gravitational collapse to the latest phases of the afterglow observations (see Fig.~\ref{gamma}). We have also determined the different regimes encountered in the relation between the laboratory time and the detector arrival time within the RSTT paradigm (see Fig.~\ref{tvsta}). We have thus determined the entire space-time grid of the GRB~991216 by giving (see Tab.~\ref{tab1}) the radial coordinate of the GRB phenomenon as a function of the four coordinate time variables. A quick glance to Tab.~\ref{tab1} shows how the extreme relativistic regimes at work lead to enormous superluminal behaviour (up to $10^5 c$!) if the classical astrophysical concepts are adopted using the arrival time as the independent variable. In turn this implies that any causal relation based on classical astrophysics and the arrival time data, as often found in the current GRB literature, is incorrect.

\begin{table}[htbp]
\ttbl{30pc}{Gamma factors for selected events and their space-time coordinates. The points marked 1,2,3,4,5,6,$P_L$,$P_A$ are the same reported in Fig.~\ref{gamma}, while the point $F$ is the endpoint of the simulation. It is particularly important to read the last column, where the apparent motion in the radial coordinate, evaluated in the arrival time at the detector, leads to an enormous ``superluminal'' behaviour, up to $9.55\times 10^4\,c$. This illustrates well the impossibility of using such a classical estimate in regimes with gamma factors up to $310.1$.}
{\tiny
\begin{tabular}{c|l|l|l|l|l|r|c}
\multicolumn{8}{c}{ }\\
 Point & $r$ (cm)& $\tau$(s) & $t$(s) & $t_a$(s) & $t_a^d$ (s)& $\gamma$ & $\begin{array}{c} {\rm ``Superluminal"} \\ v\equiv\frac{r}{t_a^d} \\ \\ \end{array}$\\
\hline \hline
\multicolumn{8}{c}{ }\\
\multicolumn{8}{c}{{\bf The Injector Phase}}\\
\hline
 & & & & & & & \\
1 & $2.354\times10^8$    & $0.0$                & $0.0$                & $0.0$                & $0.0$                & $1.000$ & $0$\\
  & $1.871\times10^9$    & $1.550\times10^{-2}$ & $5.886\times10^{-2}$ & $4.312\times10^{-3}$ & $8.625\times10^{-3}$ & $10.08$ & $7.23c$\\
  & $4.486\times10^9$    & $2.141\times10^{-2}$ & $1.463\times10^{-1}$ & $4.523\times10^{-4}$ & $9.046\times10^{-3}$ & $20.26$ & $16.5c$\\
  & $7.080\times10^9$    & $2.485\times10^{-2}$ & $2.329\times10^{-1}$ & $4.594\times10^{-3}$ & $9.187\times10^{-3}$ & $30.46$ & $25.7c$\\
  & $9.533\times10^9$    & $2.715\times10^{-2}$ & $3.148\times10^{-1}$ & $4.627\times10^{-3}$ & $9.253\times10^{-3}$ & $40.74$ & $34.4c$\\
  & $1.162\times10^{10}$ & $2.868\times10^{-2}$ & $3.845\times10^{-1}$ & $4.644\times10^{-3}$ & $9.288\times10^{-3}$ & $49.70$ & $41.7c$\\
 & & & & & & & \\
\hline
 & & & & & & & \\
2 & $1.162\times10^{10}$ & $2.868\times10^{-2}$ & $3.845\times10^{-1}$ & $4.644\times10^{-3}$ & $9.288\times10^{-3}$ & $49.70$& $41.7c$\\
  & $1.186\times10^{10}$ & $2.889\times10^{-2}$ & $3.923\times10^{-1}$ & $4.646\times10^{-3}$ & $9.292\times10^{-3}$ & $38.06$& $42.6c$\\
  & $1.234\times10^{10}$ & $2.949\times10^{-2}$ & $4.083\times10^{-1}$ & $4.655\times10^{-3}$ & $9.311\times10^{-3}$ & $24.21$& $44.2c$\\
  & $1.335\times10^{10}$ & $3.144\times10^{-2}$ & $4.423\times10^{-1}$ & $4.706\times10^{-3}$ & $9.413\times10^{-3}$ & $15.14$& $47.3c$\\
  & $1.389\times10^{10}$ & $3.279\times10^{-2}$ & $4.603\times10^{-1}$ & $4.753\times10^{-3}$ & $9.506\times10^{-3}$ & $12.94$& $48.7c$\\
  & & & & & & & \\
\hline
 & & & & & & & \\
3 & $1.389\times10^{10}$ & $3.279\times10^{-2}$ & $4.603\times10^{-1}$ & $4.753\times10^{-3}$ & $9.506\times10^{-3}$ & $12.94$& $48.7c$\\
  & $2.326\times10^{10}$ & $5.208\times10^{-2}$ & $7.733\times10^{-1}$ & $5.369\times10^{-3}$ & $1.074\times10^{-2}$ & $20.09$& $72.2c$\\
  & $6.913\times10^{10}$ & $9.694\times10^{-2}$ & $2.304$ &              $6.086\times10^{-3}$ & $1.217\times10^{-2}$ & $50.66$& $1.89\times10^2c$\\ 
  & $1.861\times10^{11}$ & $1.486\times10^{-1}$ & $6.206$ &              $6.446\times10^{-3}$ & $1.289\times10^{-2}$ & $100.1$& $4.82\times10^2c$\\
  & $9.629\times10^{11}$ & $3.112\times10^{-1}$ & $32.12$ &              $6.978\times10^{-3}$ & $1.396\times10^{-2}$ & $200.3$& $2.30\times10^3c$\\
  & $3.205\times10^{13}$ & $3.958$ &              $1.069\times10^{ 3}$ & $1.343\times10^{-2}$ & $2.685\times10^{-2}$ & $300.1$& $3.98\times10^4c$\\
  & $1.943\times10^{14}$ & $21.57$ &              $6.481\times10^{ 3}$ & $4.206\times10^{-2}$ & $8.413\times10^{-2}$ & $310.1$& $7.70\times10^4c$\\
& & & & & & & \\
\hline \hline
\multicolumn{8}{c}{ }\\
\multicolumn{8}{c}{{\bf The Beam-Target Phase}}\\
\hline
 & & & & & & & \\
4     & $1.943\times10^{14}$ & $21.57$ &             $6.481\times10^{3}$ & $4.206\times10^{-2}$& $8.413\times10^{-2}$ & $310.1$& $7.70\times10^4c$\\
      & $6.663\times10^{15}$ & $7.982\times10^{2}$ & $6.481\times10^{3}$ & $1.164$ &             $2.328$ &              $310.0$& $9.55\times10^4c$\\
      & $2.863\times10^{16}$ & $3.114\times10^{3}$ & $9.549\times10^{5}$ & $5.057$ &             $10.11$ &              $300.0$& $9.45\times10^4c$\\
      & $4.692\times10^{16}$ & $5.241\times10^{3}$ & $1.565\times10^{6}$ & $8.775$ &             $17.55$ &              $270.0$& $8.92\times10^4c$\\
$P_A$ & $5.177\times10^{16}$ & $5.853\times10^{3}$ & $1.727\times10^{6}$ & $9.933$ &             $19.87$ &              $258.5$& $8.69\times10^4c$\\
      & $5.878\times10^{16}$ & $6.791\times10^{3}$ & $1.961\times10^{6}$ & $11.82$ &             $23.63$ &              $240.0$& $8.30\times10^4c$\\
      & $6.580\times10^{16}$ & $7.811\times10^{3}$ & $2.195\times10^{6}$ & $14.03$ &             $28.06$ &              $220.0$& $7.82\times10^4c$\\
$P_L$ & $7.025\times10^{16}$ & $8.506\times10^{3}$ & $2.343\times10^{6}$ & $15.66$ &             $31.32$ &              $207.0$& $7.48\times10^4c$\\
      & $7.262\times10^{16}$ & $8.895\times10^{3}$ & $2.422\times10^{6}$ & $16.61$ &             $33.23$ &              $200.0$& $7.29\times10^4c$\\
      & $9.058\times10^{16}$ & $1.236\times10^{4}$ & $3.021\times10^{6}$ & $26.66$ &             $53.32$ &              $150.0$& $5.67\times10^4c$\\
      & $1.136\times10^{17}$ & $1.866\times10^{4}$ & $3.788\times10^{6}$ & $52.84$ &             $1.057\times10^{2}$ &  $100.0$& $3.58\times10^4c$\\
      & $1.539\times10^{17}$ & $3.819\times10^{4}$ & $5.134\times10^{6}$ & $2.000\times10^{2}$ & $4.000\times10^{2}$ &  $50.02$& $1.28\times10^4c$\\
      & $2.801\times10^{17}$ & $2.622\times10^{5}$ & $9.351\times10^{6}$ & $7.278\times10^{3}$ & $1.455\times10^{4}$ &  $10.00$& $6.42\times10^2c$\\
      & $3.624\times10^{17}$ & $6.702\times10^{5}$ & $1.213\times10^{7}$ & $3.860\times10^{4}$ & $7.719\times10^{4}$ &  $5.001$& $1.57\times10^2c$\\
      & $4.454\times10^{17}$ & $1.433\times10^{6}$ & $1.500\times10^{7}$ & $1.439\times10^{5}$ & $2.877\times10^{5}$ &  $2.998$& $51.6c$\\
 & & & & & & & \\
\hline
 & & & & & & & \\
5 & $4.454\times10^{17}$ & $1.433\times10^{6}$ & $1.500\times10^{7}$ & $1.439\times10^{5}$ & $2.877\times10^{5}$ &  $2.998$& $51.6c$\\
  & $4.830\times10^{17}$ & $1.928\times10^{6}$ & $1.635\times10^{7}$ & $2.381\times10^{5}$ & $4.762\times10^{5}$ &  $2.500$& $33.8c$\\
  & $5.390\times10^{17}$ & $2.873\times10^{6}$ & $1.844\times10^{7}$ & $4.643\times10^{5}$ & $9.285\times10^{5}$ &  $2.000$& $19.4c$\\
  & $6.422\times10^{17}$ & $5.387\times10^{6}$ & $2.271\times10^{7}$ & $1.291\times10^{6}$ & $2.581\times10^{6}$ &  $1.500$& $8.30c$\\
  & $1.034\times10^{18}$ & $2.903\times10^{7}$ & $5.002\times10^{7}$ & $1.552\times10^{7}$ & $3.103\times10^{7}$ &  $1.054$& $1.11c$\\
 & & & & & & & \\
\hline
 & & & & & & & \\
6 & $1.034\times10^{18}$ & $2.903\times10^{7}$ & $5.002\times10^{7}$ & $1.552\times10^{7}$ & $3.103\times10^{7}$ &  $1.054$& $1.11c$\\
  & $1.202\times10^{18}$ & $4.979\times10^{7}$ & $7.150\times10^{7}$ & $3.140\times10^{7}$ & $6.280\times10^{7}$ &  $1.025$& $6.38\times10^{-1}c$\\
 & & & & & & & \\
\hline
 & & & & & & & \\
$F$ & $1.248\times10^{18}$ & $5.706\times10^{7}$ & $7.894\times10^{7}$ & $3.731\times10^{7}$ & $7.461\times10^{7}$ & $1.000$& $5.58\times10^{-1}c$\\
 & & & & & & & \\
\hline
\multicolumn{3}{c}{}\\
\end{tabular}
}
\label{tab1}
\end{table}

\subsection{The explanation of the ``long bursts'' and the identification of the proper gamma ray burst(P-GRB)}

\begin{figure}[htbp]
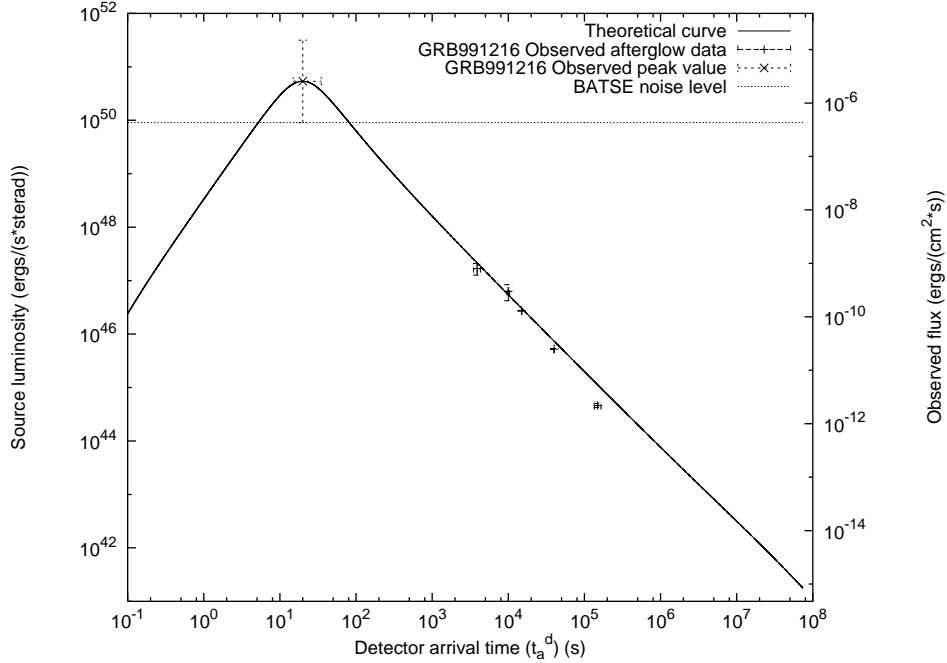

\PSFIG{ii-fig3}{\hsize}{0}
\caption{Best fit of the afterglow data of Chandra, RXTE as well as of the range of variability of the BATSE data on the major burst, by a unique afterglow curve leading to the parameter values $E_{dya}=4.83\times 10^{53}erg, B=3\times 10^{-3}$. The horizontal dotted line indicates the BATSE noise threshold. On the left axis the luminosity is given in units of the energy emitted at the source, while the right axis gives the flux as received by the detectors.}
\label{fit_1}
\end{figure}

In section~\ref{shortlongburst}, having determined the two free parameters of the EMBH theory, we analyze the theoretical predictions of this theory for the general structure of GRBs. The first striking result, illustrated in Fig.~\ref{fit_1}, shows that the peak of the afterglow emission coincides both in intensity and in arrival time ($19.87\,s$) with the average emission of the long burst observed by BATSE. For this we have introduced the new concept of {\em extended afterglow peak emission (E-APE)}. Once the proper space-time grid is given (see Tab.~\ref{tab1}) it is immediately clear that the E-APE is generated at distances of $~5\times 10^{16}$ cm from the EMBH. The long bursts are then identified with the E-APEs and are not bursts at all: they have been interpreted as bursts only because of the high threshold of the BATSE detectors (see Fig.~\ref{fit_1}). Thus the long standing unsolved problem of explaining the long GRBs\cite{wmm96,swm00,p01} is radically resolved.

\begin{figure}[htbp]
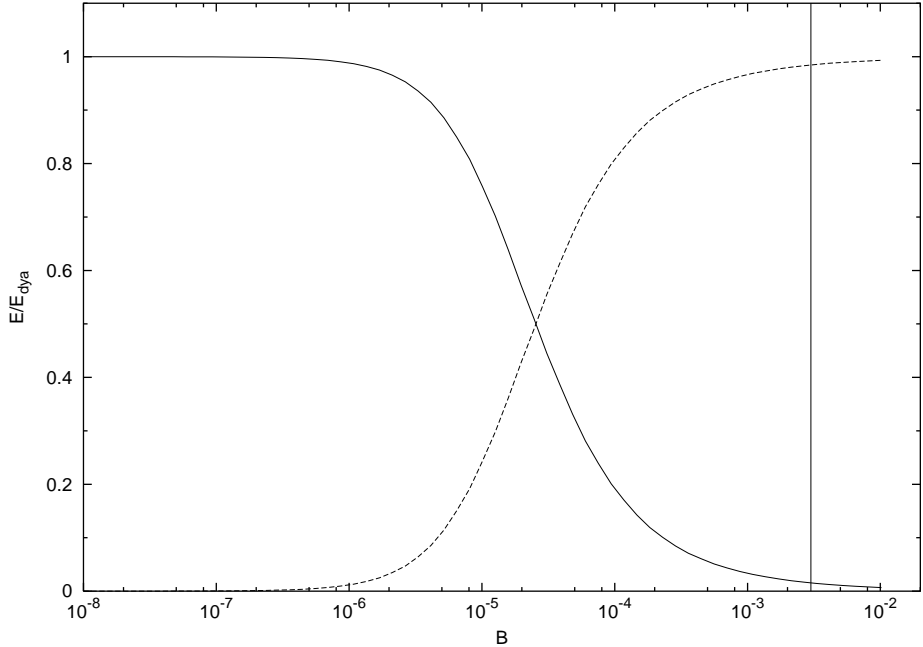

\PSFIG{ii-fig4}{\hsize}{0}
\caption{Relative intensities of the E-APE (dashed line) and the P-GRB (solid line), as predicted by the EMBH theory corresponding to the values of the parameters determined in Fig.~\ref{fit_1}, as a function of $B$. Details are given in section~\ref{shortlongburst}. The vertical line corresponds to the value $B=3\times 10^{-3}$.}
\label{crossen}
\end{figure}

\begin{figure}[htbp]
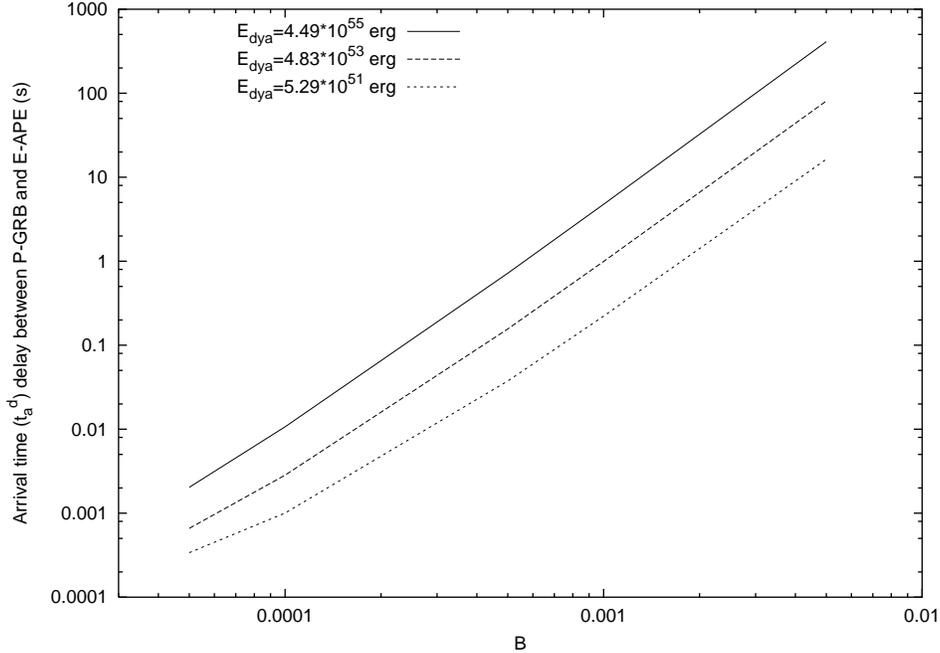

\PSFIG{dtab}{\hsize}{0}
\caption{The arrival time delay between the P-GRB and the peak of the E-APE is plotted as a function of the $B$ parameter for three selected values of $E_{dya}$.}
\label{dtab}
\end{figure}

Still in section~\ref{shortlongburst}, the search for the identification of the P-GRB in the BATSE data is described. This identification is made using the two fundamental diagrams shown in Fig.~\ref{crossen} and Fig.~\ref{dtab}. Having established the value of $E_{dya}=4.83\times 10^{53}\, erg$ and of $B=3\times 10^{-3}$, it is possible from the dashed line and the solid line in Fig.~\ref{crossen} to evaluate the ratio of the energy $E_{P\hbox{-}GRB}$ emitted in the P-GRB to the energy $E_{Baryons}$ emitted in the afterglow corresponding to the determined value of $B$, see the vertical line in Fig.~\ref{crossen}. We obtain $E_{P\hbox{-}GRB}/E_{Baryons}=1.58\times 10^{-2}$, which gives $E_{P\hbox{-}GRB}=7.54\times 10^{51}\, erg$. Having so determined the theoretically expected intensity of the P-GRB, a second fundamental observable parameter, which is also a function of $E_{dya}$ and $B$, is the arrival time delay between the P-GRB and the peak E-APE, determined in Fig.~\ref{dtab}. From Tab.~\ref{tab1}, we have that the detector arrival time of the P-GRB occurs at $8.41\times 10^{-2}\, s$, corresponding to a radial coordinate of $1.94\times 10^{14}\,cm$, a comoving time of $21.57\, s$, a laboratory time of $6.48\times 10^3\, s$ and an arrival time of $4.21\times 10^{-2}\, s$. At this point, the gamma factor is $310.1$. The peak of the E-APE occurs at a detector arrival time of $19.87\, s$, corresponding to a radial coordinate of $5.18\times 10^{16}\, cm$, a comoving time of $5.85\times 10^3\,s$, a laboratory time of $1.73\times 10^6\,s$ and an arrival time of $9.93\,s$ (see Tab.~\ref{tab1}). The delay between the P-GRB and the peak of the E-APE is therefore $19.78\,s$, see Fig.~\ref{dtab}. The theoretical prediction on the intensity and the arrival time uniquely identifies the P-GRB with the ``precursor'' in the GRB~991216 (see Fig.~\ref{grb991216}). Moreover, the hardness of the P-GRB spectra is also evaluated in this section. As pointed out in the conclusions, the fact that both the absolute and relative intensities of the P-GRB and E-APE have been predicted within a few percent accuracy as well as the fact that their arrival time has been computed with the precision of a few tenths of milliseconds, see Tab.~\ref{tab1} and Fig. \ref{final}, can be considered one of the major successes of the EMBH theory. 

\subsection{On the power-laws, beaming and temporal structures in the afterglow of GRB~991216.}

In section~\ref{approximation} a piecewise description of the afterglow by the expansion of the fundamental hydrodynamical equations given by Taub (1948)\cite{taub} and Landau \& Lifshitz\cite{ll} have allowed the determination of a power-law index for the dependence of the afterglow luminosity on the photon arrival time at the detector. It is evident that the determination of the power-law index is very sensitive to the basic assumptions made for the description of the afterglow, as well as to the relations between the different temporal coordinates which have been clarified by the RSTT paradigm.\cite{lett1} The different power-law indices obtained are compared and contrasted with the ones in the current literature (see Tab.~\ref{tab2}). As a byproduct of this analysis, see also the conclusions, there is a perfect agreement between the observational data and the theoretical predictions, implying that the assumptions adopted for the description of the afterglow are valid and therefore that there is no evidence for a beamed emission in GRB~991216.

In section~\ref{substructures} the role of the inhomogeneities in the interstellar matter has been analysed in order to explain the observed temporal substructures in the BATSE data on GRB~991216. From the data of Tab.~\ref{tab1} and the highly ``superluminal'' behaviour of the source in the region of the E-APE, it is concluded that the observed time variability in the intensity of the emission $\left(\Delta I/ \overline{I} \right)\sim 5$ can be traced to inhomogeneities in the interstellar matter: $\left(\Delta n_{ism}/n_{ism}\right)\sim 5$. The typical size of the scattering region is estimated to be $5\times 10^{16}\, cm$, and these are the typical sizes and density contrasts found in interstellar clouds. Since the emission of the E-APE occurs at typical dimensions of the order of $5\times 10^{16}\, cm$, the observed inhomogeneities are probing the structure of the interstellar medium, and have nothing to do with the ``inner engine'' of the source. These conclusions, reached in the radial approximation of the afterglow adopted in this article, have been proved to hold in the more general case when off-radial emission is taken into account.\cite{rbcfx02a_sub,rbcfx02b_beam}

\subsection{The observation of the iron lines in GRB~991216: on a possible GRB-supernova time sequence}

In section~\ref{gsts} the program of using GRBs to further explore the region surrounding the newly formed EMBH is carried one step further by using the observations of the emitted iron lines.\cite{p00} This gives us the opportunity to introduce the GRB-supernova time sequence (GSTS) paradigm and to introduce as well the novel concept of an {\em induced supernova explosion}. The GSTS paradigm reads: {\em A massive GRB-progenitor star $P_1$ of mass $M_1$ undergoes gravitational collapse to an EMBH. During this process a dyadosphere is formed and subsequently the P-GRB and the E-APE are generated in sequence. They propagate and impact, with their photon and neutrino components, on a second supernova-progenitor star $P_2$ of mass $M_2$. Assuming that both stars were generated approximately at the same time, we expect to have $M_2 < M_1$. Under some special conditions of the thermonuclear evolution of the supernova-progenitor star $P_2$, the collision of the P-GRB and the E-APE with the star $P_2$ can induce its supernova explosion}.

Using the result presented in Tab.~\ref{tab1} and in all preceding sections, the GSTS paradigm is illustrated in the case of GRB~991216. Some general considerations on the nature of the supernova progenitor star are also advanced.

Some general considerations on the EMBH formation are presented in section~\ref{gc}. The general conclusions are presented in section~\ref{conclusions}.

\begin{figure}[htbp]
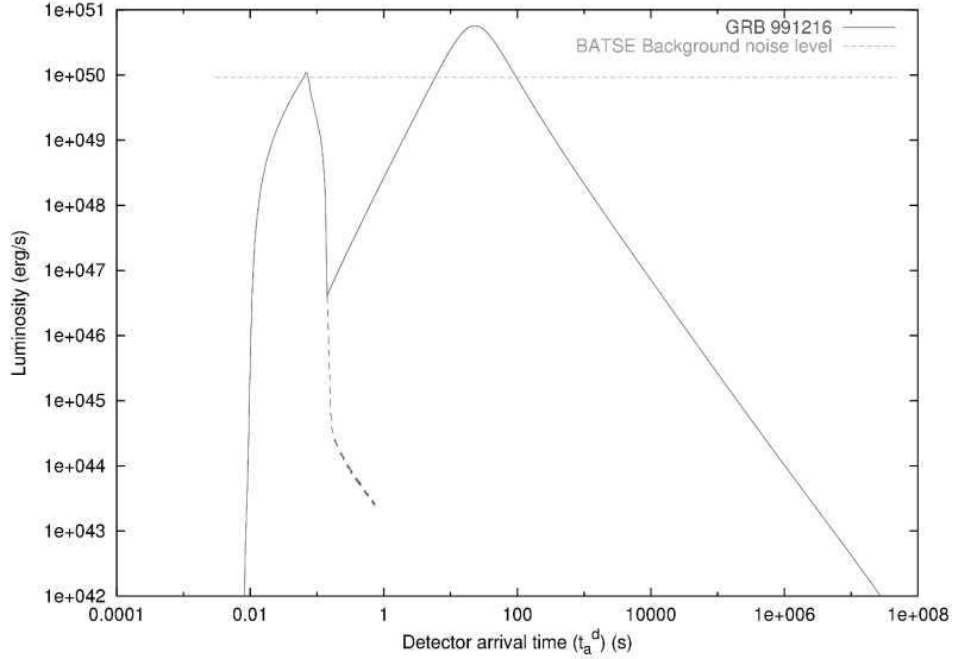

\PSFIG{final}{\hsize}{0}
\caption{A qualitative diagram showing the full picture of the model, with both P-GRB and E-APE.}
\label{final}
\end{figure}

The understanding of all these points has led to the formulation of the second part, namely the sufficient condition of the RSTT paradigm which reads: ``the necessary condition in order to interpret the GRB data, given in terms of the arrival time at the detector, is the knowledge of the {\em entire} worldline of the source from the gravitational collapse. In order to meet this condition, given a proper theoretical description and the correct governing equations, it is sufficient to know the energy of the dyadosphere and the mass of the remnant of the progenitor star''.

\section{The zeroth era: the process of gravitational collapse and the formation of the dyadosphere}\label{dyadosphere}

We first recall the three theoretical results which lie at the basis of the EMBH theory.

In 1971 in the article {\itshape ``Introducing the Black Hole''},\cite{rw71} the theorem was advanced that the most general black hole is characterized uniquely by three independent parameters: the mass-energy $M$, the angular momentum $L$ and the charge  $Q$ making it an EMBH. Such an ansatz, which came to be known as the ``uniqueness theorem'' has turned out to be one of the most difficult theorems to be proven in all of physics and mathematics. The progress in the proof has been authoritatively summarized by Carter (1997).\cite{c97} The situation can be considered satisfactory from the point of view of the physical and astrophysical considerations. Nevertheless some fundamental mathematical and physical issues concerning the most general perturbation analysis of an EMBH are still the topic of active scientific discussion.\cite{bcjr02}
 
In 1971 it was shown that the energy extractable from an EMBH is governed by the mass-energy formula,\cite{cr71}
\begin{equation}
E_{BH}^2=M^2c^4=\left(M_{\rm ir}c^2 + {Q^2\over2\rho_+}\right)^2+{L^2c^2\over \rho_+^2},\label{em}
\end{equation}
with
\begin{equation}
{1\over \rho_+^4}\left({G^2\over c^8}\right)\left( Q^4+4L^2c^2\right)\leq 1,
\label{s1}
\end{equation}
where
\begin{equation}
S=4\pi \rho_+^2=4\pi(r_+^2+{L^2\over c^2M^2})=16\pi\left({G^2\over c^4}\right) M^2_{\rm ir},
\label{sa}
\end{equation}
is the horizon surface area, $M_{\rm ir}$ is the irreducible mass, $r_{+}$ is the horizon radius and $\rho_+$ is the quasi-spheroidal cylindrical coordinate of the horizon evaluated at the equatorial plane. Extreme EMBHs satisfy the equality in Eq.(\ref{s1}). Up to 50\% of the mass-energy of an extreme EMBH can in principle be extracted by a special set of transformations: the reversible transformations.\cite{cr71}

In 1975, generalizing some previous results of Zaumen (1975)\cite{za75} and Gibbons (1975),\cite{gb75} Damour \& Ruffini (1975)\cite{dr75} showed that the vacuum polarization process {\it \`a la} Heisenberg-Euler-Schwinger\cite{he35,s51} created by an electric field of strength larger than 
\begin{equation}
{\cal E}_c=\frac{m_e^2c^3}{\hbar e}
\label{ecrit}
\end{equation}
can indeed occur in the field of a Kerr-Newmann EMBH. Here $m_e$ and $e$ are respectively the mass and charge of the electron.  There Damour and Ruffini considered an axially symmetric EMBH, due to the presence of rotation, and limited themselves to EMBH masses larger then the upper limit of a neutron star for astrophysical applications. They purposely avoided all complications of black holes with mass smaller then the dual electron mass of the electron $\left(m_e^\star=\frac{c\hbar}{G m_e}=\frac{m_{Planck}^2}{m_e}\right)$ which may lead to quantum evaporation processes.\cite{h74} They pointed out that:
\begin{enumerate}
\item The vacuum polarization process can occur for an EMBH mass larger than the maximum critical mass for neutron stars all the way up to $7.2\times 10^6 M_\odot$. 
\item The process of pair creation occurs on very short time scales, typically $\frac{\hbar}{m_e c^2}$, and is an almost perfect reversible process, in the sense defined by Christodoulou-Ruffini, leading to a very efficient mechanism of extracting energy from an EMBH.
\item The energy generated by the energy extraction process of an EMBH was found to be of the order of $10^{54}$ erg, released almost instantaneously. They concluded at the time {\itshape ``this work naturally leads to a most simple model for the explanation of the recently discovered $\gamma$-ray bursts''}.
\end{enumerate}

After the discovery of the afterglow of GRBs and the determination of the cosmological distance of their sources we noticed the coincidence between the theoretically predicted energetics and the observed ones in Damour \& Ruffini (1975):\cite{dr75} we returned to our theoretical results developing some new basic theoretical concepts,\cite{rukyoto,prxprl,prx98,rswx99,rswx00} which have led to the EMBH theory.

As a first simplifying assumption we have developed our considerations in the absence of rotation with spherically symmetric distributions. The space-time is then described by the Reissner-Nordstr\"{o}m geometry, whose spherically symmetric metric is given by
\begin{equation}
d^2s=g_{tt}(r)d^2t+g_{rr}(r)d^2r+r^2d^2\theta +r^2\sin^2\theta
d^2\phi ~,
\label{s}
\end{equation}
where $g_{tt}(r)= - \left[1-{2GM\over c^2r}+{Q^2G\over c^4r^2}\right] \equiv - \alpha^2(r)$ and $g_{rr}(r)= \alpha^{-2}(r)$.

The first new result we obtained is that the pair creation process does not occur at the horizon of the EMBH: it extends over the entire region outside the horizon in which the electric field exceeds the critical value given by Eq. \ref{ecrit}. Since the electric field in the Reissner-Nordstr\"{o}m geometry has only a radial component given by\cite{r78}
\begin{equation}
{\cal E}\left(r\right)=\frac{Q}{r^2}\, ,
\label{edir}
\end{equation}
this region extends from the horizon radius
\begin{eqnarray}
r_{+}&=&1.47 \cdot 10^5\mu (1+\sqrt{1-\xi^2})\hskip0.1cm {\rm cm}
\label{r+}
\end{eqnarray}
out to an outer radius\cite{rukyoto} 
\begin{eqnarray}
r^\star&=&\left({\hbar\over mc}\right)^{1\over2}\left({GM\over
c^2}\right)^{1\over2} \left({m_{\rm p}\over m}\right)^{1\over2}\left({e\over
q_{\rm p}}\right)^{1\over2}\left({Q\over \sqrt{G}M}\right)^{1\over2}\nonumber\\
&=&1.12\cdot 10^8\sqrt{\mu\xi} \hskip0.1cm {\rm cm},
\label{rc}
\end{eqnarray}
where we have introduced the dimensionless mass and charge parameters $\mu={M\over M_{\odot}}$, $\xi={Q\over (M\sqrt{G})}\le 1$, see Fig.~\ref{dyaon}.

The second new result has been to realize that the local number density of electron and positron pairs created in this region as a function of radius is given by
\begin{equation}
n_{e^+e^-}(r) = {Q\over 4\pi r^2\left({\hbar\over
mc}\right)e}\left[1-\left({r\over r^\star}\right)^2\right] ~,
\label{nd}
\end{equation}
and consequently the total number of electron and positron pairs in this region is
\begin{equation}
N^\circ_{e^+e^-}\simeq {Q-Q_c\over e}\left[1+{
(r^\star-r_+)\over {\hbar\over mc}}\right],
\label{tn}
\end{equation}
\begin{figure}[htbp]
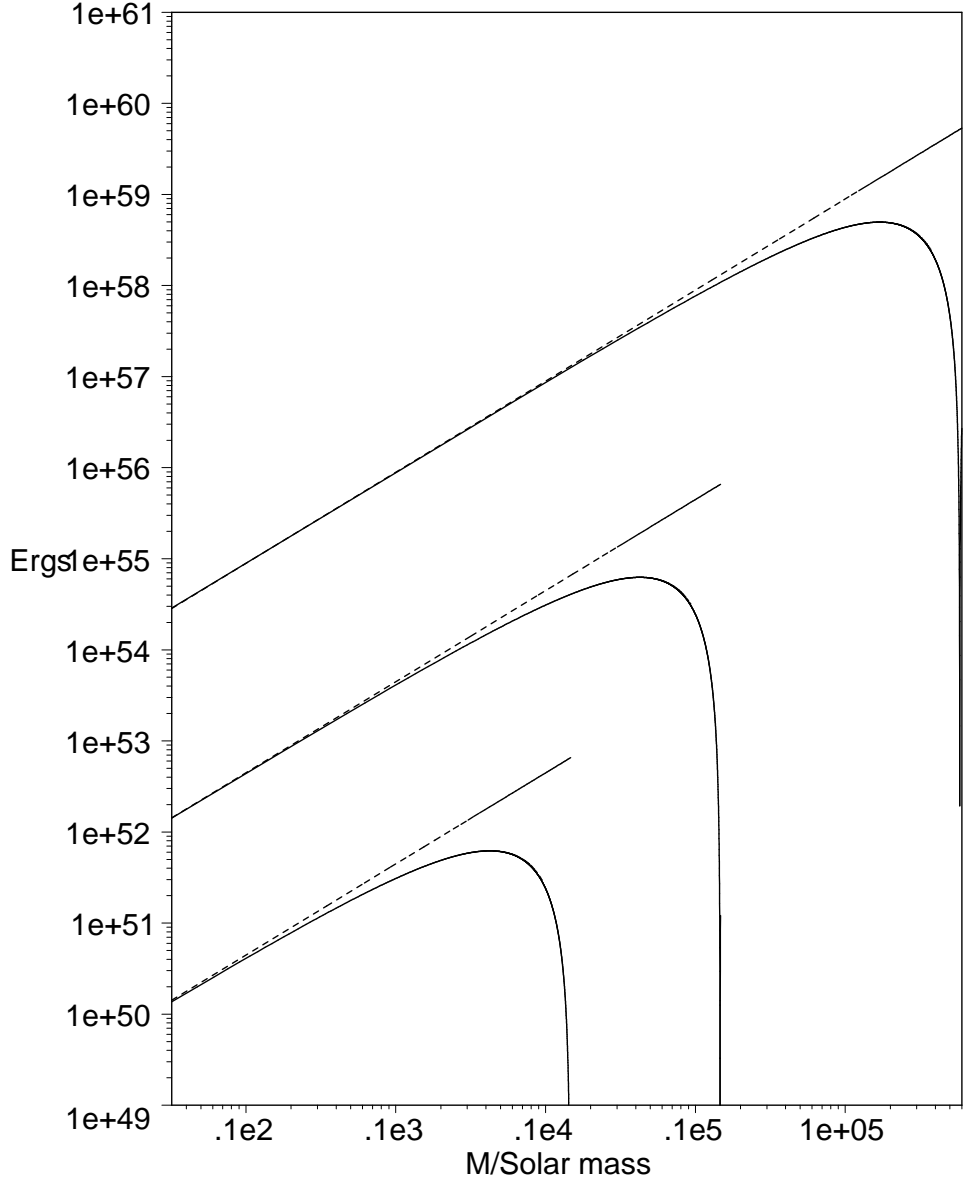

\PSFIG{prep}{\hsize}{0}
\caption{The energy extracted by the process of vacuum polarization is plotted (solid lines) as a function of the mass $M$ in solar mass units for selected values of the charge parameter $\xi=1,0.1,0.01$ (from top to bottom) for an EMBH, the case $\xi=1$ reachable only as a limiting process. For comparison we have also plotted the maximum energy extractable from an EMBH (dotted lines) given by eq.~(\ref{em}). Details in Preparata, Ruffini \& Xue (2001).$^{71}$}
\label{prep}
\end{figure}
where $Q_c={\cal E}_{\rm c}r_+^2$.

The total number of pairs is larger by an enormous factor $r^{\star}/\left(\hbar/mc\right) > 10^{18}$ than the value $Q/e$ which a naive estimate of the discharge of the EMBH would have predicted. Due to this enormous amplification factor in the number of pairs created, the region between the horizon and $r^{\star}$ is dominated by an essentially high density neutral plasma of electron-positron pairs. We have defined this region as the dyadosphere of the EMBH from the Greek duas, duadsos for pairs. Consequently we have called $r^\star$ the dyadosphere radius $r^\star \equiv r_{\rm ds}$.\cite{rukyoto,prxprl,prx98} The vacuum polarization process occurs as if the entire dyadosphere are subdivided into a concentric set of shells of capacitors each of thickness $\hbar/m_ec$ and each producing a number of $e^+e^-$ pairs on the order of $\sim Q/e$ (see Fig.~\ref{dyaon}). The energy density of the electron-positron pairs is given by
\begin{equation}
\epsilon(r) = {Q^2 \over 8 \pi r^4} \biggl(1 - \biggl({r \over
r_{\rm ds}}\biggr)^4\biggr) ~, \label{jayet}
\end{equation}
(see Figs.~2--3 of Preparata, Ruffini \& Xue, 1998a\cite{prxprl}). The total energy of pairs converted from the static electric energy and deposited within the dyadosphere is then
\begin{equation}
E_{\rm dya}={1\over2}{Q^2\over r_+}(1-{r_+\over r_{\rm ds}})\left[1-\left({r_+\over r_{\rm ds}}\right)^2\right] ~.
\label{tee}
\end{equation}

\begin{figure}[htbp]
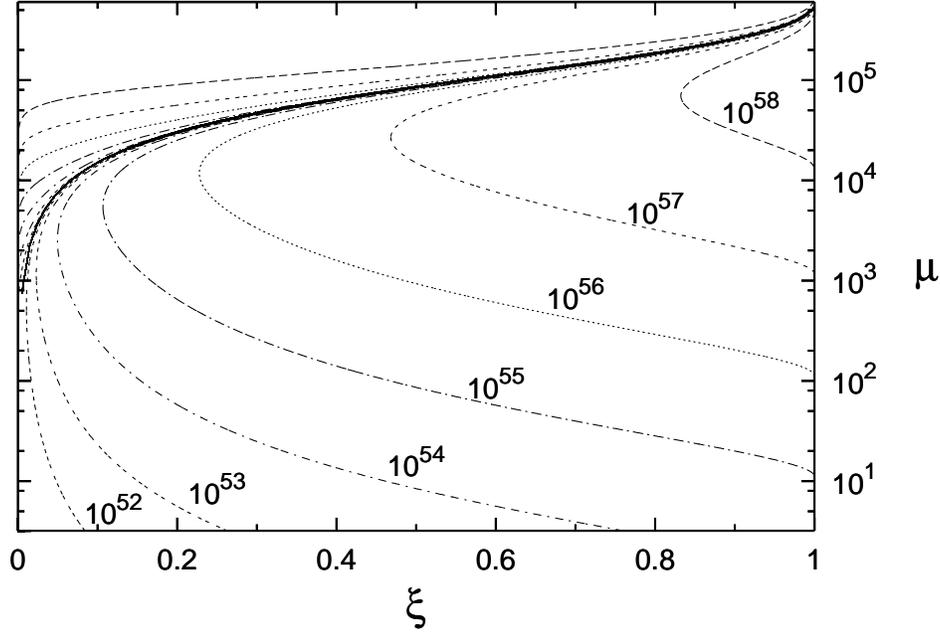

\PSFIG{muxi}{\hsize}{0}
\caption{Selected lines corresponding to fixed values of the $E_{dya}$ are given as a function of the two parameters $\mu$ $\xi$, only the solutions below the continuous heavy line are physically relevant.The configurations above the continuous heavy lines correspond to unphysical solutions with $r_{\rm ds} < r_+$  }
\label{muxi}
\end{figure}

As we will see in the following this is one of the two fundamental parameters of the EMBH theory (see Fig.~\ref{muxi}). In the limit ${r_+\over r_{\rm ds}}\rightarrow 0$, Eq.(\ref{tee}) leads to $E_{\rm dya}\rightarrow {1\over2}{Q^2\over r_+}$, which coincides with the energy extractable from EMBHs by reversible processes ($M_{\rm ir}={\rm const.}$), namely $E_{BH}-M_{\rm ir}={1\over2}{Q^2\over r_+}$,\cite{cr71} see Fig.~\ref{prep}. Due to the very large pair density given by Eq.(\ref{nd}) and to the sizes of the cross-sections for the process $e^+e^-\leftrightarrow \gamma+\gamma$, the system is expected to thermalize to a plasma configuration for which
\begin{equation}
n_{e^+}=n_{e^-} \sim n_{\gamma} \sim n^\circ_{e^+e^-},
\label{plasma}
\end{equation}
where $n^\circ_{e^+e^-}$ is the total number density of $e^+e^-$-pairs created in the dyadosphere.\cite{prxprl,prx98}

\begin{figure}[htbp]
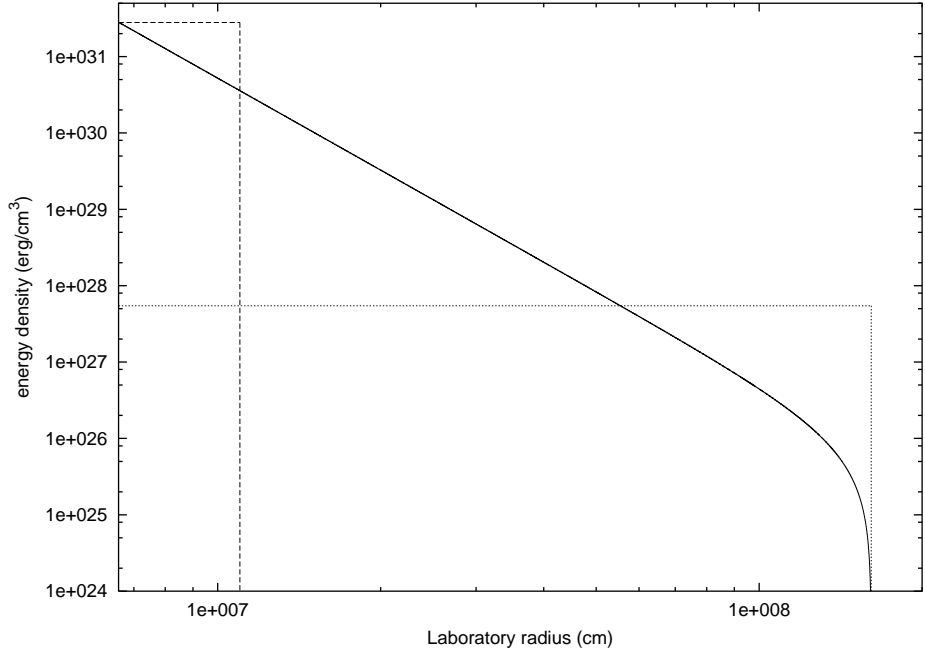

\PSFIG{dens}{\hsize}{0}
\caption{Two different approximations for the energy density profile inside the dyadosphere. The first one (dashed line) fixes the energy density equal to its peak value, and computes an ``effective'' dyadosphere radius accordingly. The second one (dotted line) fixes the dyadosphere radius to its correct value, and assumes an uniform energy density over the dyadosphere volume. The total energy in the dyadosphere is of course the same in both cases. The solid curve represents the real energy density profile.}
\label{dens}
\end{figure}

The third new result which we have introduced for simplicity is that for a given $E_{dya}$ we have assumed either a constant average energy density over the entire dyadosphere volume, or a more compact configuration with energy density equal to the peak value. These are the two possible initial conditions for the evolution of the dyadosphere (see Fig.~\ref{dens}).

These three old and three new theoretical results permit a good estimate of the general energetics processes originating in the dyadosphere, assuming an already formed EMBH. In reality, if the data become accurate enough, the full dynamical description of the dyadosphere formation mentioned above will be needed in order to follow all the general relativistic effects and characteristic time scales of the approach to the EMBH horizon\cite{crv02,rv02a,rv02b,rvx02}, see also section~\ref{gc}.

Below we shall concentrate on the dynamical evolution of the electron-positron  plasma created in the dyadosphere. We shall first examine in the next three sections the governing equations necessary to approach such a dynamical description. 

\section{The hydrodynamics and the rate equations for the plasma of $e^+e^-$-pairs}\label{hydro_pem}

The evolution of the $e^+e^-$-pair plasma generated in the dyadosphere has been treated in two papers.\cite{rswx99,rswx00} We recall here the basic governing equations in the most general case in which the plasma fluid is composed of $e^+e^-$-pairs, photons and baryonic matter. The plasma is described by the stress-energy tensor
\begin{equation}
T^{\mu\nu}=pg^{\mu\nu}+(p+\rho)U^\mu U^\nu\, ,
\label{tensor}
\end{equation}
where $\rho$ and $p$ are respectively the total proper energy density and pressure in the comoving frame of the plasma fluid and $U^\mu$ is its four-velocity, satisfying
\begin{equation}
g_{tt}(U^t)^2+g_{rr}(U^r)^2=-1 ~,
\label{tt}
\end{equation}
where $U^r$ and $U^t$ are the radial and temporal contravariant components of the 4-velocity.

The conservation law for baryon number can be expressed in terms of the proper baryon number density $n_B$
\begin{eqnarray}
(n_B U^\mu)_{;\mu}&=& g^{-{1\over2}}(g^{1\over2}n_B
U^\nu)_{,\nu}\nonumber\\
&=&(n_BU^t)_{,t}+{1\over r^2}(r^2 n_BU^r)_{,r}=0 ~.
\label{contin}
\end{eqnarray}
The radial component of the energy-momentum conservation law of the plasma fluid reduces to 
\begin{eqnarray}
&&{\partial p\over\partial r}+{\partial \over\partial t}\left((p+\rho)U^t U_r\right)+{1\over r^2} { \partial
\over \partial r}  \left(r^2(p+\rho)U^r U_r\right)\nonumber\\
&&-{1\over2}(p+\rho)\left[{\partial g_{tt}
 \over\partial r}(U^t)^2+{\partial g_{rr}
 \over\partial r}(U^r)^2\right] =0 ~.
\label{cmom2}
\end{eqnarray}
The component of the energy-momentum conservation law of the plasma fluid equation along a flow line is
\begin{eqnarray}
U_\mu(T^{\mu\nu})_{;\nu}&=&-(\rho U^\nu)_{;\nu}
-p(U^\nu)_{;\nu},\nonumber\\ &=&-g^{-{1\over2}}(g^{1\over2}\rho
U^\nu)_{,\nu} - pg^{-{1\over2}}(g^{1\over2} U^\nu)_{,\nu}\nonumber\\
&=&(\rho U^t)_{,t}+{1\over r^2}(r^2\rho
U^r)_{,r}\nonumber\\
&+&p\left[(U^t)_{,t}+{1\over r^2}(r^2U^r)_{,r}\right]=0 ~.
\label{conse1}
\end{eqnarray}

Defining the total proper internal energy density $\epsilon$ and the baryonic mass density $\rho_B$ in the comoving frame of the plasma fluid,
\begin{equation}
\epsilon \equiv \rho - \rho_B,\hskip0.5cm \rho_B\equiv n_Bmc^2 ~,
\label{cpp}
\end{equation} 
and using the law (\ref{contin}) of baryon-number conservation, from Eq. (\ref{conse1}) we have
\begin{equation}
(\epsilon U^\nu)_{;\nu} +p(U^\nu)_{;\nu}=0 ~.
\label{conse'}
\end{equation}
Recalling that ${dV\over d\tau}=V(U^\mu)_{;\mu}$, where $V$ is the comoving volume and $\tau$ is the proper time for the plasma fluid, we have along each flow line
\begin{equation}
{d(V\epsilon)\over d\tau}+p{dV\over d\tau}={dE\over d\tau}+p{dV\over
d\tau}=0 ~,
\label{f'}
\end{equation}
where $E=V\epsilon$ is the total proper internal energy of the plasma fluid. We express the equation of state by introducing a thermal index $\Gamma(\rho,T)$
\begin{equation}
\Gamma = 1 + { p\over \epsilon} ~.
\label{state}
\end{equation}

We now turn to the second set of governing equations describing the evolution of the $e^+e^-$ pairs. Letting $n_{e^-}$ and $n_{e^+}$  be the proper number densities of electrons and positrons associated with pairs and $n^b_{e^-}$ the proper number densities of ionized electrons, we clearly have
\begin{equation}
n_{e^-}=n_{e^+}=n_{\rm pair},\hskip0.5cm n^b_{e^-}=\bar Z n_B,
\label{eee}
\end{equation}
where $n_{\rm pair}$ is the number of $e^+e^-$ pairs and $\bar Z$ the average atomic number ${1\over2}<\bar Z< 1$ ($\bar Z=1$ for hydrogen atom and $\bar Z={1\over2}$ for general baryonic matter). The rate equation for electrons and positrons gives,
\begin{eqnarray}
(n_{e^+}U^\mu)_{;\mu}&=&(n_{e^+}U^t)_{,t}+{1\over r^2}(r^2 n_{e^+}U^r)_{,r}\nonumber\\
&=&\overline{\sigma v} \big[(n_{e^-}(T)+n^b_{e^-}(T))n_{e^+}(T)\nonumber\\
& - &(n_{e^-}+n^b_{e^-})n_{e^+}\big],
\label{e+contin}\\
(n_{e^-}U^\mu)_{;\mu}&=&(n_{e^-}U^t)_{,t}+{1\over r^2}(r^2 n_{e^-}U^r)_{,r}\nonumber\\
&=&\overline{\sigma v} \left[n_{e^-}(T)n_{e^+}(T) - n_{e^-}n_{e^+}\right],
\label{e-contin}\\
(n^b_{e^-}U^\mu)_{;\mu}&=&(n^b_{e^-}U^t)_{,t}+{1\over r^2}(r^2 n^b_{e^-}U^r)_{,r}\nonumber\\
&=&\overline{\sigma v} \left[n^b_{e^-}(T)n_{e^+}(T) - n^b_{e^-}n_{e^+}\right],
\label{tbe-contin}
\end{eqnarray}
where $\overline{\sigma v}$ is the mean of the product of the annihilation cross-section and the thermal velocity of the electrons and positrons, $n_{e^\pm}(T)$ are the proper number densities of electrons and positrons associated with the pairs, given by appropriate Fermi integrals with zero chemical potential, and $n^b_{e^-}(T)$ is the proper number density of ionized electrons, given by appropriate Fermi integrals with non-zero chemical potential $\mu_e$ at an appropriate equilibrium temperature $T$. These rate equations can be reduced to 
\begin{eqnarray}
(n_{e^\pm}U^\mu)_{;\mu}&=&(n_{e^\pm}U^t)_{,t}+{1\over r^2}(r^2 n_{e^\pm}U^r)_{,r}\nonumber\\
&=&\overline{\sigma v} \big[n_{e^-}(T)n_{e^+}(T)- n_{e^-}n_{e^+}\big],
\label{econtin}\\
(n^b_{e^-}U^\mu)_{;\mu}&=&(n^b_{e^-}U^t)_{,t}+{1\over r^2}(r^2 n^b_{e^-}U^r)_{,r}=0,
\label{becontin}\\
Frac&\equiv&{n_{e^\pm}\over n_{e^\pm}(T)}={n^b_{e^-}(T)\over n^b_{e^-}}.
\label{be-contin}
\end{eqnarray}
Equation (\ref{becontin}) is just the baryon-number conservation law (\ref{contin}) and (\ref{be-contin}) is a relationship satisfied by $n_{e^\pm}, n_{e^\pm}(T)$ and $n^b_{e^-}, n^b_{e^-}(T)$.

The equilibrium temperature $T$ is determined by the thermalization processes occurring in the expanding plasma fluid with a total proper energy density $\rho$ governed by the hydrodynamical equations (\ref{contin},\ref{cmom2},\ref{conse1}). We have
\begin{equation}
\rho = \rho_\gamma + \rho_{e^+}+\rho_{e^-}+\rho^b_{e^-}+\rho_B,
\label{eeq}
\end{equation}
where $\rho_\gamma$ is the photon energy density, $\rho_B\simeq m_Bc^2n_B$ is the baryonic mass density which is considered to be nonrelativistic in the range of temperature $T$ under consideration, and $\rho_{e^\pm}$ is the proper energy density of electrons and positrons pairs given by
\begin{equation}
\rho_{e^\pm}= {n_{e^\pm}\over n_{e^\pm}(T)}\rho_{e^\pm}(T),
\label{hat}
\end{equation}
where $n_{e^\pm}$ is obtained by integration of Eq.(\ref{econtin}) and $\rho_{e^\pm}(T)$ is the proper energy density of electrons(positrons) obtained from zero chemical potential Fermi integrals at the equilibrium temperature $T$. On the other hand $\rho^b_{e^-}$ is the energy density of the ionized electrons coming from the ionization of baryonic matter
\begin{equation}
\rho^b_{e^-}= {n^b_{e^-}\over n^b_{e^-}(T)}\rho^b_{e^-}(T),
\label{bhat}
\end{equation}
where $n^b_{e^-}$ is obtained by integration of Eq.(\ref{becontin}) and $\rho_{e^-}(T)$ is the proper energy density of ionized electrons obtained from an appropriate Fermi integral of non-zero chemical potential $\mu_e$ at the equilibrium temperature $T$.

Having intrinsically defined the equilibrium temperature $T$ in Eq.(\ref{eeq}), we can also analogously evaluate the total pressure 
\begin{equation}
p = p_\gamma + p_{e^+}+p_{e^-}+p^b_{e^-}+p_B,
\label{eep}
\end{equation}
where $p_\gamma$ is the photon pressure, $p_{e^\pm}$ and $p^b_{e^-}$ are given by
\begin{eqnarray}
p_{e^\pm}&=& {n_{e^\pm}\over n_{e^\pm}(T)}p_{e^\pm}(T),
\label{hat'}\\
p^b_{e^-}&=& {n^b_{e^-}\over n^b_{e^-}(T)}p^b_{e^-}(T),
\label{bhat'}
\end{eqnarray}
the pressures $p_{e^\pm}(T)$ are determined by zero chemical potential Fermi integrals, and $p^b_{e^-}(T)$ is the pressure of the ionized electrons, evaluated by an appropriate Fermi integral of non-zero chemical potential $\mu_e$ at the equilibrium temperature $T$. In Eq.(\ref{eep}), the ion pressure $p_B$ is negligible by comparison with the pressures $p_{\gamma, e^\pm, e^-}(T)$, since baryons and ions are expected to be nonrelativistic in the range of temperature $T$ under consideration. Finally using Eqs.(\ref{eeq},\ref{eep}) we compute the thermal factor $\Gamma$ of the equation of state (\ref{state}). 

It is clear that the entire set of equations considered above, namely Eqs.(\ref{contin},\ref{cmom2},\ref{conse1}) with equation of state given by Eq.(\ref{state}) and the rate equation (\ref{econtin}), have to be integrated satisfying the total energy conservation for the system. The boundary conditions adopted here are simply purely ingoing conditions at the horizon and purely outgoing conditions at radial infinity. The calculation is initiated by depositing a proper energy density (\ref{jayet}) between the Reissner-Nordstr\"{o}m horizon radius $r_+$ and the dyadosphere radius $r_{ds}$, following the approximation presented in Fig.\ref{prep}  The total energy deposited is given by Eq.(\ref{tee}).

\section{The equations leading to the relative space-time transformations}\label{arrival_time}

In order to relate the above hydrodynamic and pair equations with the observations we need the governing equations relating the comoving time to the laboratory time corresponding to an inertial reference frame in which the EMBH is at rest and finally to the time measured at the detector, which must also include the effect of the cosmological expansion. These transformations have been the object of the relative space-time transformations (RSTT) Paradigm.\cite{lett1}

For signals emitted by a pulse moving with velocity $v$ in the laboratory frame,\cite{lett1} we have the following relation between the interval of arrival time $\Delta t_a$ and the corresponding interval of laboratory time $\Delta t$ (see Fig.~\ref{ttasch}):

\begin{equation}
\Delta t_a  = \left( {t_0  + \Delta t + \frac{{R_0  - r}}{c}} \right) - \left( {t_0  + \frac{{R_0 }}{c}} \right) = \Delta t - \frac{r}{c}\, .
\label{taintr}
\end{equation}

For simplicity in what follows we indicate by $t_a$ the interval of arrival time measured from the reception of a light signal emitted at the onset of the gravitational collapse. Analogously, $t$ indicates the laboratory time interval measured from the time of the gravitational collapse. In this case, Eq.(\ref{taintr}) can be written simply as:

\begin{equation}
t_a  = t - \frac{r}{c} = t - \frac{{\int_0^t {v\left( {t'} \right)dt'}  + r_{ds} }}{c}\, ,
\label{tadef}
\end{equation}
where the dyadosphere radius $r_{ds}$ is the value of $r$ at $t=0$. We consider here only the photons emitted along the line of sight from the external surface of the pulse. The arrival time spreading due to the angular dependence and that due to the thickness of the pulse will be considered elsewhere.\cite{rbcfx02a_sub,rbcfx02b_beam} The solution of Eq.(\ref{tadef}) has the expansion:

\begin{equation}
t_a  = t - \frac{{a_1 }}{c}t - \frac{1}{2}\frac{a_2}{c}t^2  -  \ldots , 
\label{taex}
\end{equation}
so the relation between $t_a$ and $t$ is in general highly nonlinear.

If and only if the expansion of the pulse is such that $r\left(t\right)=vt$ with $v\simeq c$, Eq.(\ref{tadef}) can be written, neglecting $r_{\rm ds}$, in the following simplified form (see Fig.~\ref{tvsta}):

\begin{equation}
t_a  \simeq t\left( {1 - \frac{v}{c}} \right) = t\frac{{\left( {1 - \frac{v}{c}} \right)\left( {1 + \frac{v}{c}} \right)}}{{\left( {1 + \frac{v}{c}} \right)}} \simeq \frac{t}{{2\gamma ^2 }} .
\label{taapp}
\end{equation}

This formula has been uncritically and widely applied in all articles dealing with GRBs. It is clear, however, that the knowledge of $t_a$, which is indeed essential for any physical interpretation of GRB data, depends on the definite integral  given in Eq.(\ref{tadef}) whose integration limits in the laboratory time extend from the onset of the gravitational collapse to the time $t$ relevant for the observations. Such an integral is not generally expressible as a simple linear relation or even by any explicit analytic relation since we are dealing with processes with variable gamma factor unprecedented in the entire realm of physics (see Figs.~\ref{gamma} and Fig.~\ref{tvsta}). Any linear approximation of the kind given in Eq.(\ref{taapp}) with $\gamma$ constant or changing with time\cite{fmn96} misses a crucial feature of the GRB process and is therefore erroneous in this context.

To relate the time in the laboratory frame to the time in the detector frame we have to do one additional step: the two frames are related by a transformation which is a function of the cosmological expansion. We recall that the geometry of the space-time of the universe is described by the Robertson-Walker metric:
\begin{equation}
ds^2 = dt^2 - {\cal R}^2(t) \left(\frac{dr^2}{1-kr^2} + r^2 d\vartheta^2 + r^2 sin\vartheta^2 d\varphi^2 \right),
\label{RW}
\end{equation}
where ${\cal R}\left(t\right)$ is the cosmic scale factor and $k$ is a constant related to the curvature of the three-dimensional space ($k=0, +1, -1$ corresponds to flat, close and open space respectively). 
The wavelength of an electromagnetic wave travelling from the point $P_1(t_1, r_1, \vartheta_1, \varphi_1)$ to the point $P_{\circ}(t_{\circ}, r_{\circ}, \vartheta_{\circ}, \varphi_{\circ})$ where the observer is located is related to the red-shift parameter $z$ by
\begin{equation}
z = \frac{\lambda_{\circ}-\lambda_1}{\lambda_1},
\label{z}
\end{equation}
where $\lambda_{\circ}$ is the wavelength of the radiation for the observer and $\lambda_1$ for the emitter.
We have the following general relation:
\begin{equation}
1 + z = ( 1 + z_u ) ( 1 + z_o ) ( 1 + z_s )\, ,
\end{equation}
where  $z$ is the total redshift due to the motion of the source $ z_s$, the motion of the observer $ z_o$ and the cosmological redshift $ z_u$. In the following we will assume $ z_o << 1 $ and  $ z_s << 1 $ so $z = z_u$. In terms of the scale factor ${\cal R}\left(t\right)$ the relation (\ref{z}) gives
\begin{equation}
\frac{\lambda_{\circ}}{{ \lambda_1}}  = \frac{ {\cal R}\left(t_o\right)}{ {\cal R}\left(t_1\right)} = 1 + z=\frac{\omega_1}{\omega_0}
\label{R}
\end{equation}
where $\omega_1$ and $\omega_0$ are the frequencies associated to $\lambda_1$ and $\lambda_0$ respectively. This frequency ratio then relates the time elapsing at the source with the time elapsing at the detector due to the cosmological expansion.

We can now define the corrected arrival time $t_a^d$ measured at the detector, which is related to $t_a$ by
\begin{equation}
t_a^d = t_a \left(1+z\right),
\label{taddef}
\end{equation}
where $z$ is the cosmological redshift of the GRB source. In the case of GRB~991216 we have $z\simeq 1.00$.
 
The observed flux is the flux which crosses the surface $ 4 \pi ( {\cal R}\left(t_o\right) r)^2 $ but this flux is lower by a factor $1 + z$ due to the redshift energy of the photons and by another factor $1 + z$ due to the fact that the number of photons at reception is less than the number at emission. Thus we can define a luminosity distance by:
\begin{equation}
d_L^2 =  {\cal R}_o^2 r^2 (1 + z )^2.
\end{equation}
Then the observed flux is related to the absolute luminosity of the GRB by the following relation:
\begin{equation}
l = \frac{L }{ 4 \pi d_L^2}\, ,
\end{equation}
where the luminosity distance $ d_L$ is simply related to the proper distance $ d_p=  {\cal R}_o r $ by  $ d_L = d_p ( 1 + z ) $. The observed total fluence $ f $ is related to the total energy  E of the GRB by the following relation:
\begin{equation}
f = \frac{E (1 + z) }{ 4 \pi d_L^2} 
\end{equation}

Then the cosmological effect is taken into account by the definition of the proper distance $ {\cal R}_o r$ which depends on the cosmological parameters: the Hubble constant $ H_\circ = \dot {\cal R}\left(t_\circ\right)/{\cal R}\left(t_\circ\right)$ at time $t_\circ$ and the matter density $\rho_\circ$ or $\Omega_M = \rho_\circ / \rho_{crit}$, where $\rho_{crit}= \frac{3 H_\circ^2}{8 \pi G}$.

The computation of the proper distance is then simply given by the relation :\\
\begin{equation}
d_p = \frac{c}{H_o}  \int_0^z  \frac{dz}{F(z)}\, ,
\end{equation}
where $ F(z)  = \sqrt{ \Omega_M (1+z)^3 } $.

In the case of the Friedman flat universe, $ \Omega_M = 1$ and we have:
\begin{equation}
d_p (z) = \frac{2 c}{H_o} \left[ 1 - \frac{1}{\sqrt{1 + z}}  \right]\, .
\end{equation}

So the measurement of the redshift gives us the luminosity distance via a cosmological scenario. With the measurement of the flux we can deduce the proper luminosity of the burst and from the measurement of the total fluence the total energy so we are then able to find the $E_{dya}$.

\section{The numerical integration of the hydrodynamics and the rate equations}\label{num_int}

\subsection{The Livermore code}

A computer code\cite{wsm97,wsm98} has been used to
evolve the spherically symmetric general relativistic hydrodynamic equations starting from the dyadosphere.\cite{rswx99}

We define the generalized gamma factor $\gamma$ and the radial 3-velocity in the laboratory frame $V^r$
\begin{equation}
\gamma \equiv \sqrt{ 1 + U^r U_r},\hskip0.5cm V^r\equiv {U^r\over U^t}.
\label{asww}
\end{equation}
From Eqs.(\ref{s}, \ref{tt}), we then have
\begin{equation}
(U^t)^2=-{1\over g_{tt}}(1+g_{rr}(U^r)^2)={1\over\alpha^2}\gamma^2.
\label{rr}
\end{equation}
Following Eq.(\ref{cpp}), we also define 
\begin{equation}
E \equiv \epsilon \gamma,\hskip0.5cmD \equiv \rho_B \gamma,
\hskip0.3cm {\rm and}\hskip0.3cm\tilde\rho \equiv \rho\gamma
\label{cp}
\end{equation} 
so that the conservation law of baryon number (\ref{contin}) can then
be written as
\begin{equation}
{\partial D \over \partial t} = - {\alpha \over r^2} {
\partial \over \partial r} ({r^2 \over \alpha} D V^r).
\label{jay1}
\end{equation}
Eq.(\ref{conse1}) then takes the form,
\begin{equation}
{\partial E \over \partial t} = - {\alpha \over r^2} {
\partial \over \partial r} ({r^2 \over \alpha} E V^r) - p
\biggl[ {\partial \gamma \over \partial t} + {\alpha \over r^2}
{\partial \over \partial r} ({ r^2 \over \alpha} \gamma V^r)
\biggr].
\label{jay2}
\end{equation}
Defining the radial momentum density in the laboratory frame
\begin{equation}
S_r\equiv \alpha (p+\rho)U^tU_r = (D + \Gamma E) U_r,  
\label{mstate}
\end{equation}
we can express the radial component of the energy-momentum
conservation law given in Eq.(\ref{cmom2}) by
\begin{eqnarray}
{\partial S_r \over \partial t} &=& - {\alpha \over r^2} { \partial
\over \partial r} ({r^2 \over \alpha} S_r V^r) - \alpha {\partial p
\over \partial r}\nonumber\\ 
&-&{\alpha\over2}(p+\rho)\left[{\partial g_{tt}
\over\partial r}(U^t)^2+{\partial g_{rr} \over\partial
r}(U^r)^2\right]\nonumber\\ 
&=& - {\alpha \over r^2} { \partial \over
\partial r} ({r^2 \over \alpha} S_r V^r) - \alpha {\partial p \over
\partial r}\nonumber\\
&-& \alpha\left({M \over r^2}-{Q^2 \over r^3}\right)
\biggl({D + \Gamma E \over \gamma} \biggr) \biggl[ \left({\gamma \over
\alpha}\right)^2 + {(U^r)^2 \over \alpha^4 } \biggr] ~.
\label{jay3}
\end{eqnarray}

In order to determine the number-density of $e^+e^-$ pairs, we turn to Eq.(\ref{econtin}). 
Defining the $e^+e^-$-pair density in the laboratory frame 
$N_{e^\pm} \equiv\gamma n_{e^\pm}$ and $N_{e^\pm}(T) \equiv\gamma
n_{e^\pm}(T)$, where the equilibrium temperature $T$ has been obtained from
Eqs.(\ref{eeq}) and (\ref{hat}), and using Eq.(\ref{rr}), we rewrite the rate equation given by Eq.(\ref{econtin}) in the form
\begin{equation}
{\partial N_{e^\pm} \over \partial t} = - {\alpha \over r^2} {
\partial \over \partial r} ({r^2 \over \alpha} N_{e^\pm} V^r) +
\overline{\sigma v} (N^2_{e^\pm} (T) - N^2_{e^\pm})/\gamma^2~,
\label{jay:E:ndiff}
\end{equation}
These equations are integrated starting from the dyadosphere distributions given in Fig.~\ref{dens} and assuming as usual ingoing boundary conditions on the horizon of the EMBH.

\subsection{The Rome code}

In the following we recall a zeroth order approximation of the fully relativistic equations of the previous section:\cite{rswx99}\\
(i) Since we are mainly interested in the expansion of the $e^+e^-$ plasma away from the EMBH, we neglect the gravitational interaction.\\
(ii) We describe the expanding plasma by a
special relativistic set of equations.\\
(iii) In contrast with the previous treatment where the evolution of the density profiles given in Fig.~\ref{dens} are followed in their temporal evolution leading to a pulse-like structure, selected geometries of the pulse are a priori adopted and the correct one validated by the complete integration of the equations given by the Livermore codes.\\

In analogy to Eq.(\ref{f'}), from Eq.(\ref{contin}) we
have along each flow line in the general case in which baryonic matter is present 
\begin{equation}
{d(n_BV)\over d\tau}=0\, .
\label{f0}
\end{equation}
For the expansion
of a shell from its initial volume $\Delta V_\circ$ to the volume $\Delta  V$,
we obtain
\begin{equation}
{n_B^\circ\over n_B}= {\Delta V\over \Delta V_\circ}={\Delta {\cal V}\gamma(r)
\over \Delta {\cal V}_\circ\gamma_\circ(r)},
\label{be'_1}
\end{equation}
where $\Delta {\cal V}$ is the volume of the shell in the laboratory
frame, related to the proper volume $\Delta V$ in the comoving
frame by $\Delta V=\gamma(r) \Delta {\cal V}$, where $\gamma(r)$
defined in Eq.(\ref{asww}) is the gamma factor of the shell
at the radius $r$.

Similarly from Eq.(\ref{f'}), using the equation of state
(\ref{state}),  along the flow lines we obtain
\begin{equation}
d\ln\epsilon + \Gamma d\ln V=0.
\label{scale''}
\end{equation}
Correspondingly we obtain for the internal energy density $\epsilon$ along the flow lines 
\begin{equation}
{\epsilon_\circ\over \epsilon} = 
\left({\Delta V\over \Delta V_\circ}\right)^\Gamma=
\left({\Delta {\cal V}\over \Delta {\cal V}_\circ}\right)^\Gamma\left({\gamma(r)
\over \gamma_\circ(r)}\right)^\Gamma ~,
\label{scale'}
\end{equation}
where the thermal index $\Gamma$ given by (\ref{state}) is a
slowly-varying function with values around $4/3$. It can be computed for each value of $\epsilon,p$ as a function of $\Delta V$.

The overall energy conservation
requires that the change of the internal proper energy of a shell is compensated by a change in its  bulk kinetic energy. We then have\cite{rswx99}
\begin{equation}
dK=[\gamma(r)-1](dE+\rho_BdV).
\label{dk}
\end{equation}

In order to model the relativistic expansion of the plasma
fluid, we assume that $E$ and $D$ as defined by Eq.(\ref{cp}) are
constant in space over the volume $\Delta V$.  As a consequence the total energy conservation for the shell implies\cite{rswx99} 
\begin{equation}
(\epsilon_\circ+\rho^\circ_B)\gamma_\circ^2(r) {\Delta \cal V}_\circ =(\epsilon+\rho_B)
\gamma^2 (r){\Delta \cal V},
\label{res'}
\end{equation}
which leads the solution
\begin{equation}
\gamma (r)=\gamma_\circ(r)\sqrt{{(\epsilon_\circ+\rho^\circ_B){\Delta \cal V}_\circ
\over(\epsilon+\rho_B) {\Delta \cal V}}}.
\label{result'}
\end{equation}
Corresponding to Eq.(\ref{jay:E:ndiff}) we obtain the equation for the evolution of the $e^\pm$
number-density as seen by an observer in the laboratory frame
\begin{equation}
{\partial \over \partial t}(N_{e^\pm}) = -N_{e^\pm}{1\over\Delta {\cal V}}{\partial \Delta {\cal V}\over \partial t}+\overline{\sigma v}{1\over\gamma^2(r)}  (N^2_{e^\pm} (T) - N^2_{e^\pm})~.
\label{paira'}
\end{equation}
Eqs.(\ref{be'_1}), (\ref{scale'}), (\ref{result'}) and
(\ref{paira'}) are a complete set of equations describing the
relativistic expansion of the shell.
If we now turn from a single shell to a finite distribution of shells, we can introduce the average values of the proper internal-energy, baryon-mass, baryon-number and pair-number densities ($\bar\epsilon, \bar\rho_B,\bar n_B,\bar n_{e^\pm}$) and $\bar E\equiv\bar\gamma\bar\epsilon$, $\bar D\equiv\bar\gamma\bar\rho_B$, $\bar N_{e^\pm}\equiv\bar\gamma(r) \bar n_{e^\pm}$ for the PEM-pulse, where the average $\bar\gamma$-factor is defined by
\begin{equation}
\bar\gamma={1\over{\cal V}}\int_{\cal V}\gamma(r) d{\cal V},
\label{ga}
\end{equation}
and ${\cal V}$ is the total volume of the shell in the laboratory frame. The corresponding equations are given in Ruffini, Salmonson, Wilson \& Xue (1999).\cite{rswx99}
Having defined all its governing equations we can now return to the description of the different eras of the GRB phenomena.

\section{The era I: the PEM pulse}\label{era1}
We have assumed that, following the gravitational collapse process, a region of very low baryonic contamination exists in the dyadosphere all the way to the remnant of the progenitor star.

Recalling Eq.(\ref{nd}) the limit on such baryonic contamination, where $\rho_{B_c}$ is the mass-energy density of baryons, is given by 
\begin{equation}
\rho_{B_c}\ll m_pn_{e^+e^-}(r) = 3.2\cdot 10^8\left({r_{ds}\over r}\right)^2\left[1-\left({r\over r_{ds}}\right)^2\right](g/cm^3).
\label{nb} 
\end{equation}
Near the horizon $r\simeq r_+$, this gives
\begin{equation}
\rho_{B_c}\ll m_pn_{e^+e^-}(r) =1.86 \cdot 10^{14}\left({\xi\over\mu}\right)(g/cm^3)\, ,
\label{nb1} 
\end{equation}
and near the radius of the dyadosphere $r_{ds}$:
\begin{equation}
\rho_{B_c}\ll m_pn_{e^+e^-}(r) = 3.2\cdot 10^8\left[1-\left({r\over r_{ds}}\right)^2\right]_{r\rightarrow r_{ds}}(g/cm^3)\, .
\label{nb2} 
\end{equation}
Such conditions can be easily satisfied in the collapse to an EMBH, but not necessarily in a collapse to a neutron star. 

Consequently we have solved the equations governing a plasma composed solely of $e^+e^-$-pairs and electromagnetic radiation, starting at time zero from the dyadosphere configurations corresponding to constant density in Fig.~\ref{dens}. The Livermore code\cite{rswx99} has shown very clearly the self organization of the expanding plasma in a very sharp pulse which we have defined as the pair-electromagnetic pulse (PEM pulse), in analogy with the EM pulse observed in nuclear explosions. 
In order to further examine the structure of the PEM pulse with the simpler procedures of the Rome codes we have assumed\cite{rswx99} three alternative patterns of expansion of the PEM pulse on which to try the simplified special relativistic treatment and then compared the results with the fully general relativistic hydrodynamical results:
\begin{itemize}
\item Spherical model: we assume the radial component of the four-velocity $U_r(r)=U{r\over {\cal R}}$, where $U$ is the radial component of the four-velocity at the moving outer surface ${r=\cal R}(t)$ of the PEM pulse and the $\bar\gamma$-factor and the velocity $V_r$ are
\begin{eqnarray} 
\bar\gamma &=& {3\over 8U^3}\Big[2U(1+U^2)^{3\over2}
- U(1+U^2)^{1\over2}\nonumber\\
&-& \ln(U+\sqrt{1+U^2})\Big],\hskip0.3cm V_r={U_r\over\bar\gamma}~;
\label{sp}
\end{eqnarray}
this distribution expands keeping an uniform density profile which decreases with time similar to a portion of a Friedmann Universe.
\item Slab 1: we assume $U(r)=U_r={\rm const.}$, the constant width of the
expanding slab ${\cal D}= R_\circ$ in the laboratory frame of the PEM pulse, while $\bar\gamma$  and $V_r$ are
\begin{equation}
\bar\gamma=\sqrt{1+U_r^2},\hskip0.3cm V_r={U_r\over\bar\gamma}~;
\label{sl1_1}
\end{equation}
this distribution does not need any averaging process.
\item Slab 2: we assume a constant width $R_2-R_1=R_\circ$ of the expanding slab in the
comoving frame of the PEM pulse, while $\bar\gamma$  and  $V_r$ are
\begin{equation}
\bar\gamma=\sqrt{1+U_r^2(\tilde r)},\hskip0.3cm V_r={U_r\over\bar\gamma},
\label{sp2}
\end{equation}
This distribution needs an averaging procedure and $R_1<\tilde r <R_2$, i.e.~$\tilde r$ is an intermediate radius in the slab.
\end{itemize}

\begin{figure}[htbp]
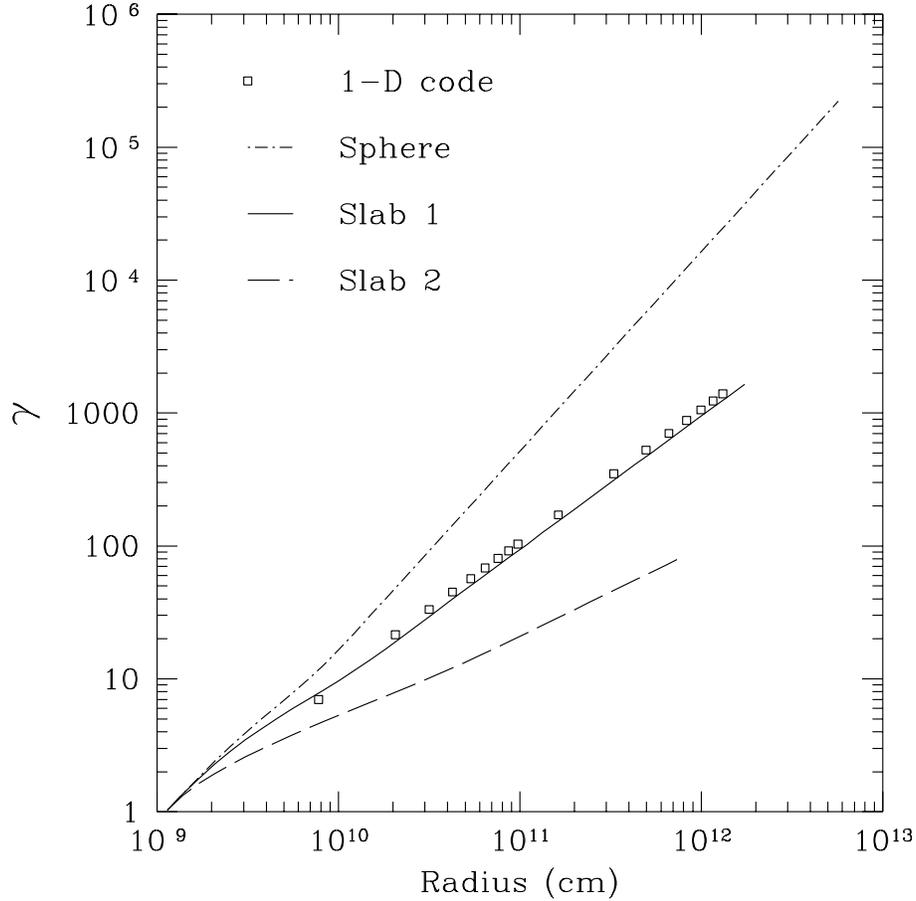

\PSFIG{shells}{\hsize}{0}
\caption{Gamma factor as a function of radius. Three models for the expansion pattern of the PEM-pulse are compared with the results of the one dimensional hydrodynamic code for an energy of dyadosphere $E_{dya}=3.1\times 10^{54}$ erg.  The 1-D code has an expansion pattern that strongly resembles that of a shell with constant thickness in the laboratory frame.
\label{figshells}}
\end{figure}

These different assumptions lead to three different distinct slopes for the monotonically increasing $\bar\gamma$-factor as a function of the radius (or time) in the laboratory frame, having assumed for the energy of dyadosphere $E_{dya}=3.1\times 10^{54}$ erg (see Fig.~\ref{figshells}). In principle, we could have an infinite number of models by defining arbitrarily the geometry of the expanding fluid in the special relativistic treatment given above. To find out which expanding pattern of PEM pulses is the physically realistic one, we need to compare and contrast the results of our simplified models (performed in Rome) with the numerical results based on the hydrodynamic Eqs.(\ref{jay1},\ref{jay2},\ref{jay3}) (obtained at Livermore).\cite{rswx99} Details of the iterative method used to solve the special relativistic equation can be found in Ruffini, Salmonson, Wilson \& Xue (1999).\cite{rswx99}

It is manifest from the results (see Fig.~\ref{figshells}) that the slab 1 approximation (constant thickness in the laboratory frame) is in excellent agreement with the Livermore results (open squares).

The remarkable validation of the special relativistic treatment of the PEM pulse,\cite{rswx99} allows us to easily estimate the related quantities of physical and astrophysical interest in the model, like the $e^+e^-$-pair densities as a function of the laboratory time, the  temperature of the plasma in the comoving and laboratory frames, the reheating ratio as a function of the $e^+e^-$-pair annihilation for a variety of initial conditions.\cite{rswx99}

\section{The era II: the interaction of the PEM pulse with the remnant of the progenitor star}\label{era2}

The PEM pulse expands initially in a region of very low baryonic contamination created  by the process of gravitational collapse. As it moves further out the baryonic remnant (see Fig.~\ref{raggi2}) of the progenitor star is encountered.
As discussed in section \ref{gc} below, the existence of such a remnant is necessary in order to guarantee the overall charge neutrality of the system: the collapsing core has the opposite charge of the remnant and the system as a whole is clearly neutral. The number of extra charges in the baryonic remnant negligibly affects the overall charge neutrality of the PEM pulse.\cite{r01mg9,rvx02}

The baryonic matter remnant is assumed to be distributed well outside the dyadosphere in a shell of thickness $\Delta$ between an inner radius $r_{\rm in}$ and an outer radius $r_{\rm out}=r_{\rm in}+\Delta$ at a distance from the EMBH at which the original PEM pulse expanding in vacuum has not yet reached transparency. For the sake of an example we choose
\begin{equation}
r_{\rm in}=100r_{\rm ds},\hskip 0.5cm \Delta = 10r_{\rm ds}.
\label{bshell_1}
\end{equation}
The total baryonic mass $M_B=N_Bm_p$ is assumed to be a  fraction of the dyadosphere initial total
energy $(E_{\rm dya})$. The total baryon-number $N_B$ is then expressed as a function of the dimensionless parameter $B$ given by 
\begin{equation}
B=\frac{N_Bm_pc^2}{E_{\rm dya}}\, ,
\label{chimical1}
\end{equation}
where  $B$ is a parameter in the range $10^{-8}-10^{-2}$ and $m_p$ is the proton mass. We shall see below the paramount importance of $B$ in the determination of the features of the GRBs. We will see in section \ref{fp} the sense in which $B$ and $E_{dya}$ can be considered to be the only two free parameters of the EMBH theory for the entire GRB family, the so called ``long bursts''. We shall see in section~\ref{new} that for the so called ``short bursts'' the EMBH theory depends on the two other parameters $\mu$, $\xi$, since in that case $B=0$. 
The baryon number density $n^\circ_B$ is assumed to be a constant
\begin{equation}
\bar n^\circ_B={N_B\over V_B},\hskip0.5cm \bar\rho^\circ_B=m_p\bar n^\circ_B c^2.
\label{bnd}
\end{equation}
 
As the PEM pulse reaches the region $r_{\rm in}<r<r_{\rm out}$, it interacts with the baryonic matter which is assumed to be at rest. In our simplified quasi-analytic model we make the following assumptions to describe this interaction: 

\begin{itemize}
\item the PEM pulse does not change its geometry during the interaction;
\item the collision between the PEM pulse and the baryonic matter is assumed to be inelastic,
\item the baryonic matter reaches thermal equilibrium with the photons and pairs of the PEM pulse.
\end{itemize}

These assumptions are valid if: (i) the total energy of the PEM pulse is much larger than the total mass-energy of baryonic matter $M_B$, $10^{-8}<B<10^{-2}$, (ii) the ratio of the comoving number density  of pairs and baryons at the moment of collision $n_{e^+e^-}/n^\circ_B$ is very high (e.g., $10^6 <n_{e^+e^-}/ n^\circ_B <10^{12}$) and (iii)  the PEM pulse has a large value of the gamma factor ($100<\bar\gamma $).  
  
In the collision between the PEM pulse and the baryonic matter at $r_{\rm out}>r>r_{\rm in}$ , we impose total conservation of energy and momentum. We consider the collision process between two radii $r_2,r_1$ satisfying 
$r_{\rm out}>r_2>r_1>r_{\rm in}$ and $r_2-r_1\ll \Delta$. The amount of baryonic mass acquired by the PEM pulse is
\begin{equation}
\Delta M = {M_B\over V_B}{4\pi\over3}(r_2^3-r_1^3) ,
\label{mcc_2}
\end{equation}
where $M_B/ V_B$ is the mean-density of baryonic matter at rest.
The conservation of total energy leads to the estimate of the corresponding quantities before (with ``$\circ$'') and after such a collision  
\begin{equation}
(\Gamma\bar\epsilon_\circ + \bar\rho^\circ_B)\bar\gamma_\circ^2{\cal V}_\circ + \Delta M = (\Gamma\bar\epsilon + \bar\rho_B + {\Delta M\over V} + \Gamma\Delta\bar\epsilon)\bar\gamma^2{\cal V},
\label{ecc_2}
\end{equation}
where $\Delta\bar\epsilon$ is the corresponding increase of internal energy due to the collision. Similarly the momentum-conservation gives
\begin{equation}
(\Gamma\bar\epsilon_\circ + \bar\rho^\circ_B)\bar\gamma_\circ U^\circ_r{\cal V}_\circ = (\Gamma\bar\epsilon + \bar\rho_B + {\Delta M\over V} + 
\Gamma\Delta\bar\epsilon)\bar\gamma U_r{\cal V},
\label{pcc_2}
\end{equation}
where the radial component of the four-velocity of the PEM pulse is $U^\circ_r=\sqrt{\bar\gamma_\circ^2-1}$ and $\Gamma$ is the thermal index. 
We then find 
\begin{eqnarray}
\Delta\bar\epsilon & = & {1\over\Gamma}\left[(\Gamma\bar\epsilon_\circ + \bar\rho^\circ_B) {\bar\gamma_\circ U^\circ_r{\cal V}_\circ \over \bar\gamma U_r{\cal V}} - (\Gamma\bar\epsilon + \bar\rho_B + {\Delta M\over V})\right],\label{heat_2}\\
\bar\gamma & = & {a\over\sqrt{a^2-1}},\hskip0.5cm a\equiv {\bar\gamma_\circ  \over  
U^\circ_r}+ {\Delta M\over (\Gamma\bar\epsilon_\circ + \bar\rho^\circ_B)\bar\gamma_\circ U^\circ_r{\cal V}_\circ}.
\label{dgamma_2}
\end{eqnarray}
These equations determine the gamma factor $\bar\gamma$ and the internal energy density $\bar\epsilon=\bar\epsilon_\circ +\Delta\bar\epsilon$ in the capture process of baryonic matter by the PEM pulse.

The effect of the collision of the PEM pulse with the remnant leads to the following results\cite{rswx00} as a function of the $B$ parameter defined in Eq.(\ref{chimical1}):\\
1) an abrupt decrease of the gamma factor given by
\begin{equation}
\gamma_{coll} = \gamma_\circ \frac{1+B}{\sqrt{ {\gamma_\circ}^2 \left(2B+B^2 \right) +1}}\, ,
\label{gamma_circ}
\end{equation}
where $\gamma_\circ$ is the gamma factor of the PEM pulse prior to the collision and $B$ is given by Eq.(\ref{chimical1}),\\
2) an increase of the internal energy in the comoving frame $E_{coll}$ developed in the collision given by
\begin{equation}
\frac{E_{coll}}{E_{dya}} =  \frac{\sqrt{ {\gamma_\circ}^2 \left(2B+B^2 \right) +1}}{\gamma_\circ} - \left(\frac{1}{\gamma_\circ} + B \right)\, ,
\label{E_int/E}
\end{equation}
3) a corresponding reheating of the plasma in the comoving frame but not in the laboratory frame, an increase of the number of $e^+e^-$ pairs and correspondingly an  overall increase of the opacity of the pulse. See details in section~\ref{at}.

\section{The era III: the PEMB pulse}\label{era3}

After the engulfment of the baryonic matter of the remnant the plasma formed of $e^+e^-$-pairs, electromagnetic radiation and baryonic matter expands again as a sharp pulse, namely the PEMB pulse. The calculation is continued as the plasma fluid expands,
cools and the $e^+e^-$ pairs recombine until it becomes optically
thin:
\begin{equation} 
\int_R dr(n_{e^\pm}+\bar
Zn_B)\sigma_T\simeq O(1),
\label{thin_1}
\end{equation}
where $\sigma_T =0.665\cdot 10^{-24}
{\rm cm^2}$ is the Thomson cross-section and the integration is over the radial interval of the PEMB pulse in the
comoving frame. 
We have first explored the general problem of the PEMB pulse evolution by integrating the general relativistic hydrodynamical equations with the Livermore codes, for a total energy in the dyadosphere of $3.1\times 10^{54}$ erg and a baryonic shell  
of thickness $\Delta =10 r_{\rm ds}$ at rest at a radius
of $100 r_{\rm ds}$ and $B\simeq 1.3\cdot 10^{-4}$. 

In total analogy with the special relativistic treatment for the PEM pulse, presented in section~\ref{era1} (see also Ruffini, Salmonson, Wilson \& Xue, 1999\cite{rswx99}), we obtain for the adiabatic expansion of the PEMB pulse in the constant-slab approximation described by the Rome codes the following hydrodynamical equations with $\rho_B\not=0$
\begin{eqnarray}
{\bar n_B^\circ\over \bar n_B}&=& { V\over  V_\circ}={ {\cal V}\bar\gamma
\over {\cal V}_\circ\bar\gamma_\circ},
\label{be'}\\
{\bar\epsilon_\circ\over \bar\epsilon} &=& 
\left({V\over V_\circ}\right)^\Gamma=
\left({ {\cal V}\over  {\cal V}_\circ}\right)^\Gamma\left({\bar\gamma
\over \bar\gamma_\circ}\right)^\Gamma,
\label{scale1'}\\
\bar\gamma &=&\bar\gamma_\circ\sqrt{{(\Gamma\bar\epsilon_\circ+\bar\rho^\circ_B){\cal V}_\circ
\over(\Gamma\bar\epsilon+\bar\rho_B) {\cal V}}},
\label{result1'}\\
{\partial \over \partial t}(N_{e^\pm}) &=& -N_{e^\pm}{1\over{\cal V}}{\partial {\cal V}\over \partial t}+\overline{\sigma v}{1\over\bar\gamma^2}  (N^2_{e^\pm} (T) - N^2_{e^\pm}).
\label{paira'_2}
\end{eqnarray}
In these equations ($r>r_{\rm out}$) the comoving baryonic mass- and number densities are $\bar\rho_B=M_B/V$ and $\bar n_B=N_B/V$, where $V$ is the comoving volume of the PEMB pulse.

We compare and contrast (see Fig.~\ref{twocodecompare}) the bulk gamma factor as computed from the Rome and Livermore codes, where excellent agreement has been found. This validates the constant-thickness approximation in the case of the PEMB pulse as well. On this basis we easily estimate a variety of physical quantities for an entire range of values of $B$.

For the same EMBH we have considered five different cases: a shell of baryonic mass with (1) $B\simeq 1.3\cdot 10^{-4}$; (2) $B\simeq 3.8\cdot 10^{-4}$; (3) $B\simeq 1.3\cdot 10^{-3}$; (4) $B\simeq 3.8\cdot 10^{-3}$; (5) $B\simeq 1.3\cdot 10^{-3}$). The results of the integration given in detail in Ruffini, Salmonson, Wilson \& Xue (2000)\cite{rswx00} show that for the first parameter range the PEMB pulse propagates as a sharp pulse of constant thickness in the laboratory frame, but already for $B\simeq 1.3\cdot 10^{-2}$ the expansion of the PEMB pulse becomes much more complex and the constant-thickness approximation ceases to be valid; see Ruffini, Salmonson, Wilson \& Xue (2000)\cite{rswx00} for details.

\begin{figure}[htbp]
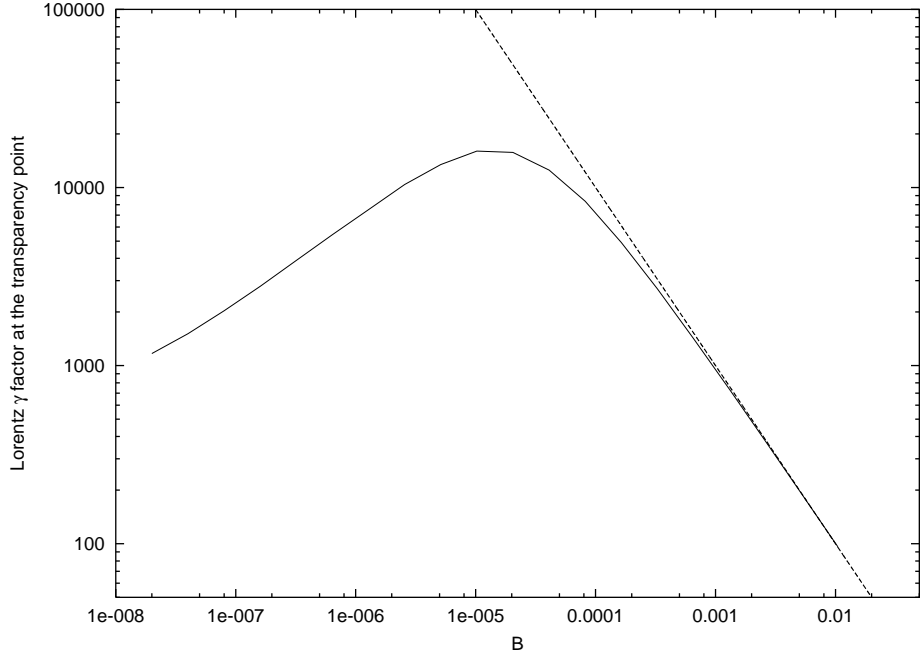

\PSFIG{h2047f9}{\hsize}{0}
\caption{The gamma factor (the solid line) at the transparent point is plotted as a function of the $B$ parameter. The asymptotic value (the dashed line) $E_{\rm dya}/ (M_Bc^2)$ is also plotted.}
\label{fgamma}
\end{figure}

It is particularly interesting to evaluate the final value of the gamma factor of the PEMB pulse when the transparency condition given by Eq.(\ref{thin_1}) is reached as a function of $B$, see Fig.~\ref{fgamma}. For a given EMBH, there is a {\em maximum} value of the gamma factor at transparency. By further increasing the value of $B$ the entire $E_{dya}$ is transferred into the kinetic energy of the baryons; see also section~\ref{new}.  
Details are given in Ruffini, Salmonson, Wilson \& Xue (2000).\cite{rswx00}

\begin{figure}[htbp]
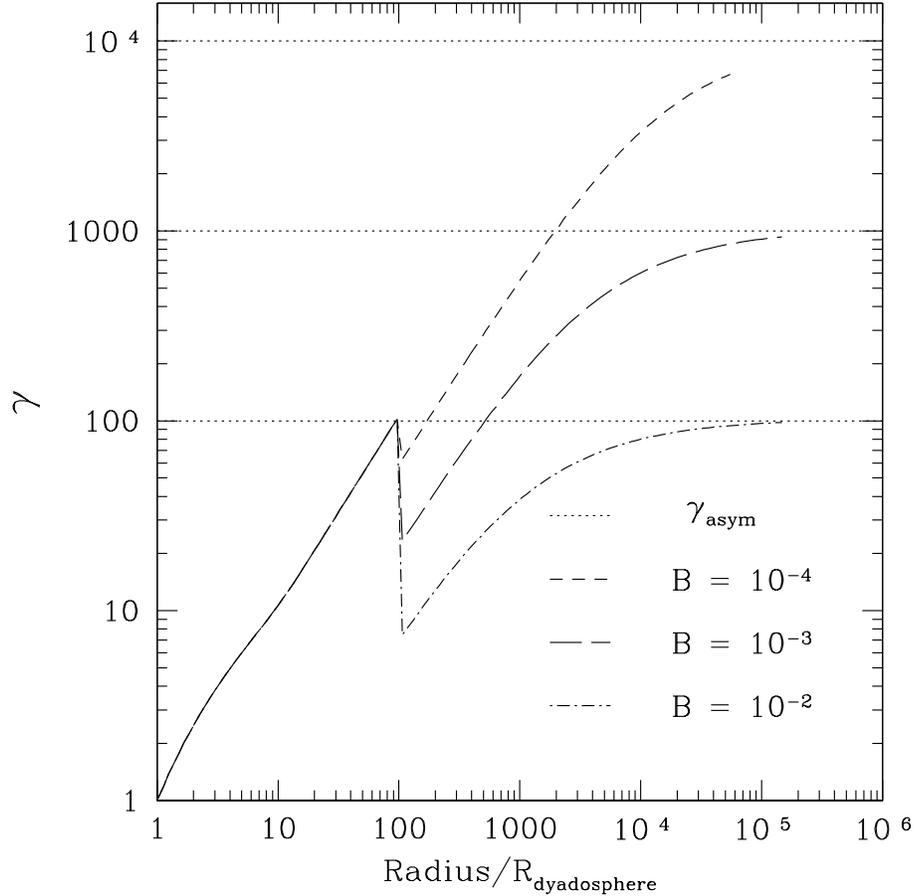

\PSFIG{h2047f8}{\hsize}{0}
\caption{The gamma factors are given as functions of the radius in units of the dyadosphere radius for selected values of $B$ for the typical case $E_{dya}=3.1\times 10^{54}$ erg. The asymptotic values $\gamma_{\rm asym} = E_{\rm dya}/(M_Bc^2)=10^4,10^3,10^2$ are also plotted. The collision of the PEM pulse with the baryonic remnant occurs at $r/r_{ds}=100$ where the jump occurs and the PEMB pulse starts.}
\label{3gamma}
\end{figure}

In Fig.~\ref{3gamma} we plot the gamma factor of the PEMB pulse versus the radius for different amounts of baryonic matter. The diagram extends to values of the radial coordinate at which the transparency condition given by Eq.(\ref{thin_1}) is reached. The ``asymptotic'' gamma factor
\begin{equation}
\bar\gamma_{\rm asym}\equiv {E_{\rm dya}\over M_B c^2}
\label{asymp}
\end{equation}
is also shown for each curve. The closer the gamma value approaches the ``asymptotic'' value (\ref{asymp}) at transparency, the smaller the intensity of the radiation emitted in the burst and the larger the amount of kinetic energy left in the baryonic matter.

\section{The identification of the free parameters of the EMBH theory}\label{fp}

Within the approximation presented in section~\ref{dyadosphere} the EMBH is characterized by two parameters: $\mu$ and $\xi$. The energy of the dyadosphere is expressed in terms of these two parameters by Eq.(\ref{tee}).

\begin{figure}[htbp]
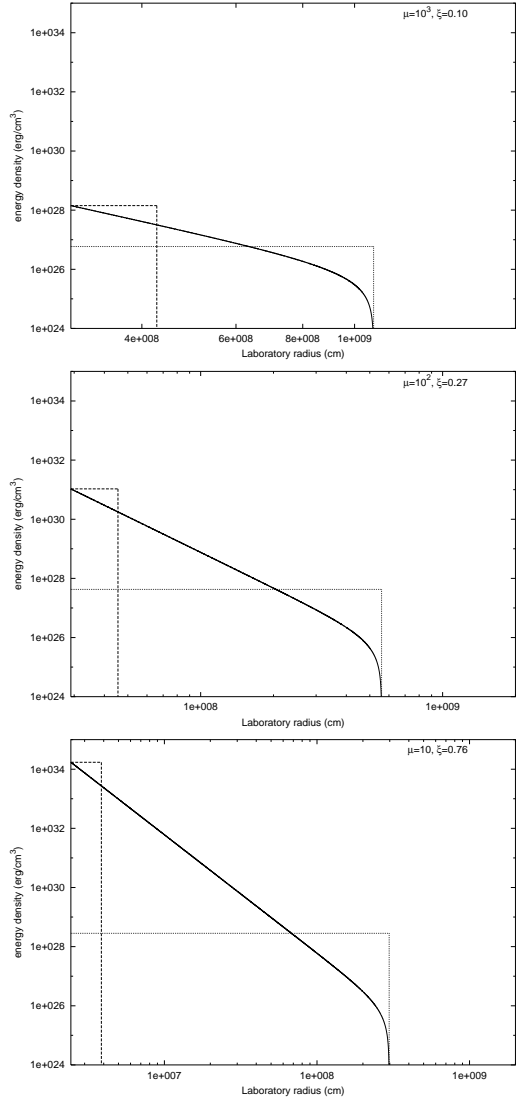

\PSFIG{3dens}{7cm}{0}
\caption{Three different dyadospheres corresponding to the same value of $E_{dya}=3.1\times 10^{54}$ erg and with different values of the two parameters $\mu$ and $\xi$ are given. The three different configurations are markedly different in their spatial extent as well as in their energy-density distribution.}
\label{3dens}
\end{figure}

There is an entire family of EMBH solutions with different values of $\mu$ and $\xi$ corresponding to the same value of $E_{dya}$ (see Fig.~\ref{muxi}). These solutions are physically different with respect to the density of electron-positron pair distributions given by Eq.(\ref{nd}), as well as to their energy density given by Eq.(\ref{jayet}). A clear example of such a degeneracy is given in Fig.~\ref{3dens} where the two limiting energy density profiles approximating the dyadosphere as introduced in Fig.~\ref{dens} are given for three different EMBH configurations corresponding to the same value of $E_{dya}=3.1\times 10^{54}$ erg. The three configurations correspond respectively to the three different pairs $\left(\mu,\xi\right)$: $\left(10,0.76\right)$, $\left(10^2,0.27\right)$, $\left(10^3,0.10\right)$.

\begin{figure}[htbp]
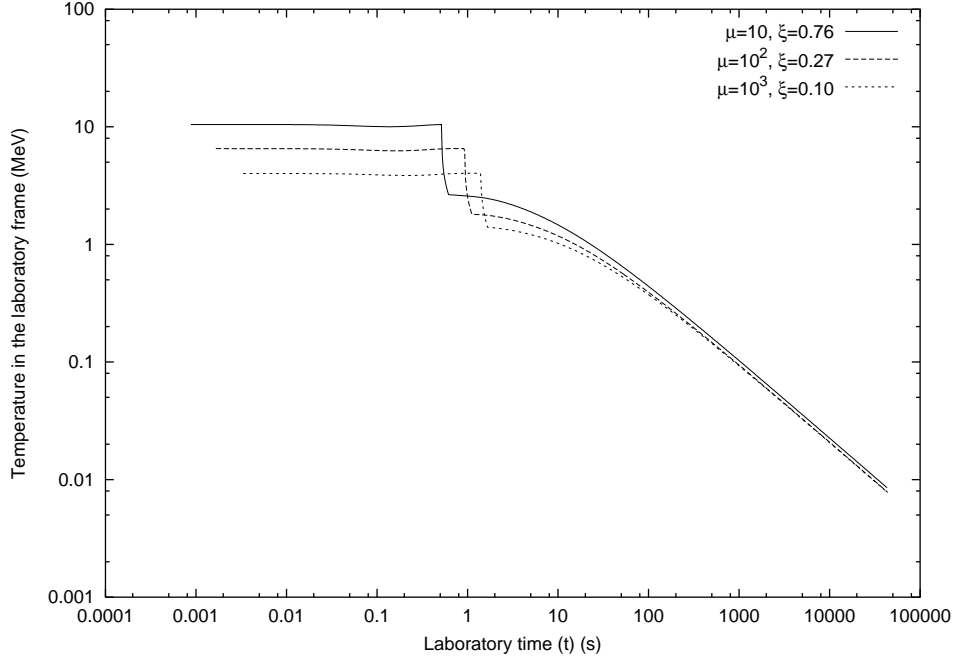

\PSFIG{3tem}{\hsize}{0}
\caption{ The temperature of the plasma during the PEM pulse and PEMB pulse eras, measured in the laboratory frame, corresponding to the three configurations presented in Fig. \ref{3dens} is given as a function of the laboratory time. The three different curves converge to a common one in the PEMB pulse era, which is therefore only a function of the $E_{dya}$ and $B$. The difference among the three curves in the early part of the PEMB pulse follows from having located the baryonic matter at a distance of $50(r_{\rm ds}-r_+)$, which is different in the three cases. Such difference  become negligible at  large distances in the later phases of the evolution. }
\label{3tem}
\end{figure}

The corresponding dynamical evolution of the PEM pulse introduced in section~\ref{era1} and Ruffini, Salmonson, Wilson \& Xue (1999)\cite{rswx99} is clearly different in the three cases. It is  remarkable that when the collision with the remnant of the progenitor star is considered all these differences disappear. As usual (see section~\ref{era2}) we describe the baryonic content of the remnant by the parameter $B$. The PEMB pulse generated after the collision with the baryonic matter depends uniquely on the two parameters $E_{dya}$ and $B$. In Fig.~\ref{3tem} the temperature in the laboratory frame is given for the PEM pulse and the PEMB pulse corresponding to the three configurations of Fig.\ref{3dens} and $B=4\times 10^{-3}$. It is clear that while for the PEM pulse era the three configurations are markedly different, they do converge to a common behaviour in the PEMB pulse era.

\begin{figure}[htbp]
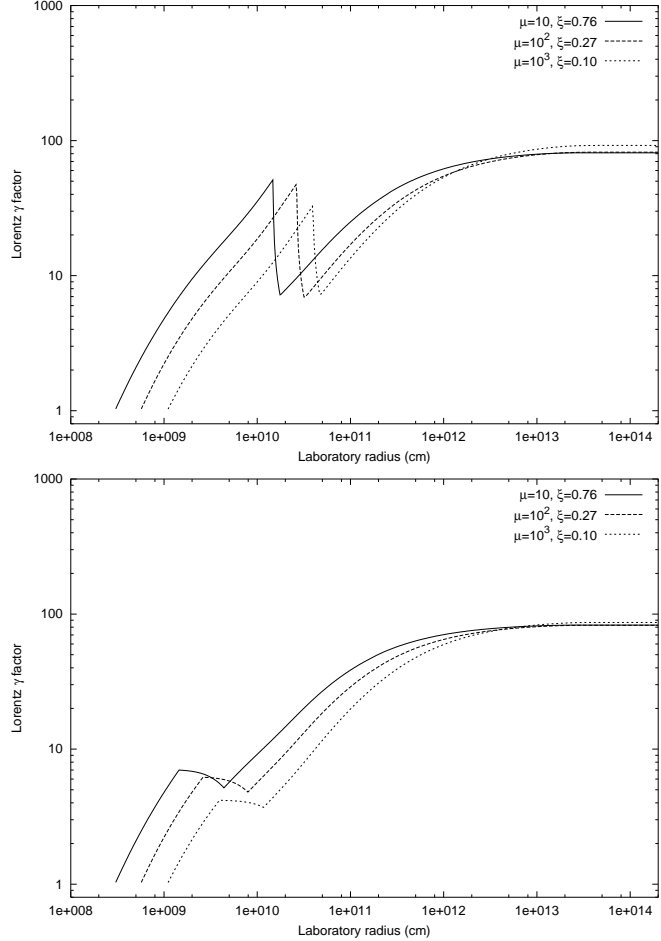

\PSFIG{3gamma2b}{9cm}{0}
\caption{The gamma factors for the three configurations considered in Fig. \ref{3dens} are given as a function of the radial coordinate in the laboratory frame. The two figures correspond to a baryonic remnant positioned respectively at $r_{in}=50(r_{\rm ds} - r_+)$ (above) and at $r_{in}=5(r_{\rm ds} - r_+)$. Again the convergence to a common behaviour, uniquely a function of $E_{dya}$ and $B$ for the late stages of the PEMB pulse, is manifest. }
\label{3gamma2b}
\end{figure}

If we turn now to the effect of the distance between the EMBH and the baryonic remnant, we see that this degeneracy is further extended: while the three PEM pulse eras are quite different, the common PEMB pulse era is largely insensitive to the location of the baryonic remnant, see Fig.~\ref{3gamma2b}. We have plotted the three gamma factors in the PEM pulse era corresponding to the different configurations of Fig.~\ref{3dens} and $B=10^{-2}$, in the two cases the baryonic remnant is positioned at different distances from the EMBH. 

If the PEM pulse has reached extreme relativistic regimes, the common value $\gamma_{coll}$ to which the three gamma factors drop in the collision with the baryonic matter of the remnant can be simply expressed by the large gamma limit of Eq.(\ref{gamma_circ})
\begin{equation}
\gamma_{coll}=\frac{B+1}{\sqrt{B^2+2B}}\, ,
\label{gammacollb}
\end{equation}
while the internal energy $E_{coll}$ developed in that collision is simply given by the corresponding limit of Eq.(\ref{E_int/E})
\begin{equation}
\frac{E_{coll}}{E_{dya}}=-B+\sqrt{B^2+2B}\, .
\label{eintcollb}
\end{equation}
This approximation applies when the final gamma factor at the end of the PEM pulse era is larger than $\gamma_{coll}$, upper panel in Fig.~\ref{3gamma2b}.

Turning from these general considerations to the GRB data, this degeneracy in the PEMB pulse eras and their dependence on only two parameters $E_{dya}$ and $B$ has far reaching astrophysical implications for the identification of the source of GRBs. As we will see in the conclusions all the information obtainable from GRBs with a large value of the parameter $B$  will lead to the determination of the above two parameters. An entire family of degenerate astrophysical solutions in the range of charges and masses given in Fig.~\ref{muxi} are possible. The direct knowledge of the mass and charge of the EMBH can only be gained from the PEM pulse or from GRBs with very small values of $B$ --- the so called ``short bursts'', see section~\ref{new} and the conclusions.

\section{The approach to transparency: the thermodynamical quantities}\label{at}

As the condition of transparency expressed by Eq.(\ref{thin_1}) is reached the {\em injector phase} terminates. The electromagnetic energy of the PEMB pulse is released in the form of free-streaming photons --- the proper GRB. The remaining energy of the PEMB pulse is released as an accelerated-baryonic-matter (ABM) pulse.

We now proceed to the analysis of the approach to the transparency condition. It is then necessary to turn from the pure dynamical description of the PEMB pulse described in the previous sections to the relevant thermodynamic parameters. Also such a description at the time of transparency needs the knowledge of the thermodynamical parameters in all previous eras of the GRB. 

As above we shall consider as  a typical case an EMBH of $E_{dya}=3.1\times 10^{54}$ erg and $B=10^{-2}$. The considerations will refer to a dyadosphere configuration described by the two limiting approximations shown in Fig.~\ref{dens}.

\begin{figure}[htbp]
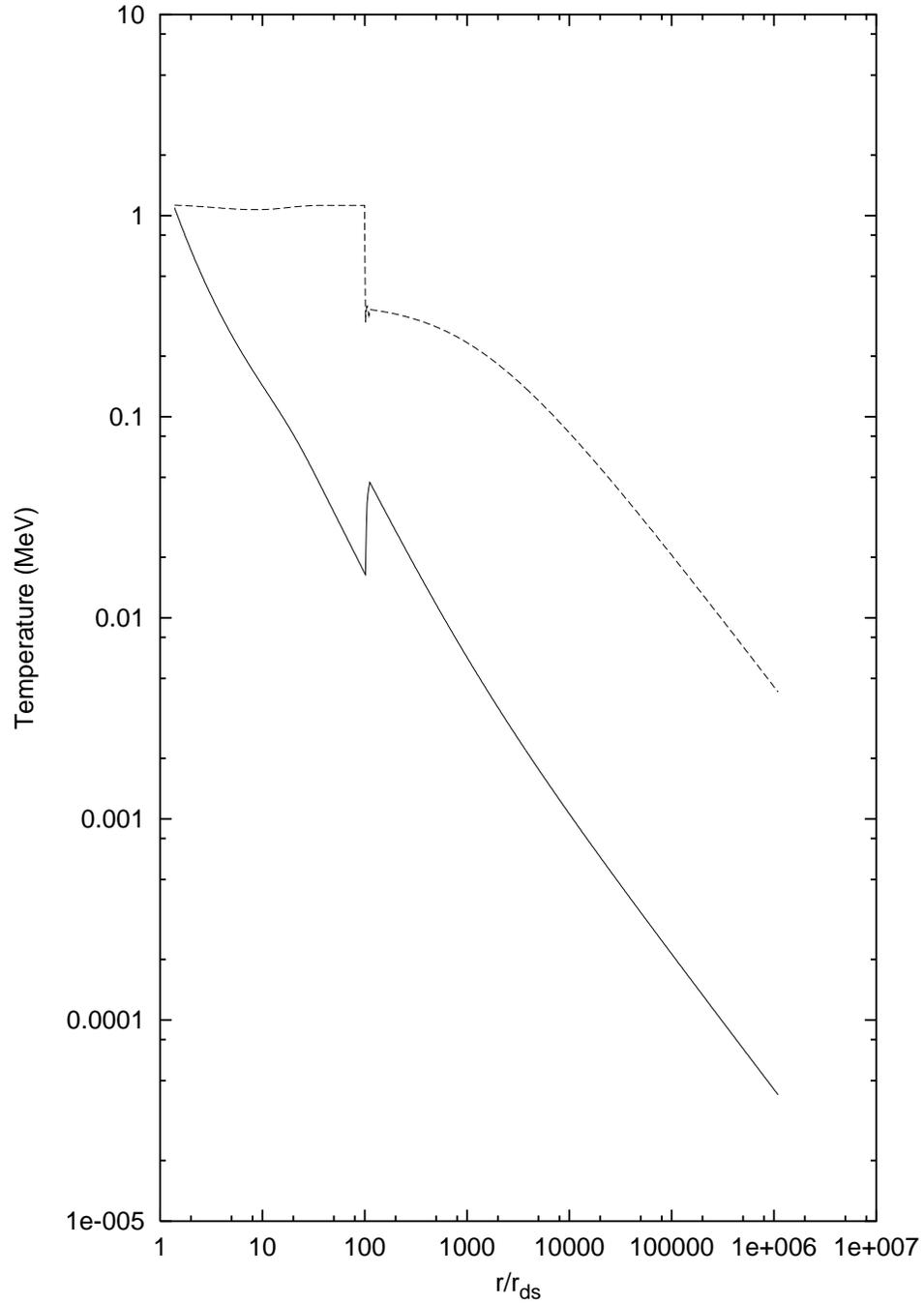

\PSFIG{h2047f12}{\hsize}{0}
\caption{The temperature of the plasma in the comoving frame $T'$(MeV) (the solid line) and in the laboratory frame $\bar\gamma T'$ (the dashed line) are plotted as functions of the radius in the unit of the dyadosphere radius $r_{\rm ds}$.}
\label{tem}
\end{figure}

One of the key thermodynamical parameters is represented by the temperature of the PEM and PEMB pulses. It is given as a function of the radius both in the comoving and in the laboratory frames in Fig.~\ref{tem}. 
Before the collision the PEM pulse expands keeping its temperature in the laboratory frame constant while its temperature in the comoving frame falls.\cite{rswx99}. In fact Eqs.(\ref{res'},\ref{result'}) are equivalent to
\begin{equation}
{d(\epsilon\gamma^2{\cal V})\over dt}=0,
\label{tc}
\end{equation}
where the baryon mass-density is $\rho_B=0$ and the thermal energy-density of photons and $e^+e^-$-pairs is $\epsilon=\sigma_B T^4(1+f_{e^+e^-})$, $\sigma_B$ is the Boltzmann constant and $f_{e^+e^-}$ is the Fermi-integral for $e^+$ and $e^-$. This leads to
\begin{equation}
\epsilon\gamma^2{\cal V}=E_{\rm dya},\hskip0.3cm T^4\gamma^2{\cal V}={\rm const.}
\label{econ}
\end{equation}
Since $e^+$ and $e^-$ in the PEM pulse are extremely relativistic, we have the equation of state $p\simeq\epsilon/3$ and the thermal index (\ref{state}) $\Gamma\simeq 4/3$ in the evolution of PEM pulse. Eq.(\ref{econ}) is thus equivalent to
\begin{equation}
T^3\bar\gamma {\cal V}\simeq {\rm const.}
\label{encon}
\end{equation}
These two equations (\ref{tc}) and (\ref{encon}) result in the constancy of the laboratory temperature $T\bar\gamma$ in the evolution of the PEM pulse.

It is interesting to note that Eqs.(\ref{econ}) and (\ref{encon}) hold as well in
the cross-over region where $T\sim m_ec^2$ and $e^+e^-$ annihilation takes place.  
In fact from the conservation of entropy it follows that asymptotically we have
\begin{equation}
      \frac{(V T^3)_{T<m_ec^2}}{(V T^3)_{T>m_ec^2}}  =\frac{11}{4}\ ,
\label{reheat}
\end{equation}
exactly for the same reasons and physics scenario discussed in the cosmological framework by Weinberg, see e.g. Eq.~(15.6.37) of Weinberg (1972). 
The same considerations when
repeated for the conservation of the total energy 
$\epsilon\gamma V=\epsilon\gamma^2{\cal V}$
following from Eq.~(\ref{tc}) then lead to
\begin{equation}
      \frac{(V T^4 \gamma)_{T<m_ec^2}}{(V T^4 \gamma)_{T>m_ec^2}}  
             =\frac{11}{4}\ .
\end{equation}
The ratio of these last two quantities gives asymptotically 
\begin{equation}
      T_\circ= (T \gamma)_{T>m_ec^2}= (T \gamma)_{T<m_ec^2},
\label{rt}
\end{equation}
where $T_\circ$ is the initial average temperature of the dyadosphere at rest.

\begin{figure}[htbp]
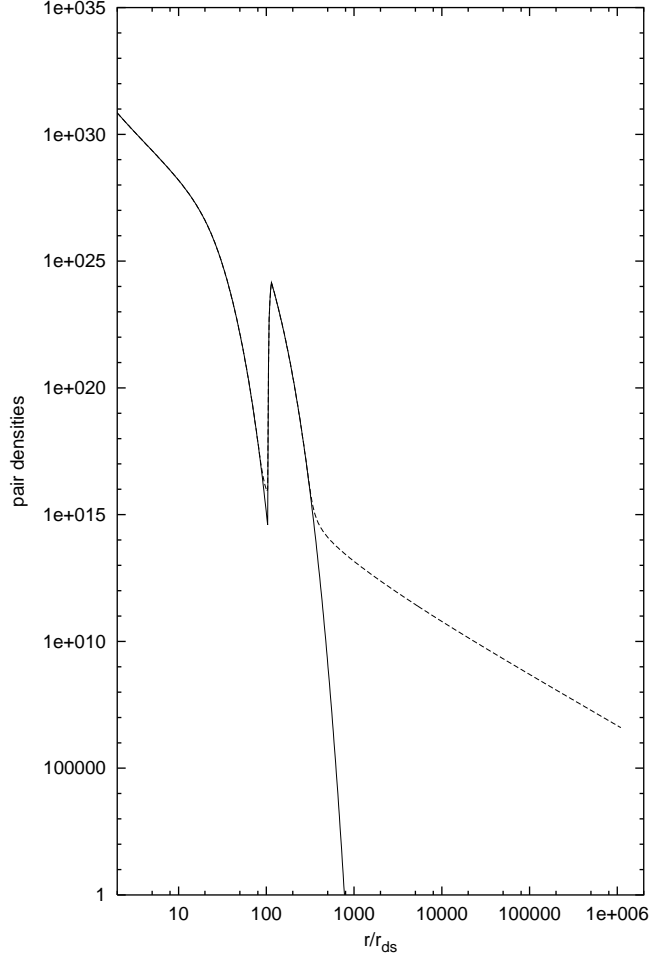

\PSFIG{h2047f14}{9cm}{0}
\caption{The number densities $n_{e^+e^-}(T)$ (the solid line) computed by the Fermi integral and $n_{e^+e^-}$ (the dashed line) computed by the rate equation (see section~\ref{hydro_pem}) are plotted as functions of the radius. $T'\ll m_ec^2$, two curves strongly divergent due to $e^+e^-$-pairs frozen out of the thermal equilibrium. The peak at $r\simeq100r_{\rm ds}$ is due to the internal energy developed in the collision.}
\label{pair}
\end{figure}

During the collision of the PEM pulse with the remnant we have an increase in the number density of $e^+e^-$ pairs (see Fig.~\ref{pair}). This transition corresponds to an {\em increase} of the temperature in the comoving frame and a {\em decrease} of the temperature in the laboratory frame as a direct effect of the dropping of the gamma factor (see Fig.~\ref{3gamma}). 

\begin{figure}[htbp]
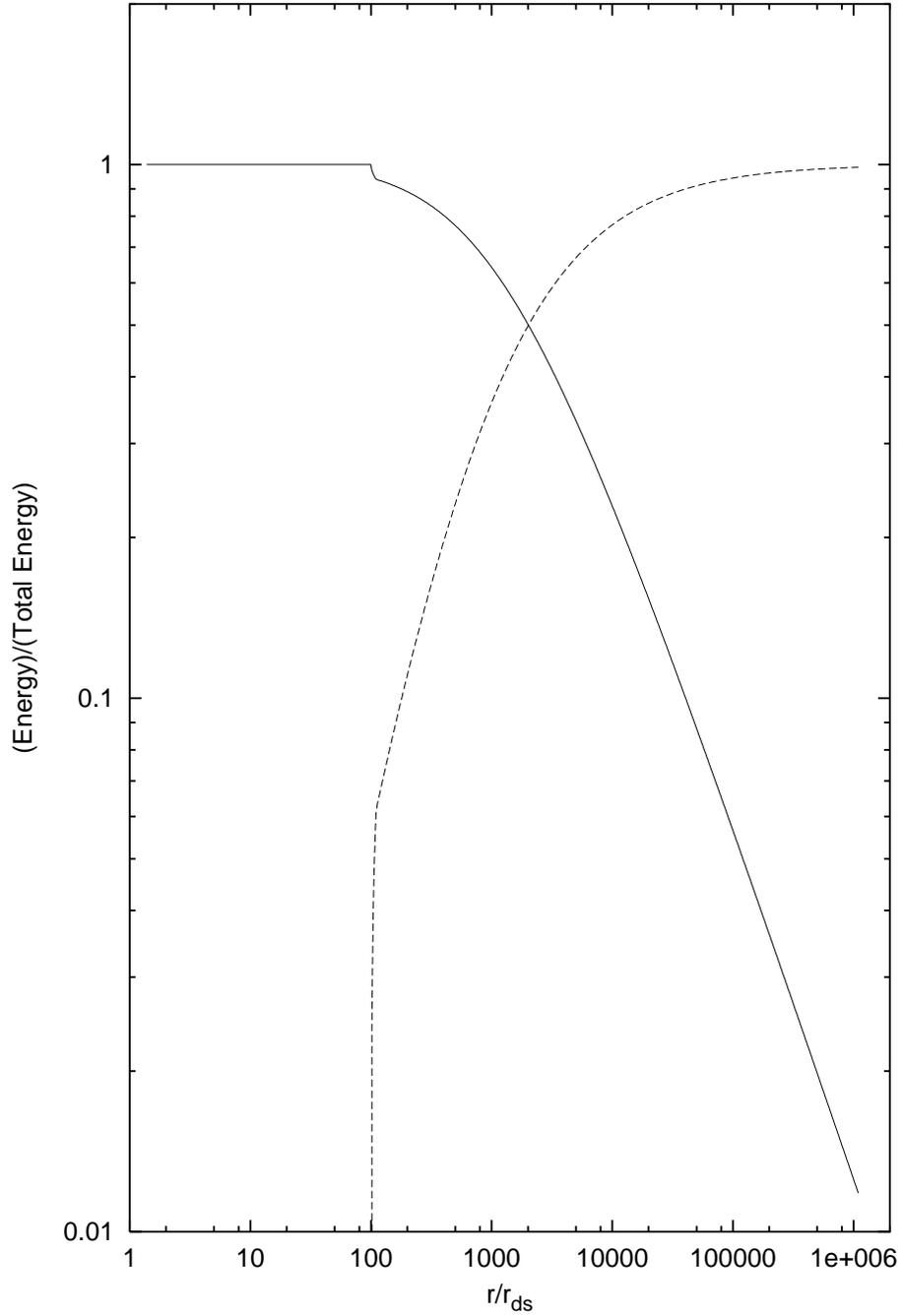

\PSFIG{h2047f13}{\hsize}{0}
\caption{The energy of the non baryonic components of the PEMB pulse (the solid line) and the kinetic energy of the baryonic matter (the dashed line) in unit of the total energy are plotted as functions of the radius in the unit of the dyadosphere radius $r_{\rm ds}$.}
\label{intkin}
\end{figure}

After the collision we have the further acceleration of the PEMB pulse (see Fig.~\ref{3gamma}). The temperature now decreases both in the laboratory and the comoving frame (see Fig.~\ref{tem}). Before the collision the total energy of the $e^+e^-$ pairs and the photons is constant and equal to $E_{\rm dya}$. After the collision
\begin{equation}
E_{\rm dya}=E_{\rm Baryons}+E_{e^+e^-}+E_{\rm photons},
\label{Etotal}
\end{equation}
which includes both the total energy $E_{e^+e^-}+E_{\rm photons}$ of the nonbaryonic components
and the kinetic energy $E_{\rm Baryons}$ of the baryonic matter
\begin{equation}
E_{\rm Baryons}=\bar\rho_B V(\bar\gamma -1).
\label{kinetic}
\end{equation}
In Fig.~\ref{intkin} we plot both the total energy $E_{e^+e^-}+E_{\rm photons}$ of the nonbaryonic components and the kinetic energy $E_{\rm Baryons}$ of the baryonic matter as functions of the radius for the typical case $E_{\rm dya}=3.1\times 10^{54}$ erg and $B=10^{-2}$. Further details are given in Ruffini, Salmonson, Wilson \& Xue (2000)\cite{rswx00}.

\section{The P-GRBs and the ``short bursts''. The end of the injector phase.}\label{new}

We now analyze the approach to the transparency condition given by Eq.(\ref{thin_1}). For selected values of $B$ we give the energy $E_{P\hbox{-}GRB}$ of the P-GRB, and $E_{\rm Baryons}$ of the ABM pulse. We clearly have
\begin{equation}
E_{dya}= E_{P\hbox{-}GRB} + E_{\rm Baryons}\, .
\label{esum}
\end{equation}

Taking into account the results shown in Figs. \ref{tem}--\ref{intkin}, we can  repeat all the considerations for selected values of $B$. We shall examine values of $B$ ranging from $B=10^{-8}$ only up to $B=10^{-2}$: for larger values of $B$ our constant slab approximation breaks down. We will see in the following that this range does indeed cover the most relevant observational features of the GRBs. 

As clearly shown in Fig.~\ref{3gamma} both the final value of the gamma factor and the radial coordinate at which the transparency condition is reached depend very strongly on $B$. Therefore a strong dependence on $B$ is also found in the relative values of $E_{P\hbox{-}GRB}$ and $E_{\rm Baryons}$.

We are now finally ready to give in Fig.~\ref{fintkin} the crucial diagram representing the values of $E_{P\hbox{-}GRB}$ and $E_{\rm Baryons}$ in units of the $E_{dya}$ as functions of $B$. This diagram, a universal one, is very important and is essential for the understanding of the GRB structure. 

We find that for small values of $B$ (around $10^{-8}$) almost all the $E_{dya}$ is emitted in the P-GRB (see also our previous paper Ruffini, Salmonson, Wilson \& Xue, 1999\cite{rswx99}) and very little energy is left in the baryons. While for $B\simeq10^{-2}$ roughly only $10^{-2}$ of the total initial energy of the dyadosphere is radiated away in the P-GRB and almost all energy is transferred to the baryons.

This behaviour is at the heart of the fundamental difference between the so called {\em short bursts} and {\em long bursts}. We have proposed\cite{lett2} that the {\em short bursts} must be identified with the P-GRBs in the case of very small $B$. There are a variety of reasons supporting this identification:

\begin{enumerate}
\item For small values of $B$, $E_{\rm Baryons}$ is negligible, see Fig.~\ref{fintkin}, and consequently the intensity of the afterglow is also negligible and the entire energy $E_{dya}$ is released into the P-GRB. This is clearly consistent with the absence of observed afterglows in the short bursts.
\item The temperature of the P-GRB in the laboratory frame $\bar\gamma T$ at the transparency point is a strongly decreasing function of $B$, see Fig.~\ref{energypeak}. $\bar\gamma T$ is related to the energy corresponding to the peak of the photon-number spectrum, as described in Ruffini, Salmonson, Wilson \& Xue (1999)\cite{rswx99}. This is also in very good agreement with the observed decrease of the hardness ratio between the {\em short bursts} and the {\em long bursts}.\cite{ka93}
\item The time $T_{90}$, the duration of 90\% of the energy emission as used in the current literature and discussed in Ruffini, Salmonson, Wilson \& Xue (2000)\cite{rswx00} is plotted in Fig.~\ref{t90bs} for selected values of $E_{dya}$ and for different values of $B$.

\begin{figure}[htbp]
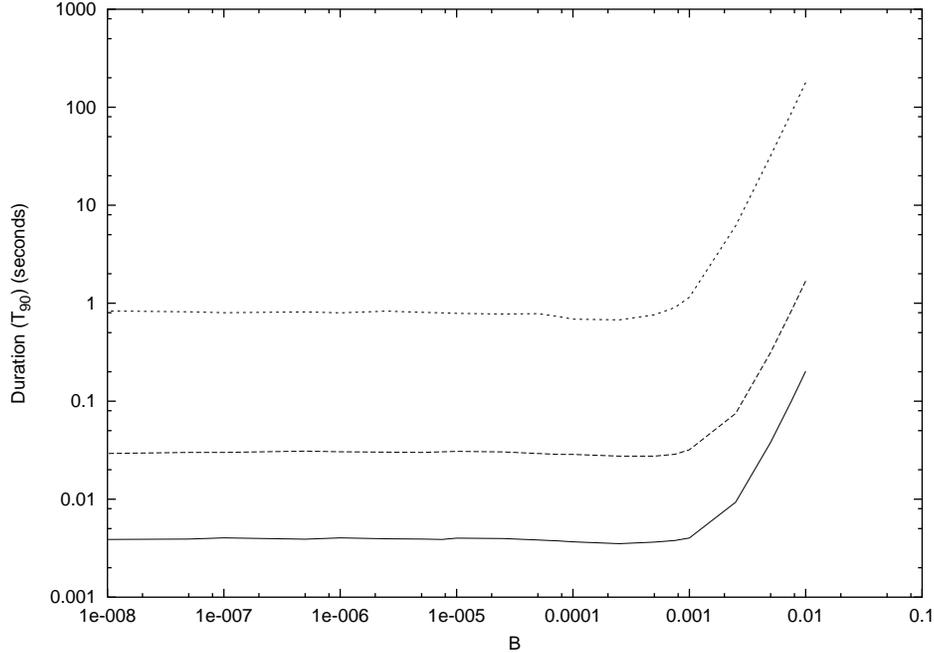

\PSFIG{t90bs}{\hsize}{0}
\caption{The duration computed with the $T_{90}$ criterion is represented as a function of the $B$ parameter for three selected EMBH respectively with $E_{dya}=4.4\times 10^{52}$ erg, $E_{dya}=3.1\times 10^{54}$ erg, $E_{dya}=4.1\times 10^{58}$ erg going from the lower curve to the upper one.}
\label{t90bs}
\end{figure}
\end{enumerate}

Before concluding a word of caution is needed about how to use the above results: all these considerations are based on the drastic approximations in the description of the dyadosphere presented in section \ref{dyadosphere}, see also Fig.~\ref{3dens}. This treatment is very appropriate in estimating the general dependence of the energy of the P-GRB, the kinetic energy of the ABM pulse and consequently the intensity of the afterglow. Especially powerful is the establishment of the dependence of $E_{P\hbox{-}GRB}$ and $E_{\rm Baryons}$ on $B$ (see Fig.~\ref{fintkin}). As we will see in the next sections, this approximation is similarly powerful in determining the overall time structure of the GRB and especially the time of the release of the P-GRB with respect to the moment of gravitational collapse and the afterglow. 

If, however, we turn to the detailed temporal structure of the P-GRB and its detailed spectral distribution, it is clear that the approximations given in section \ref{dyadosphere} is no longer valid. The detailed description of the formation of the dyadosphere as qualitatively expressed in Fig.~\ref{dyaform} is now needed in all mathematical rigour with the full development of all its governing equations. Progress in this direction is being made at this moment.\cite{crv02,rv02a,rv02b,rvx02} This situation, however, provides a unique opportunity to follow in real time the general relativistic effects of the approach to the EMBH horizon as it occurs. In other words all direct general relativistic effects of the GRBs are encoded in the fine structure of the P-GRB. For the reasons given in section \ref{fp} the information on the EMBH mass and charge can only come from the short bursts.

This terminates the {\em injector phase}. We now turn to the {\em Beam-Target phase} in which the ABM pulse collides with the interstellar medium target and the afterglow is generated. We shall in the following sections review the basic theoretical treatment necessary for the description of these remaining eras and proceed then to the confrontation of the EMBH theory with the data. 

\section{The era IV: the ultrarelativistic and relativistic regimes in the afterglow}\label{era4}

In the introduction we have already expressed the basic assumptions which we have adopted for the description of the collision of the ABM pulse with the ISM. In analogy and by extension of the results obtained for the PEM and PEMB pulse cases, we also assume that the expansion of the ABM pulse through the ISM occurs keeping its width constant in the laboratory frame, although the results are quite insensitive to this assumption. We assume then that this interaction can be represented by a sequence of inelastic collisions of the expanding ABM pulse with a large number of thin and cold ISM spherical shells at rest with respect to the central EMBH. Each of these swept up shells of thickness $\Delta r$ has a mass $\Delta M_{\rm ism}$ and is assumed to be located between two radial distances $r_1$ and $r_2$ (where $r_2-r_1 = \Delta r \ll r_1$) in the laboratory frame. These collisions create an internal energy $\Delta E_{\rm int}$.

We indicate by $\Delta\epsilon$ the increase in the proper internal energy density due to the collision with a single shell and by $\rho_B$ the proper energy density of the swept up baryonic matter. This includes the baryonic matter composing the remnant around the central EMBH, already swept up in the PEMB pulse formation, and the baryonic matter from the ISM swept up by the ABM pulse:

\begin{equation}
\rho_B=\frac{\left(M_B+M_{\rm ism}\right)c^2}{V}.
\label{rhob}
\end{equation}

Here $V$ is the ABM pulse volume in the comoving frame, $M_B$ is the mass of the baryonic remnant and $M_{\rm ism}$ is the ISM mass swept up from the transparency point through the $r$ in the laboratory frame:
\begin{equation}
M_{\rm ism}=m_pn_{\rm ism}{4\pi\over3}\left(r^3-{r_\circ}^3\right)\, ,
\label{dgm1}
\end{equation}
where $m_p$ the proton mass and $n_{\rm ism}$ the number density of the ISM in the laboratory frame.

The energy conservation law in the laboratory frame at a generic step of the collision process is given by

\begin{equation}
\rho_{B_1} {\gamma_1}^2{\cal V}_1 + \Delta M_{\rm ism} c^2 = \left(\rho_{B_1}\frac{V_1}{V_2} + {{\Delta M_{\rm ism} c^2}\over V_2} + \Delta\epsilon \right){\gamma_2}^2{\cal V}_2,
\label{ecc}
\end{equation}
where the quantities with the index ``$1$'' are calculated before the collision of the ABM pulse with an elementary shell of thickness $\Delta r$ and the quantities with ``$2$'' after the collision, $\gamma$ is the gamma factor and ${\cal V}$ the volume of the ABM pulse in the laboratory frame so that $V=\gamma {\cal V}$.
 
The momentum conservation law in the laboratory frame is given by
\begin{equation}
\rho_{B_1} \gamma_1 U_{r_1} {\cal V}_1 = \left(\rho_{B_1}\frac{V_1}{V_2} + {\Delta M_{\rm ism} c^2\over V_2} + 
\Delta\epsilon \right)\gamma_2 U_{r_2} {\cal V}_2,
\label{pcc}
\end{equation}
where $U_r=\sqrt{{\gamma}^2 - 1}$ is the radial covariant component of the four-velocity vector\cite{rswx99,rswx00} (see Eq.(\ref{asww})).

We thus obtain
\begin{eqnarray}
\Delta\epsilon & = & \rho_{B_1} {\gamma_1 U_{r_1} {\cal V}_1 \over \gamma_2 U_{r_2} {\cal V}_2} - \left(\rho_{B_1}\frac{V_1}{V_2} + {\Delta M_{\rm ism} c^2\over V_2} \right) ,\label{heat}\\
\gamma_2 & = & {a\over\sqrt{a^2-1}},\hskip0.5cm a\equiv {\gamma_1  \over  
U_{r_1}}+ {\Delta M_{\rm ism} c^2\over \rho_{B_1} \gamma_1 U_{r_1} {\cal V}_1}.
\label{dgamma}
\end{eqnarray}

We can use for $\Delta \varepsilon$ the following expression
\begin{equation}
\Delta \varepsilon = \frac{E_{{\rm int}_2}}{V_2}-\frac{E_{{\rm int}_1}}{V_1} = \frac{E_{{\rm int}_1}+\Delta E_{\rm int}}{V_2}-\frac{E_{{\rm int}_1}}{V_1} = \frac{\Delta E_{\rm int}}{V_2}
\label{deltaexp}
\end{equation}
because we have assumed a ``fully radiative regime'' and so $E_{{\rm int}_1}=0$.
Substituting Eq.(\ref{dgamma}) in Eq.(\ref{heat}) and applying Eq.(\ref{deltaexp}), we obtain:
\begin{eqnarray}
 \Delta E_{\rm int}  = \rho_{B_1} {V_1}\sqrt {1 + 2\gamma_1 \frac{{\Delta M_{\rm ism} c^2 }}{{\rho_{B_1} V_1 }} + \left( {\frac{{\Delta M_{\rm ism} c^2 }}{{\rho_{B_1} V_1 }}} \right)^2 } \nonumber \\ 
  - \rho_{B_1}{V_1} \left( 1 + \frac{{\Delta M_{\rm ism} c^2 }}{\rho_{B_1} V_1} \right)\, , \label{heat2}
\end{eqnarray}
\begin{equation}
 \gamma_2  = \frac{{\gamma_1  + \frac{{\Delta M_{\rm ism} c^2 }}{{\rho_{B_1} V_1 }}}}{{\sqrt {1 + 2\gamma_1 \frac{{\Delta M_{\rm ism} c^2 }}{{\rho_{B_1} V_1 }} + \left( {\frac{{\Delta M_{\rm ism} c^2 }}{{\rho_{B_1} V_1 }}} \right)^2 } }}\, . \label{dgamma2} 
\end{equation}
These relativistic hydrodynamic (RH) equations have to be numerically integrated.

These are the actual set of equations we have integrated in the EMBH theory. In order to compare and contrast our results with the ones in the current literature, in section \ref{approximation} we have introduced the continuous limit of our  equations and proceeded to have piecewise approximate power law solutions.
We examine as well in section \ref{substructures} still under the above assumptions, the effects of a possible departure from homogeneity in the interstellar medium, still keeping the average density $n_{ism}=const$. Although these inhomogeneities are not relevant for the overall behaviour of the afterglow which we address here, they are indeed important for the actual observed flux and its temporal structures\cite{lett5}. Also these considerations are affected by the angular spreading.\cite{rbcfx02a_sub}

\section{The era V: the approach to the nonrelativistic regimes in the afterglow}\label{era5}

The only reason for addressing this last era is that the issue of the approach to nonrelativistic behaviour has been extensively discussed in the literature. In our treatment these results do not show any particular problems and the relativistic equations of the previous section continue to hold. In the specific example of GRB~991216 we will present in section \ref{approximation} some analytic asymptotic expansions of these equations.

This concludes the exposition of the different eras of the EMBH theory. It goes without saying that for the description of each era, all the preceding eras must necessarily be known in order to determine the space-time grid in the laboratory frame and its relation to the arrival times as seen by a distant observer. This is the basic message expressed in the RSTT paradigm.

We can now turn to the comparison of the EMBH theory with the observational data.

\section{The best fit of the EMBH theory to the GRB~991216: the global features of the solution}\label{bf} 

For reasons already explained in the introduction, we use the GRB~991216 as a prototype. We will then later apply the EMBH theory to other GRBs. The relevant data of GRB~991216 are reproduced in Fig.~\ref{grb991216}: the data on the burst as recorded by BATSE\cite{brbr99} and the data on the afterglow from the RXTE satellite\cite{cs00} and the Chandra satellite,\cite{p00} see also Halpern et al. (2000).\cite{ha00}

The data fitting procedure relies on three basic assumption:
\begin{enumerate}
\item In the E-APE region, the source luminosity is mainly in the energy band 50--300~KeV, so we consider the flux observed by BATSE a good approximation of the total flux.
\item In the decaying part of the afterglow, we assume that during the R-XTE and Chandra observations the source luminosity is mainly in the energy band 2--10~KeV, so we can again assume that the flux observed by these satellites is a good approximation of the total one.
\item We have neglected in this paper the optical and radio emissions, since they are always negligible with respect to the X-ray and $\gamma$-ray fluxes. In fact, even in the latest afterglow phases up to where the X-ray data are available, they are one order of magnitude smaller then the X-ray flux.
\end{enumerate}
These assumptions were initially adopted for the sake of simplicity, but have now also been justified on the basis of the spectral description of the afterglow.\cite{rbcfx02c_spectrum}

As already emphasized in the previous sections, in the EMBH theory there are only two free parameters characterising the afterglow: the energy of the dyadosphere, $E_{dya}$, and the baryonic matter in the remnant of the progenitor star, parametrized by the dimensionless parameter $B$. The location of the remnant has been assumed $\sim 10^{10}$ cm. As discussed in Ruffini et al. (2001a)\cite{lett1} and section \ref{fp}, the results are rather insensitive to the actual density and location of the baryonic component but they are very sensitive to the value of $B$.\cite{rswx00}

In Fig.~\ref{ii-fig2} we present the actual first results of fitting our EMBH theory to the data from the R-XTE and Chandra satellites, corresponding to selected values of $E_{dya}$ and $B$. There are three distinct features which are clearly evident as a function of the arrival time at the detector: an initial rising part in the afterglow luminosity which reaches a peak followed by a monotonically decreasing part. 

\begin{figure}[htbp]
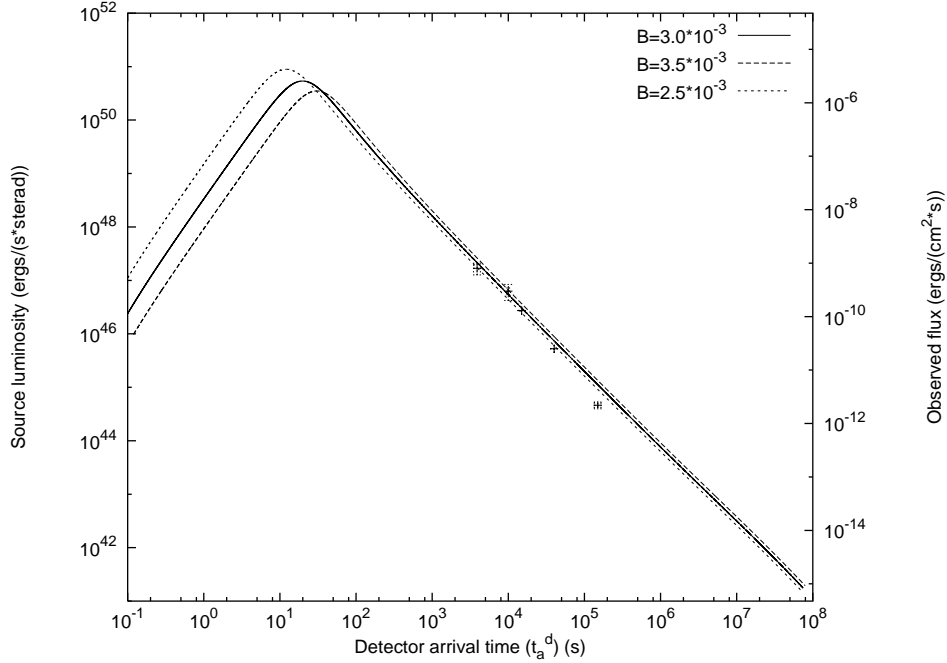

\PSFIG{fabv}{\hsize}{0}
\caption{Fine tuning of the best fit of the afterglow data of Chandra, RXTE as well as of the range of variability of the BATSE data on the major burst by a unique afterglow curve leading to the parameter values $E_{dya}=4.83\times 10^{53} erg, B=3\times 10^{-3}$.}
\label{fit_var}
\end{figure}

We have then proceeded to fine tune the two parameters in Fig.~\ref{fit_var}. The main conclusions from our model are the following:

1) The slope of the afterglow in the region where the experimental data are present is $n=-1.6$ and is in perfect agreement with the observational data. The index $n$ in this region is rather insensitive to the values of the parameters $E_{dya}$ and $B$. The physical reason for this universality of the slope is rather remarkable since it depends on a variety of factors including the ultrarelativistic energy of the baryons in the ABM pulse, the assumption of constant average density in the ISM, the ``fully radiative'' conditions leading predominantly to X-ray emission, as well as all the different relativistic effects described in the RSTT paradigm (see also section \ref{approximation}).

2) The afterglow fit does not depend directly on the parameters $\mu, \xi$ but only through their combination $E_{dya}$.  Thus there is a 1-parameter family of values of the pair $(\mu,\xi)$ allowed by a given viable value of $E_{dya}$ (see Fig~\ref{muxi} and section \ref{fp}).

3) By fine tuning the parameters of the best fit of the luminosity profile and time evolution of the afterglow the following parameters have been found:

\begin{equation}
E_{dya}=4.83\times 10^{53}erg,\; \; B=3\times 10^{-3} \ .
\label{values}
\end{equation}

After fixing in Eq.(\ref{values}) the two free parameters of the EMBH theory, modulo the mass-charge relationship which fixes $E_{dya}$,  we can derive all the space-time parameters of the GRB~991216 (see Tab.~\ref{tab1}) as well as the explicit dependence of the gamma factor as a function of the radial coordinate (see Fig.~\ref{gamma}).

Of special interest is the fundamental diagram of Fig.~\ref{tvsta}. Its role is essential in interpreting all quantities measured in arrival time (the time of an observer in an inertial frame at the detector) and their relations to the ones measured in the laboratory time by an observer in an inertial frame at the GRB source. The two times are clearly related by light signals (see Fig.~\ref{ttasch}) and expressed by the integral Eq.(\ref{tadef}) and are also affected by the cosmological expansion (see section \ref{arrival_time}).

\section{The explanation of the ``long bursts'' and the identification of the proper gamma ray burst (P-GRB)}\label{shortlongburst}

\begin{figure}[htbp]
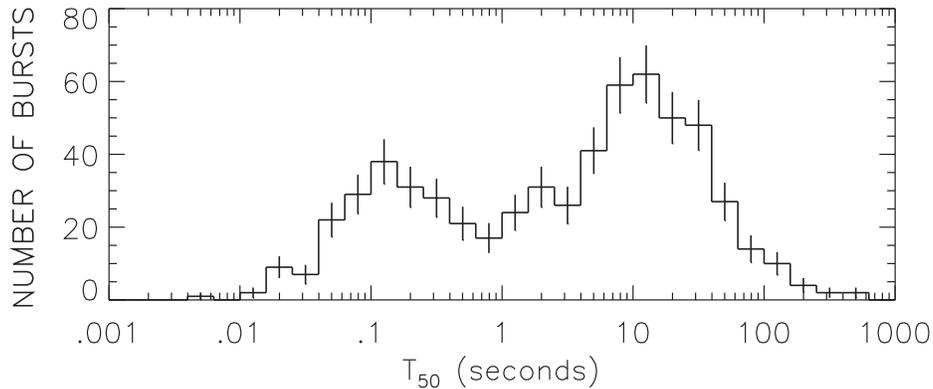

\PSFIG{t50}{\hsize}{0}
\caption{The distribution of the burst durations clearly shows two different classes of events: the ``short bursts'' and the ``long bursts'' (reproduced from Paciesas et al., 1999$^{83}$).}
\label{t50}
\end{figure}

Having determined the two free parameters of the EMBH theory, any other feature is a new prediction. An unexpected result soon became apparent, namely that the average luminosity of the main burst observed by BATSE can be fit by the afterglow curve (see Fig.~\ref{fit_1}). This led us to the identification of the long bursts observed by BATSE with the extended afterglow peak emission (E-APE). The peak of this E-APE occurs at $\sim 19.87\, s$ and its intensity and time scale are in excellent agreement with the BATSE observations.\cite{lett5} It is clear that this E-APE is {\em not} a burst, but is seen as such by BATSE due to its high noise threshold.\cite{lett5} Thus the outstanding unsolved problem of explaining the long GRBs\cite{wmm96,swm00,p01} is radically resolved: the so called ``long bursts'' do not exist, they are just E-APEs (see Fig.~\ref{t50}).

We now turn to the most cogent question to be asked: where does one find the burst which is emitted when the condition of transparency against Thomson scattering is reached? We have referred to this as the proper gamma ray burst (P-GRB) in order to distinguish it from the global GRB phenomena.\cite{lett1,brx00} We are guided in this search by two fundamental diagrams (see Fig.~\ref{crossen} and Fig.~\ref{dtab}):
\begin{enumerate}
\item In Ruffini, Salmonson, Wilson \& Xue (2000)\cite{rswx00} it is shown that for a fixed value of $E_{dya}$ the value of $B$ uniquely determines the energy $E_{P\hbox{-}GRB}$ of the P-GRB and the kinetic energy $E_{Baryons}$ of the ABM pulse which gives origin to the afterglow  (see Fig.~\ref{crossen}). For the particular values of the parameters given in Eq.~(\ref{values}), we find
\begin{equation}
E_{P\hbox{-}GRB}=7.54\times 10^{51}erg\, , \quad E_{Baryons}= 9.43\times 10^{52}erg \label{fittedvalues}
\end{equation}
and then:
\begin{equation}
\frac{E_{P\hbox{-}GRB}}{E_{Baryons}}=1.58\times 10^{-2}\, .
\label{fittedvalues2}
\end{equation}
\item One important additional piece of information comes from the differences in arrival time between the P-GRB and the peak of the E-APE, see Fig.~\ref{dtab}. Using the results of this figure and the numerical values given in Tab.~\ref{tab1}, we can retrace the P-GRB by reading off the time parameters of point 4 in Fig.~\ref{gamma}. Transparency is reached at $21.57\, s$ in comoving time at a radial coordinate $r=1.94\times 10^{14}$ cm in the laboratory frame and at $8.41\times 10^{-2}\, s$ in arrival time at the detector.
\end{enumerate}

All this, namely the energy predicted in Eq.(\ref{fittedvalues}) for the intensity of the burst and its time of arrival, leads to the unequivocal identification of the P-GRB with the apparently inconspicuous initial burst in the BATSE data. We have estimated from the BATSE data the ratio of the P-GRB to the E-APE over the noise threshold to be $\sim 10^{-2}$, in excellent agreement with the result in Eq.~(\ref{fittedvalues2}), see Fig.~\ref{final}.

\begin{figure}[htbp]
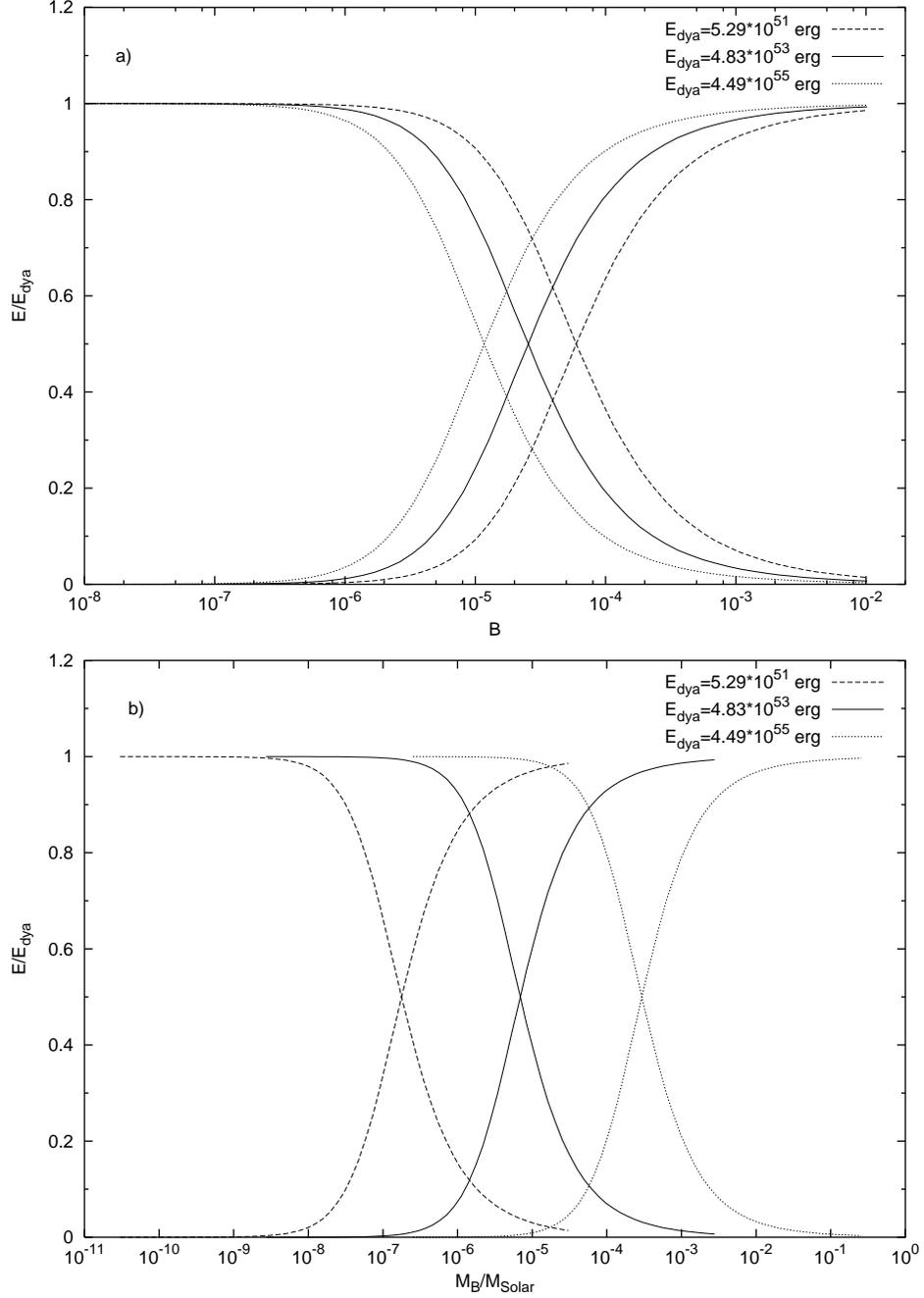

\PSFIG{b_multi2}{\hsize}{0}
\caption{{\bf a)} The same diagram of Fig.~\ref{fintkin} is plotted for three different $E_{dya}$ values: $E_{dya}=5.29\times 10^{51}$ erg (dashed lines), $E_{dya}=4.83\times 10^{53}$ erg (solid lines) and $E_{dya}=4.49\times 10^{55}$ erg (dotted lines). {\bf b)} Same as in a) but plotted as a function of the baryonic mass $M_B$ in units of solar masses instead of $B$.}
\label{b_multi2}
\end{figure}

It is important to emphasize that the diagrams in Fig.~\ref{fintkin} and Fig.~\ref{crossen} are not universal, but depend on the dyadosphere energy. The corresponding diagrams for three selected $E_{dya}$ values ($E_{dya}=5.29\times 10^{51}$ erg, $E_{dya}=4.83\times 10^{53}$ erg and $E_{dya}=4.49\times 10^{55}$ erg) are given in Fig.~\ref{b_multi2}a where we have plotted the energy of the P-GRB and of the E-APE as a function of $B$. The crossing of the intensity of P-GRB and E-APE occurs respectively at $B_1=6.0\times 10^{-5}$, $B_2=2.5\times 10^{-5}$ and $B_3=1.2\times 10^{-5}$ where $B_1>B_2>B_3$. In Fig.~\ref{b_multi2}b the same quantities are plotted as a function of the baryon mass $M_B$ in units of solar masses and the opposite dependence occurs: $M_1<M_2<M_3$.

\begin{figure}[htbp]
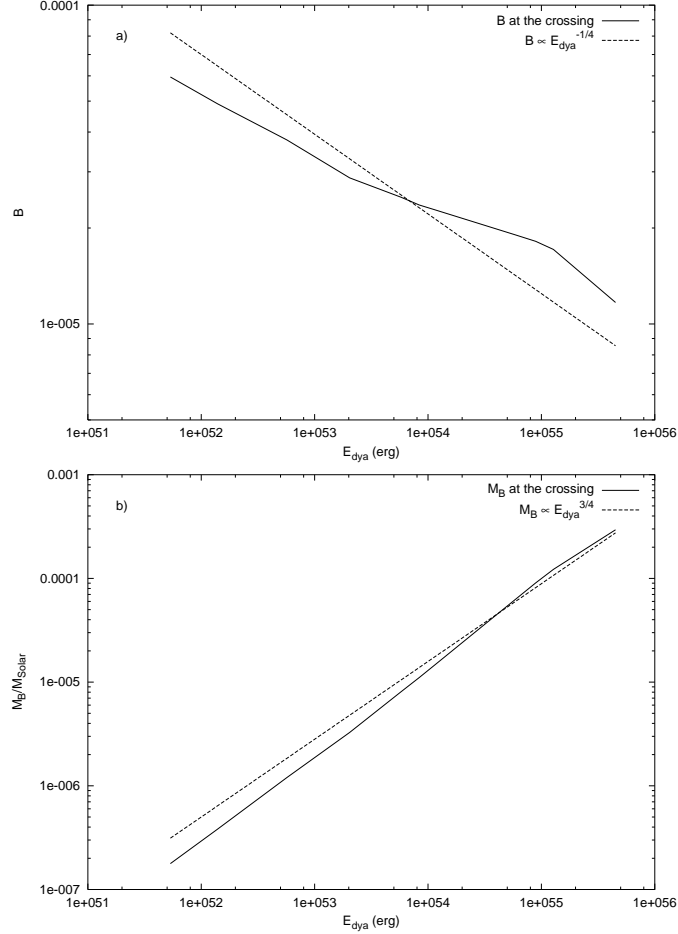

\PSFIG{b_cross2}{9cm}{0}
\caption{{\bf a)} The $B$ values corresponding to the crossings in Fig.~\ref{b_multi2}a are plotted versus $E_{dya}$ (solid line). The function $B\propto E_{dya}^{-1/4}$ obtained from a qualitative theoretical estimate (see Eq.(\ref{bmbboh})) is also plotted (dashed line). {\bf b)} The $M_B$ values corresponding to the crossings in Fig.~\ref{b_multi2}b are plotted versus $E_{dya}$ (solid line). The function $M_B\propto E_{dya}^{3/4}$ obtained from a qualitative theoretical estimate (see Eq.(\ref{bmbboh})) is also plotted (dashed line).}
\label{b_cross2}
\end{figure}

The physical reasons beyond these results is the following. We recall that the kinetic energy $E_{Baryons}$ and mass $M_B$ of PEMB pulse are
\begin{equation}
E_{Baryons}=(\gamma -1 )M_B\quad M_B\equiv BE_{dya}
\end{equation}
at the crossing point defined by
\begin{equation}
E_{Baryons}=E_{P\hbox{-}GRB}={1\over2}E_{dya}.
\end{equation}
From these two equations, we obtain
\begin{equation}
B={1\over2(\gamma_\circ-1)}\simeq {1\over2\gamma_\circ},
\end{equation}
$\gamma_\circ$ is the Lorentz gamma factor of the PEMB pulse at the transparency point, where (see section~\ref{at})
\begin{equation}
(n_{pair}+n_B)\sigma_T\simeq n_B\sigma_T=1,\hskip0.5cm n_B={M_B\over 4\pi r_\circ^2\Delta\gamma_\circ},
\end{equation}
$\Delta_t$ is the PEMB pulse thickness and $r_\circ$ the radial position at the transparency point. In addition, from the total energy conservation, we have
\begin{equation}
(\epsilon+n_B)\gamma^2_\circ4\pi r_\circ^2\Delta =const.,
\end{equation}
where $\epsilon$ is the thermal energy of the PEMB pulse. In the regime $n_B\gg\epsilon$, we have 
\begin{equation}
\gamma_\circ\simeq {E_{dya}\over M_B},
\end{equation}
and in the regime $n_B\ll\epsilon$, we have 
\begin{equation}
\gamma_\circ\sim r_\circ.
\label{g0rtapp}
\end{equation}
Considering the crossing point to occur in the second regime, we obtain at the crossing point
\begin{equation}
B\sim (E_{dya})^{-{1\over4}}, \quad M_B\sim (E_{dya})^{3\over4}.
\label{bmbboh}
\end{equation}
These results are plotted in Figs.~\ref{b_cross2}a--b. The agreement with the computed results is quite satisfactory. The differences can be attributed to the approximation adopted in Eq.(\ref{g0rtapp}) which is modified for high $B$ values.

The conclusion is that for increasing $E_{dya}$ also the baryonic mass corresponding to the cross increases, but in percentage it increases less than $E_{dya}$.

\section{Considerations on the P-GRB spectrum and the hardness of the short bursts}

\begin{figure}[htbp]
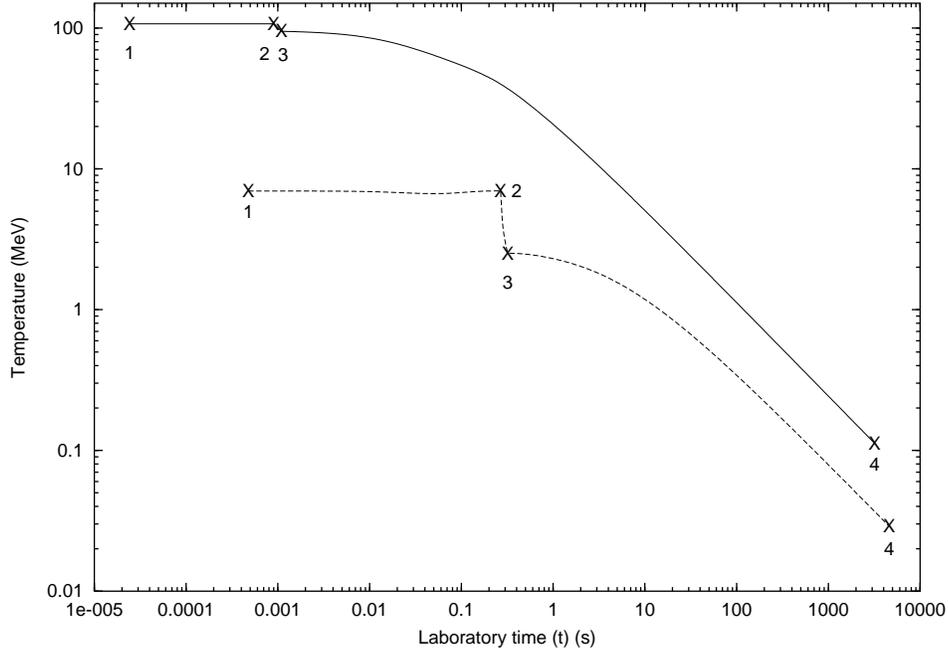

\PSFIG{ii-fig5}{\hsize}{0}
\caption{The temperature of the pulse in the laboratory frame for the first three eras of Fig.~1 of Ruffini et al. (2001a)$^1$ is given as a function of the laboratory time. The numbers 1, 2, 3, 4 represent the beginning and end of each era. The two curves refer to two extreme approximations adopted in the description of the dyadosphere. Details are given in Ruffini, Salmonson, Wilson \& Xue (2000)$^{50}$ and in section~\ref{fp}.}
\label{temp}
\end{figure}

Regarding the P-GRB spectrum, the initial energy of the electron-positron pairs and photons in the dyadosphere for given values of the parameters can be easily computed following the work of Preparata, Ruffini \& Xue (1998).\cite{prx98} We obtain respectively $T=1.95$ MeV and $T=29.4$ MeV in the two approximations we have used for the average energy density of the dyadosphere (see section~\ref{fp}). It is then possible to follow in the laboratory frame the time evolution of the temperature of the electron-positron pairs and photons through the different eras, see Fig.~\ref{temp}. The condition of transparency is reached at temperatures in the range of $\sim 15-55$ KeV at the detector, in agreement with the BATSE results. We emphasize that in the limit of $B$ going to $10^{-8}$ in which the P-GRB coincides with the ``short bursts'' the spectrum of the P-GRB becomes harder in agreement with the observational data\cite{na86,ba93,dbc99,fa00} (see Fig.~\ref{energypeak}).

All the above are average values derived from the two approximations used in Fig.~\ref{dens}. If one wishes to compare the EMBH theoretical results with the fine temporal details of the observational data on the P-GRB, a departure from this average approach will be needed and the fully time varying relativistic analysis outlined in Fig.~\ref{dyaform} applies as will be further discussed in section~\ref{gc}.

\section{Approximations and power laws in the description of the afterglow}\label{approximation}

In addition to the BATSE data, there is also clearly perfect agreement with the decaying part of the afterglow data from the RXTE and Chandra satellites. 

We can also establish at this point a first set of conclusions on the luminosity power law index ``$n$'' which is a function depending strongly on the transformation $t \rightarrow  t_a \rightarrow  t_a^d$ (see Fig.~\ref{tvsta}). In the current literature such transformations and the corresponding $n$ values are incorrect. Our theoretical value $n_{theo} = -1.6$ obtained for spherical symmetry for fully radiative conditions and constant density of the ISM is in agreement with observed $n_{obs} = -1.616 \pm 0.067$. No evidence of beaming is found in GRB~991216. We shall return to this point in the conclusions.

An extremely large number of papers in the literature deal with the power law index in the afterglow era. This issue has been particularly debated in connection with the aim of decreasing the energy requirements of GRBs by the effect of beaming.\cite{my94,dbpt94} It is currently very popular to infer the existence of beaming from the direct observations of breakings in the power-law index of the afterglow.\cite{mr97a,r97,mrw98,pmr98,dc98,sph99,pm99,r99,ha00,gdhl01} Our aim here is to underline an often neglected point that the power law index of the afterglow is the result of a variety of factors including the very different regimes in the relation between the laboratory time $t$ and the detector arrival time $t_a^d$ presented in Fig.~\ref{tvsta}. No meaningful statements on the values of the power-law index of the afterglow can be made having neglected these necessary considerations expressed in the RSTT paradigm. This becomes particularly transparent from the power law expansion in the semianalytic treatments we present below. It is therefore not so surprising, as we will show in the next session, that the results obtained in the EMBH theory differ from the ones in the current literature.

\subsection{The approximate expression of the hydrodynamic equations}

We proceed to a first approximation and expand Eqs.(\ref{heat2}, \ref{dgamma2}) to second order in the quantity
\begin{equation}
\frac{\Delta M_{\rm ism} c^2}{\rho_{B_1} V_1} \ll 1\,.
\label{expansion1}
\end{equation}
We obtain the following expressions:
\begin{equation}
\Delta E_{{\rm int}}  = \left( {\gamma _1  - 1} \right)\Delta M_{{\rm ism}} c^2 - \frac{1}{2}\frac{{\gamma _1^2  - 1}}{{M_B  + M_{{\rm ism}} }}\left( {\Delta M_{{\rm ism}} } \right)^2 c^2\, , \label{Eint2}
\end{equation}
\begin{equation} 
\Delta \gamma  =  - \frac{{\gamma _1^2  - 1}}{{M_B  + M_{{\rm ism}} }}\Delta M_{{\rm ism}} + \frac{3}{2}\gamma _1 \frac{{\gamma _1^2  - 1}}{{\left( {M_B  + M_{{\rm ism}} } \right)^2 }}\left( {\Delta M_{{\rm ism}} } \right)^2\, , \label{gammadecel2}
\end{equation}
where we set $\Delta \gamma \equiv \gamma_2-\gamma_1$ and have used the fact that $\rho_{B_1}V_1\equiv \left(M_B+M_{\rm ism}\right)c^2$. In the limit $\Delta E_{\rm int} \rightarrow dE_{\rm int}$, $\Delta \gamma\rightarrow d\gamma$,  and $\Delta M_{\rm ism}\rightarrow dM_{\rm ism}$, neglecting also second order terms, where
\begin{equation}
dM_{\rm ism}=4\pi r^2m_p n_{\rm ism}dr=4\pi r^2m_pn_{\rm ism}v dt,\hskip0.3cm v={dr\over dt},
\label{dm}
\end{equation}
and where the ISM number density $n_{\rm ism}$ is assumed for simplicity to be $n_{\rm ism}=1\, {\rm cm}^{-3}$, we obtain:
\begin{equation}
dE_{\rm int} = \left(\gamma - 1\right) dM_{\rm ism} c^2\,,
\label{Eint}
\end{equation}
\begin{equation}
d\gamma = - \frac{{\gamma}^2 - 1}{M_B + M_{\rm ism}} dM_{\rm ism}\,.
\label{gammadecel}
\end{equation} 
Eqs.(\ref{Eint}, \ref{gammadecel}) are limiting cases of Taub's hydrodynamical equations.\cite{taub,bor02,ll} They have been at times referred into the GRB literature as the Blandford-McKee equations.\cite{bm76} It is clear that the application of these equations holds if Eq.(\ref{expansion1}) applies. The behaviour of $\frac{\Delta M_{\rm ism} c^2}{\rho_{B_1} V_1}$ as a function of the radius when $M_{\rm ism} \ll M_B$ is:
\begin{equation}
\frac{\Delta M_{\rm ism} c^2}{\rho_{B_1} V_1} \sim \frac{r^2\Delta r}{M_B}.
\label{expansion1a}
\end{equation}
\begin{figure}[htbp]
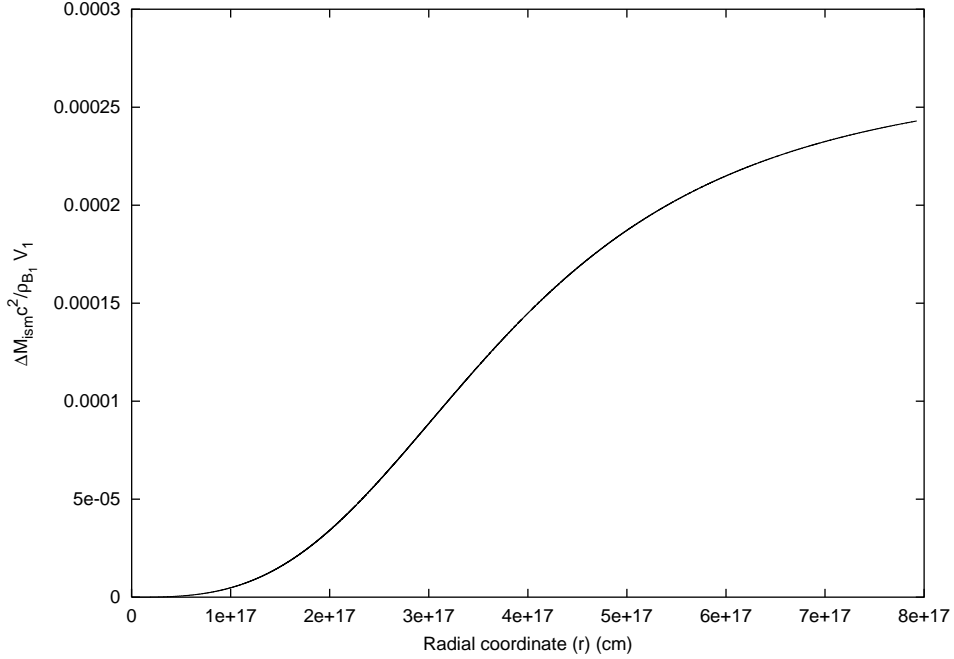

\PSFIG{delta_m}{\hsize}{0}
\caption{The factor $\frac{\Delta M_{\rm ism} c^2}{\rho_{B_1} V_1}$ is represented as a function of the radial coordinate. It is manifestly an increasing function.}
\label{dm_m_fig}
\end{figure}
The condition $M_{\rm ism} \ll M_B$ holds for GRB~991216 during the entire evolution of the system and so Eq.(\ref{expansion1}) is valid (see Fig.~\ref{dm_m_fig}).

Eqs.(\ref{Eint},\ref{gammadecel}) can be simply solved analytically (see e.g. Blandford \& McKee 1976).\cite{bm76} We then have:

\begin{equation}
\gamma={(M_B+M_{\rm ism})^2+C\over (M_B+M_{\rm ism})^2-C},
\label{dg}
\end{equation}
where

\begin{equation}
C={M_B}^2{\gamma_\circ-1\over\gamma_\circ +1},
\label{dgm2}
\end{equation}
where we recall that $r_\circ$ and 
$\gamma_\circ$ are the radial coordinate and the gamma factor at the transparency point and $M_B$ is the initial baryonic mass of the ABM pulse.

Eq.(\ref{dg}) is a differential equation for $r\left(t\right)$, namely

\begin{equation}
1 - \left( {\frac{{dr}}{{cdt}}} \right)^2  = \left[{(M_B+M_{\rm ism})^2+C\over (M_B+M_{\rm ism})^2-C}\right]^{-2}\,,
\label{dgeq}
\end{equation}
which can be integrated analytically with solution\cite{intbook}

\begin{eqnarray}
2c\sqrt C \left( {t - t_\circ } \right) = \left( {M_B  - m_i^\circ } \right)\left( {r - r_\circ } \right) \label{analsol} \\ \nonumber
+\frac{1}{4}m_i^\circ r_\circ \left[ {\left( {\frac{r}{{r_\circ }}} \right)^4  - 1} \right] \\ \nonumber
+\frac{{Cr_\circ }}{{6m_i^\circ B^2 }} \ln \left[ {\frac{{\left( {B + \frac{r}{{r_\circ }}} \right)^3 }}{{B^3  + \left( {\frac{r}{{r_\circ }}} \right)^3 }}\frac{{B^3  + 1}}{{\left( {B + 1} \right)^3 }}} \right] \\ \nonumber
+\frac{{Cr_\circ }}{{3m_i^\circ B^2 }} \left[\sqrt 3 \arctan \frac{{2\frac{r}{{r_\circ }} - B}}{{B\sqrt 3 }} - \sqrt 3 \arctan \frac{{2 - B}}{{B\sqrt 3 }}\right],
\end{eqnarray}
where $m_i^\circ=\frac{4}{3}\pi m_p n_{\rm ism} r_\circ^3$, $B=\left(\frac{M_B-m_i^\circ}{m_i^\circ}\right)^{1/3}$ and we recall that $t_\circ$ is the laboratory time at the transparency point. Clearly the fulfilment of Eq.(\ref{expansion1}) has to be checked to ensure the validity of this solution.

\subsection{The approximate expression of the emitted flux}\label{app_expr_em_flux}

\begin{figure}[htbp]
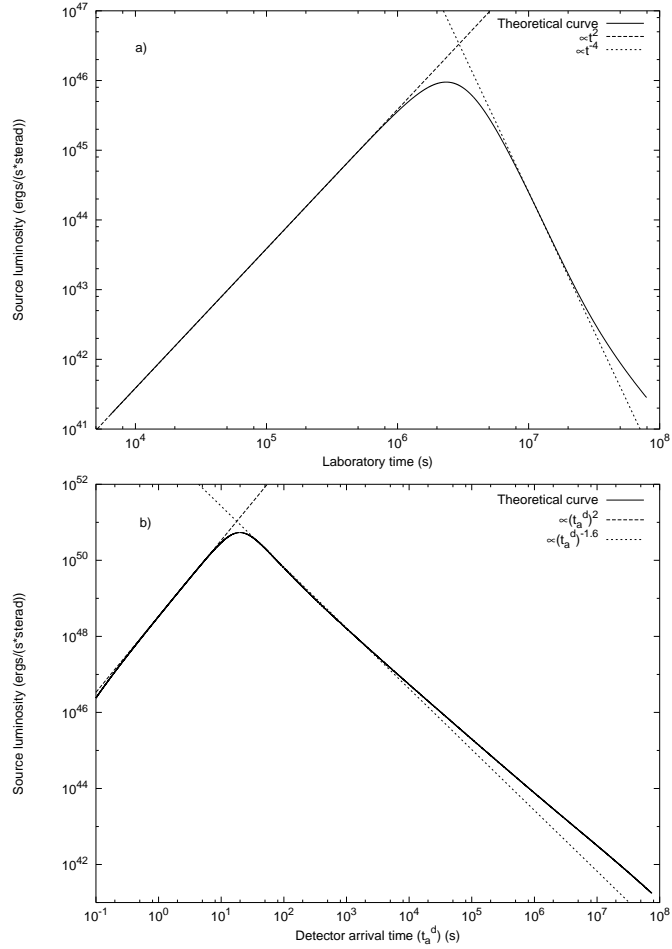

\PSFIG{flux_slo}{9cm}{0}
\caption{{\bf a)} The GRB flux emitted in laboratory time. {\bf b)} the flux emitted in the arrival time, measured by an observer at rest with respect to the detector (see section~\ref{approximation}).}
\label{fluxes}
\end{figure}

From Eqs.(\ref{Eint},\ref{gammadecel}), it follows that the emitted flux in the laboratory frame is given by (see Fig.~\ref{fluxes}a)

\begin{equation}
\frac{{dE}}{{dt}} = 4\pi r^2 n_{{\rm ism}} m_p v\gamma \left( {\gamma  - 1} \right)c^2, 
\label{fluxgen}
\end{equation}
and the corresponding flux in detector arrival time (see Fig.~\ref{fluxes}b) by

\begin{eqnarray}
\frac{{dE}}{{dt_a^d}} =\left[\frac{dt}{dt_a^d}\frac{dE}{dt}\right]_{t=t\left(t_a^d\right)} \\ \nonumber = 4\pi n_{{\rm ism}} m_p c^2 \left[vr^2 \gamma \left( {\gamma  - 1} \right)\frac{dt}{dt_a^d}\right]_{t=t\left(t_a^d\right)}.
\label{fluxgenarr}
\end{eqnarray}

For the solution of these equations we distinguish four different phases 
(A--D). The first two correspond to era V.

\subsection*{Phase A}

Just after the transparency condition is reached, the ISM matter involved is so small that we can approximately neglect the $M_{\rm ism}$ term in Eq.(\ref{dg}) and we have:

\begin{equation}
\gamma\simeq\gamma_\circ .
\label{gA}
\end{equation}
In the specific case of GRB 991216 we have $\gamma_\circ = 310.1$, $r_\circ = 1.94\times 10^{14}\, cm$, $t_\circ = 6.48\times 10^3\, s$, $t_{a_\circ}\simeq 4.21\times 10^{-2}\, s$ and $t_{a_\circ}^d\simeq 8.41\times 10^{-2}\, s$, where the index ``$\circ$'' refers to the quantities at the transparency point. We can then establish the following equation describing the ABM pulse motion in this phase: $r\left(t\right)=vt$ with $v\simeq c$. We can than use the following relation between laboratory time and arrival time:

\begin{equation}
t = 2 {\gamma_\circ}^2 t_a = \frac{2 {\gamma_\circ}^2 }{1+z}t_a^d ,
\label{appA}
\end{equation}
which is in perfect agreement with the full numerical computation (see Fig.~\ref{tvsta}).

We can substitute these equations into Eqs.(\ref{fluxgen},\ref{fluxgenarr}), obtaining:

\begin{equation}
\frac{{dE}}{{dt}} \propto \gamma _\circ^2 n_{{\rm ism}} t^2 
\label{fluxEA}
\end{equation}
in laboratory time and

\begin{equation}
\frac{{dE}}{{dt_a^d }} \propto \frac {\gamma _\circ^8 n_{{\rm ism}}}{\left(1+z\right)^3} {\left({t_a^d}\right)}^2 
\label{fluxAA}
\end{equation}
in arrival time, assuming $\gamma\left(\gamma-1\right)\simeq \gamma^2$. The results of the numerical integration of Eqs.(\ref{heat},\ref{dgamma}) are in perfect agreement with these approximations (see Fig.~\ref{fluxes}).

\subsection*{Points P -- the two maxima of the energy flux}

Since the contribution of the ISM mass in Eqs.(\ref{dg}--\ref{dgm2}) can no longer be neglected, the value of $\gamma$ starts to significantly decrease (see Fig.~\ref{gamma}) and the flux reaches a maximum value. We integrate Eq.(\ref{fluxgen}) and Eq.(\ref{fluxgenarr}) using Eq.(\ref{dg}) for $\gamma$, assuming $r\left(t\right)=vt$ with $v\simeq c$ and Eq.(\ref{appA}) for the relation between the laboratory time and the arrival time (see Figs.~\ref{rvst}--\ref{tvsta}). We can now obtain the point where the emitted flux reaches its maximum. In general, the location of the maximum of the flux, point $P$ in Ruffini et al. (2001a),\cite{lett1} will occur at different events, if considered in the arrival time $\left(P_A\right)$ or in the laboratory time $\left(P_L\right)$. In this second case, the point $P_L$ is determined by equating to zero the first derivative of Eq.(\ref{fluxgen}), and we have:

\begin{equation}
\gamma_{P_L}\simeq\frac{2}{3}\gamma_\circ, \quad 
\left. {\frac{M_B}{M_{\rm ism}}} \right|_{P_L}\simeq 2\gamma_\circ ,
\label{LabB}
\end{equation}
which in the case of GRB 991216 gives $\gamma_{P_L} = 206.7$ and 
$\left. {\frac{M_B}{M_{\rm ism}}} \right|_{P_L}\simeq 620.2$. 
The maximum of the observed flux is determined by equating to zero the first derivative of Eq.(\ref{fluxgenarr}). We obtain:

\begin{equation}
\gamma_{P_A}\simeq\frac{5}{6}\gamma_\circ, \quad 
\left. {\frac{M_B}{M_{\rm ism}}} \right|_{P_A}\simeq 5\gamma_\circ ,
\label{ArrB}
\end{equation}
which in the case of GRB 991216 gives $\gamma_{P_A}\simeq 258.4$ and $\left. {\frac{M_B}{M_{\rm ism}}} \right|_{P_A}\simeq 1550.5$.

The results of the numerical integration of Eqs.(\ref{heat},\ref{dgamma}) are in perfect agreement with these approximations (see Fig.~\ref{fluxes}).

\subsection*{Phase B -- the ``golden value'' $n=-1.6$}

In this phase $\gamma$ can no longer be considered constant and strongly decreases (see Fig.~\ref{gamma}). $M_{\rm ism}$ is increasing, but $v$ is still almost constant, equal to $c$. As a consequence, we can still say that $r\left(t\right)=vt$ with $v=c$, but the relation between laboratory time and arrival time given in Eq.(\ref{appA}) is no longer valid, and also Eq.(\ref{taapp}) is no longer applicable in this phase (see Fig.~\ref{tvsta}). We can instead write the following ``effective'' relation:

\begin{equation}
t\propto {\left(t_a^d\right)}^{0.20} ,
\label{appC}
\end{equation}
which is a result of a best fit of the numerical data in this region. Expanding the squares in Eq.(\ref{dg}), neglecting $M_{\rm ism}^2$ with respect to $M_B^2$ but retaining the terms in $M_{\rm ism}$ and assuming $\gamma_\circ \gg 1$ we obtain:

\begin{equation}
\gamma\sim\frac{M_B}{M_{\rm ism}}\sim 
\gamma_{P_L}\frac{r_{P_L}^3}{r^3}=\gamma_{P_L}\frac{t_{P_L}^3}{t^3} ,
\label{gammaC}
\end{equation}
where $r_{P_L}$ and $t_{P_L}$ are the values of $r$ and $t$ at point $P_L$. Substituting this result into Eqs.(\ref{fluxgen}), we obtain the emitted flux in the laboratory frame, given by
\begin{equation}
\frac{{dE}}{{dt}} \propto \gamma _P^2 t_P^6 n_{{\rm ism}} t^{-4} \, ,
\label{fluxEC}
\end{equation}
and this is in good agreement with the full numerical computation (see Fig.~\ref{fluxes}).

To obtain an analytic formula for the observed flux on the detector, we can still try to use the approximate relation between $t$ and $t_a^d$ given by Eq.(\ref{taapp}):
\begin{equation}
t = 2 {\gamma\left(t\right)}^2 t_a = \frac{2 {\gamma\left(t\right)}^2 }{1+z}t_a^d ,
\label{appC1}
\end{equation}
where $\gamma\left(t\right)$ is given by Eq.(\ref{gammaC}). We obtain:
\begin{equation}
t=\left(\frac{2\gamma_{P_L}^2t_{P_L}^6}{1+z}t_a^d\right)^{1/7}\, .
\label{appC2}
\end{equation}

Using this formula in Eq.(\ref{fluxgenarr}), we finally obtain:
\begin{equation}
\frac{{dE}}{{dt_a^d }} \propto \frac{ \gamma_P^{\frac{8}{7}} t_P^{\frac{24}{7}} n_{{\rm ism}}}{\left(1+z\right)^{-\frac{17}{7}}} {\left(t_a^d\right)}^{-\frac{10}{7}} 
\label{fluxACW}
\end{equation}
where we again assumed $\gamma\left(\gamma-1\right)\simeq \gamma^2$. This results are not in agreement with the observational data, because the power-law index for the observed flux is $-10/7\simeq -1.43$, instead of the observed value $-1.6$.

This is a confirmation that Eq.(\ref{appC1}) cannot be applied in this phase, as instead has been done by many authors in the current literature. We instead have to use Eq.(\ref{appC}). In fact, doing so we obtain the correct value:
\begin{equation}
\frac{{dE}}{{dt_a^d }} \propto n_{{\rm ism}} {\left(t_a^d\right)}^{-1.6} ,
\label{fluxAC}
\end{equation}
The results of the numerical integration of Eqs.(\ref{heat},\ref{dgamma}) are in perfect agreement with these approximations (see Fig.~\ref{fluxes}), which implies that the approximate Eq.(\ref{Eint},\ref{gammadecel}) can still be used in this regime, but not Eq.(\ref{taapp}), which has to be replaced by an ``effective'' local power-law behaviour (see Eq.(\ref{appC})).

\subsection*{Phase C}

This new phase begins when $\gamma$ has decreased so much that the approximation $r=ct$ is no longer valid (see Fig.~\ref{rvst}). In the case of GRB~991216 this happens when $\gamma\simeq 3.0$, $t\simeq 1.5\times 10^7$ s, $t_a^d\simeq 2.9\times 10^5$ s and $r\simeq 4.4\times 10^{17}$ cm. In this entire phase, $r\left(t\right)$ manifests the following behaviour typical of damped motion:

\begin{figure}[htbp]
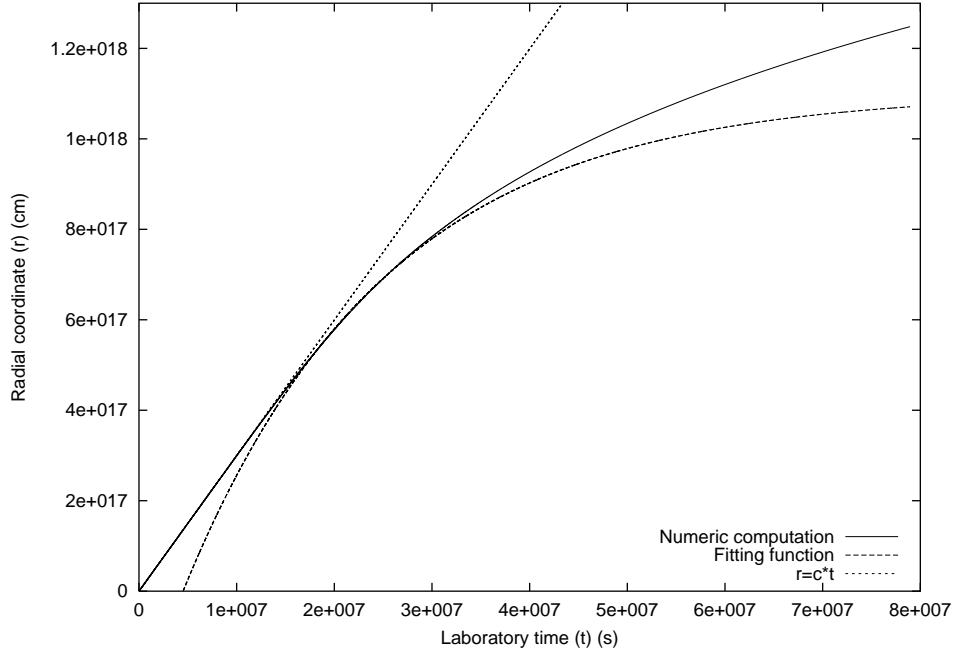

\PSFIG{fit_r_all}{\hsize}{0}
\caption{The exact numerical solution for $r\left(t\right)$ (solid line), together with the line $r=ct$ (dotted line) and the fitting function given in Eq.(\ref{rdit}) (dashed line).}
\label{rvst}
\end{figure}

\begin{equation}
r\left( t \right) = \hat r \left( {1 - e^{ - \frac{{t - t^ \star }}{\tau }} } \right),
\label{rdit}
\end{equation}
where $\hat r$, $t^\star$ and $\tau$ are constants that can be determined by the best fit of the numerical solution. In the present case of GRB 991216 we obtain:

\begin{equation}
\hat r\simeq 1.101\times 10^{18} cm,\quad 
\tau\simeq 2.072\times 10^7 s,\quad 
t^\star\simeq 4.52\times 10^6 s.
\label{fitted}
\end{equation}

It is important to note that this interesting behaviour, typical of a damped motion, does not lead to any power-law relationship for the emitted flux as a function of the laboratory time (see Fig.~\ref{fluxes}). However, if we look at the observed flux as a function of the detector arrival time, we see that a power-law relationship still can be established, fitting the numerical solution. The result is:
\begin{equation}
\frac{{dE}}{{dt_a^d }} \propto {\left(t_a^d\right)}^{-1.36}.
\label{fluxCfit}
\end{equation}
This quite unexpected result can be explained because the relation between $t$ and $t_a^d$ depends on $r\left(t\right)$ in a nonpower-law behaviour. This fact balances the complex behaviour of the emitted flux as a function of the laboratory time, leading finally again to a power-law behaviour arrival time.

In this last phase, however, the flux decreases markedly, and from the point of view of the GRB observations, the most relevant regions are phases {\em A} and {\em B} described above, as well as the peak separating them.

\subsection*{Phase D}

This last phase starts when the system approaches a Newtonian regime. In the case of GRB~991216 this occurs when $\gamma\simeq 1.05$, $t\simeq 5.0\times 10^7$ s, $t_a^d\simeq 3.1\times 10^7$ s and $r\simeq 1.0\times 10^{18}$ cm. In this phase $r\left(t\right)$ is again approaching a linear behaviour, due to the velocity decreasing less steeply than in Phase C. The emitted flux as a function of the laboratory time still does not show a power-law behaviour, while the observed flux as a function of detector arrival time does, with an index $n=-1.45$ (see Fig.~\ref{fluxes}).

\section{The power-law index of the afterglow and inferences on beaming in GRBs}\label{power-law}

\begin{table}
\ttbl{30pc}{We compare and contrast the results on the power-law index {\em n} of the afterglow in the EMBH theory with other treatments in the current literature, in the limit of high energy and fully radiative conditions. The differences between the values of $-10/7\sim -1.43$ (Dermer) and the results $-1.6$ in the EMBH theory can be retraced to the use of the two different approximation in the arrival time versus the laboratory time given in Fig.~\ref{tvsta}. See details in section~\ref{approximation}.}
{\tiny
\begin{tabular}{c|c|c|c|c|c}
\multicolumn{6}{c}{ }\\
&& Chiang \& Dermer (1999)\cite{cd99} & Piran (1999)\cite{p99} &&\\
& EMBH theory & Dermer, Chiang \& B\"ottcher (1999)\cite{dcb99} & Sari \& Piran (1999)\cite{sp99} & Vietri (1997)\cite{v97} & Halpern et al. (2000)\cite{ha00}\\
&& B\"ottcher \& Dermer (2000)\cite{bd00} & Piran (2001)\cite{p01} &&\\
\hline \hline
&&&&&\\
Ultra-relativistic & $\displaystyle{\gamma=\gamma_\circ}$  & $\displaystyle{\gamma=\gamma_\circ}$ & $\displaystyle{\gamma=\gamma_\circ}$ & &\\
&&&&&\\
& $\gamma_\circ=310.1$ &&&& \\
&&&&&\\
 & $n=2$ & $n=2$ & $n\simeq2$ && \\
&&&&&\\
\hline
&&&&&\\
Relativistic & $\displaystyle{\gamma \simeq r^{-3}}$ & $\displaystyle{\gamma \sim r^{-3}}$ & $\gamma \sim r^{-3}$  && 
$n>-1.47$\\
&&&&&\\
& $3.0<\gamma<258.5$ &&&&  \\
&&&&&\\
 & $n=-1.6$ & $n=-\frac{10}{7}=-1.43$ & $n=-\frac{5.5}{4}=-1.375$ && \\
&&&&&\\
\hline
&&&&&\\
Non-relativistic & $n=-1.36$ &&& $n=-1.7$ & \\
&&&&&\\
& $1.05<\gamma<3.0$ &&&&\\
&&&&&\\
\hline
&&&&&\\
Newtonian & $n=-1.45$ &&&&\\
&&&&&\\
& $1<\gamma<1.05$ &&&&\\
&&&&&\\
\hline
\multicolumn{3}{c}{}\\
\end{tabular}
}
\label{tab2}
\end{table}

The results obtained in the previous sections have emphasized the relevance of the proper application of the RSTT paradigm to the determination of the power-law index of the afterglow. Particularly interesting is the subtle interplay between the different regimes in the relation between the laboratory time and the arrival time at the detector clearly expressed by Fig.~\ref{tvsta} and the corresponding different regimes encountered in the first order expansion of the relativistic hydrodynamic equations of Taub (1948)\cite{taub} (see section~\ref{approximation}). It is interesting to compare and contrast our treatment with selected results of the current literature, in order to illustrate some relevant points (see Tab.~\ref{tab2}). We will consider the results in the literature only with reference to the limiting case which we address in our work: the condition of fully radiative emission.

The first line of Tab.~\ref{tab2} describes the ultrarelativistic regime, corresponding to an increasing energy flux of the afterglow as a function of the arrival time (phase A in previous section). Our treatment and the results in the literature by Dermer et al.\cite{dcb99,cd99,bd00} coincide. They agree as well with the results by Piran et al.\cite{p01,p99,sp99}

The second line corresponds to the relativistic regime, in which the energy flux of the afterglow, after having reached the maximum (point P in previous section), monotonically decreases (phase B in previous section). The dependence we have found of the gamma factor on the radial coordinate of the expanding ABM pulse does coincide with the one given by Dermer et al. and Piran et al. Our power law index $n$ in this regime, which perfectly fits the data, however, is markedly different from the others. Particularly interesting is the difference between our results and those of Dermer et al: the two treatments coincide up to the last relation between the laboratory time and the arrival time at the detector. As explained in Eqs.(\ref{fluxACW}-\ref{fluxAC}), the two treatments differ in the approximation adopted in relating the laboratory time to the arrival time at the detector, illustrated in Fig.~\ref{tvsta}. Dermer et al. incorrectly adopted the approximation represented by the lower curve in Fig.~\ref{tvsta} and consequently they do not find agreement with the observational data. We have not been able to retrace in the treatment by Piran et al. the steps which have led to their different results. Special mention must be made of a result stated by Halpern et al. (2000),\cite{ha00} the last entry in line 2, that an absolute lower limit for the power-law index $n-1.47$ can be established on theoretical grounds. Such a result, clearly not correct also on the basis of our analysis, has been erroneously used ti support the existence of beaming in GRBs, as we will see below.

The third line in Tab.~\ref{tab2} is also interesting, treating the nonrelativistic limit (Phase C in previous section). This regime has been analysed by Vietri (1997),\cite{v97} avoiding the exact integration of the equations and relying on simple qualitative arguments. These results are not confirmed by the integration of the equations we have performed. This is an interesting case to be examined for its pedagogical consequences. Having totally neglected the relation between the laboratory time and the time of arrival at the detector, which we have illustrated in Fig.~\ref{tvsta}, and identifying $t_a^d\equiv t$, Vietri reaches a very different power law from our. Moreover, his solution brings to an underestimation of the radial coordinate: he estimated a radial coordinate of $1.1\times 10^{15}\,cm$ at $t_a^d=3.5\times 10^4\,s$, while the exact computation shows a result greater than $3.0\times 10^{17}\,cm$ (see Tab.~\ref{tab1}). On the other hand if one assumes, from the above mentioned identity $t_a^d\equiv t$, $t=3.5\times 10^4\,s$, one obtains a gamma factor of $\sim 300$ (see Tab.~\ref{tab1}) in total disagreement with the nonrelativistic approximation adopted by Vietri. Quite apart from this pedagogical value, this nonrelativistic phase is of little interest from the observational point of view, due to the smallness of the flux emitted. 

For completeness, we have also shown our estimates of the index $n$ as the Newtonian phase approaches in the last line of Tab.~\ref{tab2}.

The perfect agreement between our theoretically predicted value for the power-law index, $n_{theo}$, and the observed one, $n_{obs}$,
\begin{equation}
n_{theo}=-1.6,\quad n_{obs}=-1.616\pm0.067,
\label{nembh}
\end{equation}
confirms the validity of our major assumptions:
\begin{enumerate}
\item The fully radiative regime.
\item The constant average density of the ISM ($n_{ism}=1\, proton/cm^3$).
\item The spherical symmetry of the emission and the absence of beaming in GRB~991216.
\end{enumerate}

After the work of Mao \& Yi (1994)\cite{my94} pointing to the possibility of introducing beaming to reduce the energetics of GRBs and after the discovery of the afterglow, many articles have appeared trying to obtain theoretical and observational evidence for beamed emission in GRBs. The observations have ranged from radio\cite{rh99,f00} to optical\cite{ga00,ha00,sa00,s00} all the way to X-rays. Particular attention has been devoted to relating the existence of beaming to possible breaks in the light curve slope, generally expected at a value of the gamma factor
\begin{equation}
\gamma=\frac{1}{\vartheta_0},
\label{beam}
\end{equation}
where $\vartheta_0$ is the beam opening angle. There are many articles on this subject; to mention only the most popular ones, we recall\cite{r97,r97b,r99,mrw98,pm99,sph99}. Far from having reached a standard formulation, these approaches differ from each other in the expected time at which the break should take place up to a factor of $20$.\cite{sph99}. They differ as well for the opening angle of the beam, up to a factor of $3$.\cite{sph99} Disagreement still exists on the number of breaking points: two in the case of,\cite{pm99} one in the case of,\cite{sph99} one again in the case of\cite{r97,r97b,r99} but differing in position from the one of.\cite{sph99} It has also been noticed that other authors have shown through numerical simulations that such a transition, if visible at all, is not very sharp.\cite{ha00}

Ample observational data have been obtained for the GRB~991216, in addition to the X-ray band, also in the optical and radio. For the reason mentioned at the beginning of section~\ref{bf}, we only address in this article the problem of the $\gamma$- and the X-ray emission. In that respect, the main article addressing the issue of beaming in the X-rays for GRB~991216 is the one of.\cite{ha00} The key argument is based on the theoretical inequality claimed to exist for the power-law index $n>-1.47$ (see above). The fact that the observed X-ray decay rate is found to be $n_{obs}=1.616\pm 0.067$ is interpreted by the authors as evidence for beaming. Moreover, the fact that the decay rate $n=-1.6$ has been observed before a steepening in the optical decay occurred at approximately 1 day of arrival time authorized an even more extreme proposal of a narrower beam in the X-rays within the optical beam.

It is clear from the entire treatment which we have presented and the results of the EMBH theory given by $n_{theo}=-1.6$ that there is no evidence for such a beaming, as already stated above. The motivation by Halpern et al. (2000)\cite{ha00} stems from the incorrect theoretical assumption of the existence of a lower limit in the afterglow power-law index $n>-1.47$. From our theoretical analysis the existence of $n=-1.6$ is clear proof of isotropic emission in the GRB~991216 and a clear test of the complete relativistic treatment of the source. The fact that the break in the index should be ``achromatic'' and the absence of beaming in the X-rays imply an absence of beaming also in the optical and radio bands. The observed steepening in the optical decay has to find an alternative explanation. Although this is not the subject of our present work for the above mentioned reasons, we have found interesting the considerations by Panaitescu \& Kumar (2001)\cite{pk01}, which find that ``there are some major difficulties to apply a jet model to GRB~991216''. They also state, still for GRB~991216, that ``the steepening of the optical decay of a few days is not due to a jet effect, as suggested by Halpern et al. (2000),\cite{ha00} but to the passage of a spectral break''.

Concerning our own position on the possibility of beaming in GRBs, we would like just to remark that, from a preliminary analysis of beamed emission within the EMBH model, we have found some new features which are not encompassed by the results in the current literature, and they could become a distinctive signature for the discrimination of the existence or nonexistence of beaming.\cite{rbcfx02b_beam} The study of the steepening in the optical and radio decay is addressed within the EMBH theory in a forthcoming paper.\cite{rbcfx02c_spectrum}

\section{Substructures in the E-APE due to inhomogeneities in the Interstellar medium}\label{substructures}

The afterglow is emitted as the ABM pulse plows through the interstellar matter engulfing new baryonic material. In our previous articles we were interested in explaining the overall energetics of the GRB phenomena and in this sense, we have adopted the very simplified assumption that the interstellar medium is a constant density medium with $n_{ism}=1/cm^3$. Consequently, the afterglow emission obtained is very smooth in time.  We are now interested in seeing if in this framework we can also explain most of the time variability observed by BATSE, all of which except for the P-GRB should correspond to the beam-target phase in the IBS paradigm. 

We pursue this treatment still neglecting the angular spreading due to off-axis scattering in the radiation of the afterglow.

Our goal is to focus in this simplified model on the basic energetic parameters as well as on the drastic consequences of the space-time variables expressed in the RSTT paradigm.

\begin{figure}[htbp]
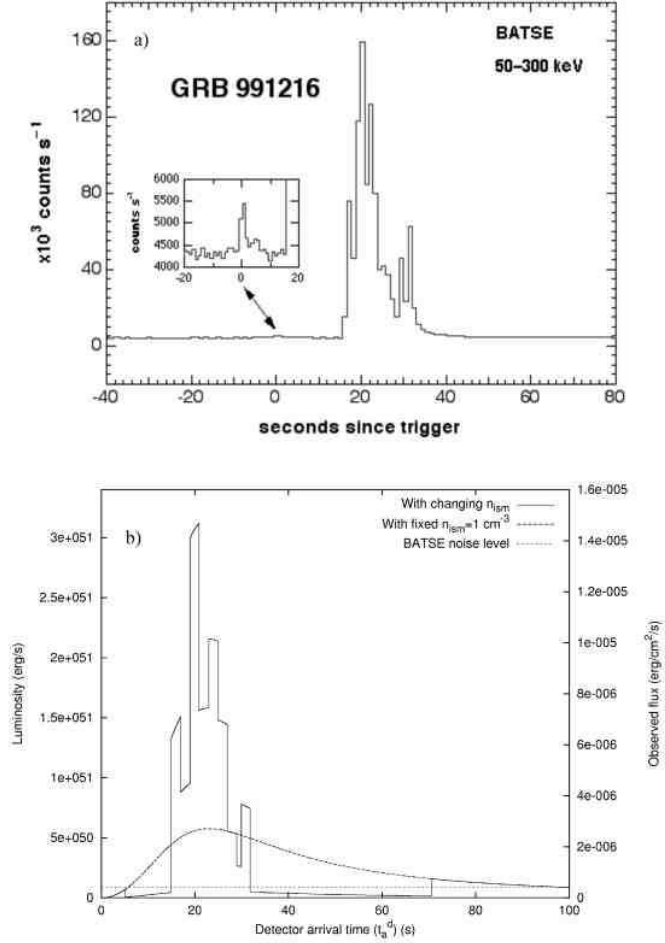

\PSFIG{v-fig3}{9cm}{0}
\caption{a) Flux of GRB 991216 observed by BATSE. The enlargement clearly shows the P-GRB.$^2$ b) Flux computed in the collision of the ABM pulse with an ISM cloud with the density profile given in Fig.~\ref{denstity_prof}. The dashed line indicates the emission from an uniform ISM with $ n = 1 cm^{-3} $. The dotted line indicates the BATSE noise level.}
\label{fit_subs_1}
\end{figure}

Having obtained the two results presented in Fig.~\ref{gamma} and Fig.~\ref{fluxes}, we can proceed to attack the specific problem of the time variability observed by BATSE.

The fundamental point is that in both regimes {\em the flux observed in the arrival time is proportional to the interstellar matter density}: any inhomogeneity in the interstellar 
medium $\Delta n_{ism}/ \overline{n}_{ism}$ will lead correspondingly to a proportional variation in the intensity  $\Delta I/ \overline{I}$ of the afterglow. This result has been erroneously interpreted in the current literature as a burst originating in an unspecified ``inner engine''.
 
\begin{figure}[htbp]
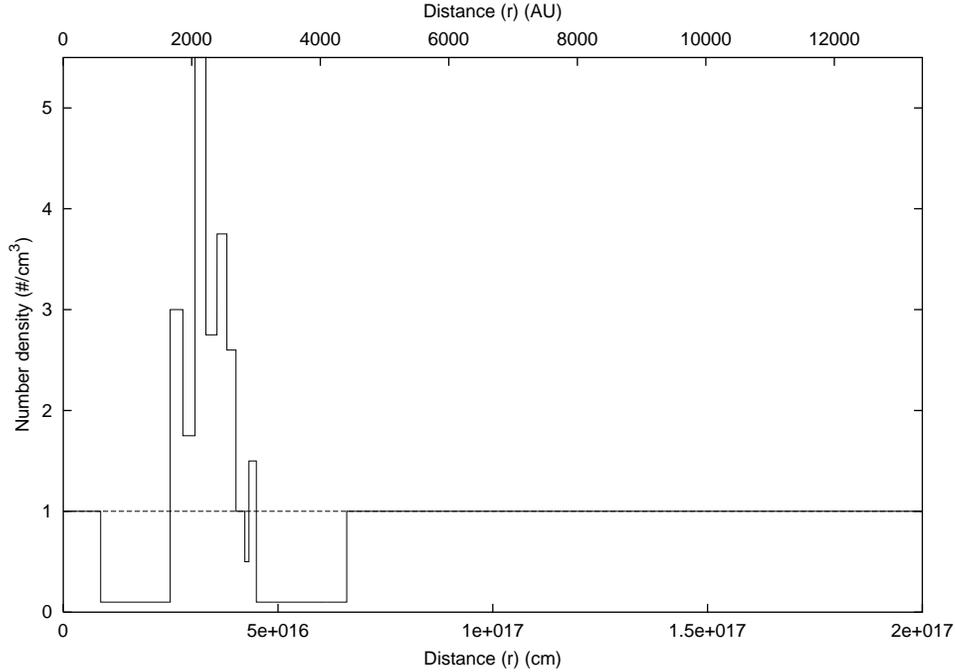

\PSFIG{v-fig4}{\hsize}{0}
\caption{The density contrast of the ISM cloud profile introduced in order to fit the observation of 
the burst of GRB991216. The dashed line indicates the average uniform density $ n = 1 cm^{-3} $.}
\label{denstity_prof}
\end{figure}

\begin{figure}[htbp]
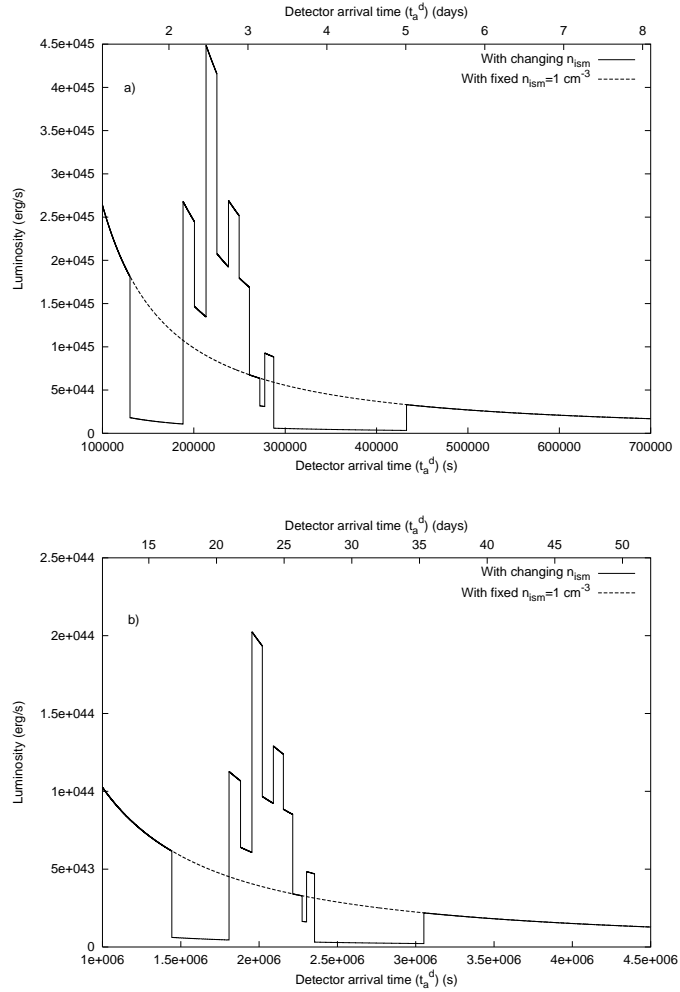

\PSFIG{v-fig5}{9cm}{0}
\caption{(a) Same as Fig.~\ref{fit_subs_1}b with the ISM cloud located at a distance of $3.17\times 10^{17} cm$ from the EMBH, the time scale of the burst now extends to $\sim 1.58 \times 10^5 s $. (b) Same as (a) with the ISM cloud at a distance of $4.71\times 10^{17} cm$ from the EMBH, the time scale of the burst now extends to $\sim 1.79 \times 10^6 s $.}
\label{fit_subs_2}
\end{figure}

In particular, for the main burst observed by BATSE (see Fig.~\ref{fit_subs_1}a) we have
\begin{equation}
 \left(\Delta 
I/ \overline{I} \right) =\left(\Delta n_{ism}/n_{ism}\right) \sim 5 .
\label{disui}
\end{equation}
There are still a variety of  physical circumstances which may lead to such density inhomogeneities.

The additional crucial parameter in understanding the physical nature of such inhomogeneities is 
the time scale of the burst observed by BATSE. Such a burst lasts $ \Delta t_a \simeq 20 s $ and shows substructures on a time scale of $ \sim 1s$ (see Fig.~\ref{fit_subs_1}a). In order to infer the nature of the structure emitting such a burst we must express these times scales in the laboratory time.\cite{lett1} Since we are at the peak of the GRB we have $ \gamma_{P_A} \sim 258.5 $ (see Eq.(\ref{ArrB})) and $ \Delta t_a $ corresponds in the laboratory time to an interval
\begin{equation}
\Delta t \sim 1.0\times 10^6 s ,
\label{deltatl}
\end{equation}
which determines the characteristic size of the inhomogeneity creating the burst $\Delta L\sim 5.0 \times 10^{16} cm$ (see Tab.~\ref{tab1} and Fig.~\ref{tvsta}).

It is immediately clear from Eq.(\ref{disui}) and Eq.(\ref{deltatl}) that these are the typical 
dimensions and density contrasts corresponding to a small interstellar cloud. As an explicit example we have shown in Fig.~\ref{denstity_prof} the density contrasts and dimensions of an interstellar cloud with an {\em average density} $<n>=1/cm^3$. Such a cloud is located at a distance of $\sim 8.7\times 10^{15}cm$ from the EMBH, gives rise to a signal similar to the one observed by BATSE (see Fig.~\ref{fit_subs_1}b).

It is now interesting to see the burst that would be emitted, if our present approximation would still apply, by the interaction of the ABM pulse 
with the same ISM cloud encountered at later times during the evolution of the afterglow. Fig.~\ref{fit_subs_2}a shows the expected structure of the burst at a distance $4.1\times 10^{17}cm$, corresponding to an arrival time delay of $\sim 2$ days, where the gamma factor is now $\gamma_\star \sim 3.6$. It is interesting that the overall intensity would be smaller, the intensity ratio of the burst relative to the average emission would remains consistent with Eq.(\ref{disui}), but the time scales of the burst would be longer by a factor $ \left( \frac{\gamma_{P_A}}{\gamma_\star} \right)^2 \simeq  5 \times 10^3$. Fig.~\ref{fit_subs_2}b shows the corresponding quantities for the same ISM cloud located at a distance $6.4\times 10^{17}cm$ from the EMBH, corresponding to an arrival time delay of $\sim 1$ month, where the gamma factor is $\sim 1.5$.

We return in future work\cite{rbcfx02a_sub} to examine the angular spreading effects pointing out how they improve the results presented here: the explanation of the time variability observed in the so called ``long bursts'' in the BATSE classification of GRBs is confirmed. The smoothness, namely the absence of the above mentioned substructures, observed in the latest phases of the afterglow finds as well a most natural explanation.

\section{The observation of the iron lines in GRB~991216: on a possible GRB-Supernova time sequence}\label{gsts}

We have seen in the previous sections how the time structure of the E-APE gives information on the composition of the interstellar matter at distances of the order of $5\times 10^{16}\, cm$ from the source. We would like now to point out that the data on the iron lines from the Chandra satellite on the GRB~991216\cite{p00} and similar observations from other sources\cite{p99b,a00,p00} make it possible to extend this analysis to a larger distance scale, possibly all the way out to a few light years, and consequently probe the distribution of stars in the surroundings of the newly formed EMBH.

Most importantly, these considerations lead to a new paradigm for the interpretation of the supernova-GRB correlation.\cite{lett3} Indeed a correlation between the occurrence of GRBs and supernova events exists and has been established by the works.\cite{b99,g98b,g98c,g00,k98,p98a,p99,r99,vp00}

Such an association has been assumed to indicate that GRBs are generated by supernova explosions.\cite{k98} In turn, such a point of view has implied further consequences: the optical and radio data of the supernova have been attributed to the GRB afterglow, and many theorists have tried to encompass these data and explain them as a genuine component of the GRB scenario.

We propose instead an alternative point of view implying a very clear distinction between the GRB phenomenon and the supernova: if relativistic effects presented in the RSTT paradigm are properly taken into account, then a kinematically viable explanation can be given of the supernova-GRB association. We still use GRB~991216 as a prototypical case.

The GRB-Supernova Time Sequence paradigm, which we have indicated for short as GSTS paradigm,\cite{lett3} states that: {\em A massive GRB-progenitor star $P_1$ of mass $M_1$ undergoes gravitational collapse to an EMBH. During this process a dyadosphere is formed and subsequently the P-GRB and the E-APE are generated in sequence. They propagate and impact, with their photon and neutrino components, on a second supernova-progenitor star $P_2$ of mass $M_2$. Assuming that both stars were generated approximately at the same time, we expect to have $M_2 < M_1$. Under some special conditions of the thermonuclear evolution of the supernova-progenitor star $P_2$, the collision of the P-GRB and the E-APE with the star $P_2$ can induce its supernova explosion}.

Especially relevant to our paradigm are the following data from the Chandra satellite:\cite{p00}
\begin{enumerate}
\item At the arrival time of 37 hr after the initial burst there is evidence of
iron emission lines for GRB~991216.
\item The emission lines are present during the entire observation period of $10^4$ s. The iron lines could also have been produced earlier, before Chandra was observing. Thus the times used in these calculations are not unique: they do serve to provide an example of the scenario.
\item The emission lines appear to have a peak at an energy of $3.49 \pm 0.06$ keV which, at a redshift $z=1.00 \pm 0.02$ corresponds to an hydrogen-like iron line at 6.97 keV at rest. This source does not appear to have any significant motion departing from the cosmological flow. The iron lines have a width of 0.23 keV consistent with a radial velocity field of $0.1c$. The iron lines are only a small fraction of the observed flux.
\end{enumerate}

On the basis of the explicit computations of the different eras presented in the above sections, we make three key points:
\begin{enumerate}
\item An arrival time of $37$ hr in the detector frame corresponds to a radial distance from the EMBH travelled by the ABM pulse of $3.94\times10^{17}$ cm in the laboratory frame (see Tab.~\ref{tab1}).
\item It is likely that a few stars are present within that radius as members of a cluster. It has 
become evident from observations of dense clusters of star-forming regions that a stellar average density of typically $ 10^2 \mbox{pc}^{-3} $ \cite{btk00} should be expected. There is also the distinct possibility for this case and other systems that the stars $P_1$ and  $P_2$ are members of a binary system.
\item The possible observations at different wavelengths of the supernova crucially depend on the relative intensities between the GRB and the supernova as well as on the value of the distance and the redshift of the source. In the present case of GRB~991216, the expected optical and radio emission from the supernova are many orders of magnitude smaller than the GRB intensity. The opposite situation will be encountered in GRB~980425.\cite{rbcfx02d_supernova}
\end{enumerate}

\begin{figure}[htbp]
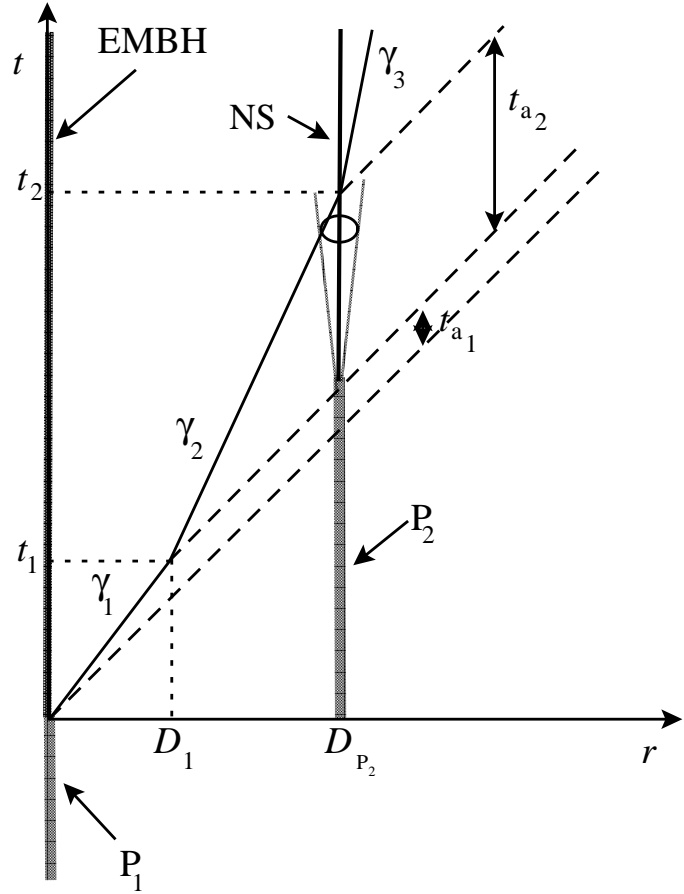

\PSFIG{iii-fig1}{9cm}{0}
\caption{A qualitative simplified space-time diagram (in arbitrary units) illustrating the GSTS paradigm. The EMBH, originating from the gravitational collapse of a massive GRB-progenitor star $P_1$, and the massive supernova-progenitor star $P_2$-neutron star ($P_2$-NS) system, separated by a radial distance $D_{P_2}$, are assumed to be at rest in in the laboratory frame. Their worldlines are represented by two parallel vertical lines. The supernova shell moving at $0.1c$ generated by the $P_2$-NS transition is represented by the dotted line cone. The solid line represents the motion of the pulse, as if it would move with an ``effective'' constant gamma factor $\gamma_1$ during the eras reaching the condition of transparency. Similarly, another ``effective'' constant gamma factor $\gamma_2<\gamma_1$ applies during era IV up to the collision with the $P_2$-NS system. A third ``effective'' constant gamma factor $\gamma_3 < \gamma_2$ occurs during era V after the collision as the nonrelativistic regime of expansion is reached. The dashed lines at 45 degrees represent signals propagating at speed of light.}
\label{iii-fig1}
\end{figure}

In order to reach an intuitive understanding of these complex computations we present a schematic very simplified diagram (not to scale) in Fig.~\ref{iii-fig1}.

We now describe the sequence of events and the specific data corresponding to the GSTS paradigm:
\begin{enumerate}
\item The two stars $P_1$ and $P_2$ are separated by a distance $D_{P_2} = 3.94\times 10^{17}$ cm in the laboratory frame, see Fig.~\ref{iii-fig1}. Both stars are at rest in the inertial laboratory frame. At laboratory time $t=0$ and at comoving time $\tau=0$, the gravitational collapse of the GRB-progenitor star $P_1$ occurs, and the initial emission of gravitational radiation or a neutrino burst from the event then synchronizes this event with the arrival times $t_a=0$ at the supernova-progenitor star $P_2$ and $t_a^d=0$ for the distant observer at rest with the detector. The electromagnetic radiation emitted by the gravitational collapse process is instead practically zero, due to the optical thickness of the material at this stage,\cite{brx00} see Tab.~\ref{tab1}.

\item From Tab.~\ref{tab1}, at laboratory time $t_1 = 6.48 \times 10^3\,s$ and at a distance from the EMBH of $D_1=1.94\times 10^{14}\,cm$, the condition of transparency for the PEMB pulse is reached and the P-GRB is emitted (see section~\ref{era3}). This time is recorded in arrival time at the detector $t_{a_1}^d = 8.41\times 10^{-2}\,s$, and, at $P_2$, at $t_{a_1}=4.20\times 10^{-2}\, s$. The fact that the PEMB pulse in an arrival time of $8.41\times 10^{-2}\,s$ covers a distance of $1.94\times 10^{14}\,cm$ gives rise to an apparent ``superluminal'' effect. This apparent paradox can be straightforwardly explained by introducing an ``effective'' gamma factor.\cite{lett3}

\item At laboratory time $t = 1.73\times 10^6\,s$ and at a distance from the EMBH of $5.18\times 10^{16}\,cm$ in the laboratory frame, the peak of the E-APE is reached which is recorded at the arrival time $t_a=9.93\,s$ at $P_2$ and $t_a^d=19.87\,s$ at the detector. This also gives rise to an apparent ``superluminal'' effect.

\item At a distance $D_{P_2} = 3.94 \times 10^{17}\,cm $, the two bursts described in the above points 2) and 3) collide with the supernova-progenitor star $P_2$ at arrival times $t_{a_1}=4.20\times 10^{-2}\,s$ and $ t_a  = 9.93\,s $ respectively. They can then induce the supernova explosion of the massive star $P_2$.

\item The associated supernova shell expands with velocity $0.1c$.

\item The expanding supernova shell is reached by the ABM pulse generating the afterglow with a delay of $t_{a_2}=18.5\,hr$ in arrival time following the arrival of the P-GRB and the E-APE. This time delay coincides with the interval of laboratory time separating the two events, since the $P_2$ is at rest in the inertial laboratory frame.\cite{lett3} The ABM pulse has travelled in the laboratory frame a distance $D_{P_2}-D_1\simeq D_{P_2}=3.94\times 10^{17}\,cm$ in a laboratory time $t_2-t_1 \simeq t_2=1.32\times 10^7\,s$ (neglecting the supernova expansion).
\end{enumerate}

The collision of the pulse with the supernova shell occurs at $\gamma\simeq 4.0$. By this time the supernova shell has reached a dimension of $1.997\times 10^{14}\,cm$, which is consistent with the observations from the Chandra satellite.

In these considerations on GRB~991216 the supernova remnant has been assumed to be close to but not exactly along the line of sight extending from the EMBH to the distant observer. If such an alignment should exist for other GRBs, it would lead to an observation of iron absorption lines as well as to an increase in the radiation observed in the afterglow corresponding to the crossing of the supernova shell by the ABM pulse. In fact, as the ABM pulse engulfs the baryonic matter of the remnant, above and beyond the normal interstellar medium baryonic matter, the conservation of energy and momentum implies that a larger amount of internal energy is available and radiated in the process (see section~\ref{era4}). This increased energy-momentum loss will generally affect the slope of the afterglow decay, approaching more rapidly a nonrelativistic expansion phase (details are given in section~\ref{approximation}).

It is quite clear that as soon as the relativistic transformations of the RSTT paradigm are duly taken into account, the sequence of events between the supernova and the GRB occurrences are exactly the opposite of the one postulated in the so-called ``supranova'' scenario.\cite{vs98,vs99,va99} This can be considered a very appropriate pedagogical example of how classical nonrelativistic applied to ultrarelativistic regimes can indeed subvert the very causal relation between events.

If we now turn to the possibility of dynamically implementing the scenario, there are at least three different possibilities:
\begin{enumerate}
\item Particularly attractive is the possibility that a massive star $P_2$ has rapidly evolved during its thermonuclear evolution to a white dwarf.\cite{c78} It it then sufficient that the P-GRB and the E-APE implode the star sufficiently as to reach a central density above the critical density for the ignition of thermonuclear burning. Consequently, the explosion of the star $P_2$ occurs, and a significant fraction of a solar mass of iron is generated. These configurations are currently generally considered precursors of some type I supernovae (see e.g. Filippenko, 1997,\cite{f97} and references therein).
\item Alternatively, the massive star $P_2$ can have evolved to the condition of being close to the point of gravitational collapse, having developed the formation of an iron-silicon core, type II supernovae. The above transfer of energy momentum from the P-GRB and the E-APE may enhance the capture of the electrons on the iron nuclei and consequently decrease the Fermi energy of the core, leading to the onset of gravitational instability (see e.g. Bethe, 1991,\cite{b91} p. 270 and followings). Since the time for the final evolution of a massive star with an iron-silicon core is short, this event requires a well tuned coincidence.
\item The pressure wave may trigger massive and instantaneous nuclear burning process, with corresponding changes in the chemical composition of the star, leading to the collapse.
\end{enumerate}

The GSTS paradigm has been applied to the case of the GRB 980425 - SN1998bw which, with a red shift of 0.0083, is one of the closest and weaker GRBs observed. In this case, the radio and the optical emission of the supernova is distinctively observed. For this particular case, the EMBH appears to have a significantly lower value of the parameter $\xi$ and the validity of the GSTS paradigm presented here is confirmed.\cite{rbcfx02d_supernova}

\section{General considerations on the EMBH formation}\label{gc}

Before concluding let us consider the problem of the EMBH formation. Such a problem has been debated for many years since the earliest discussions in 1970 in Princeton and has been finally clarified and addressed in general terms to justify the plausibility of the hypothesis in.\cite{r01mg9} There has been a basic change of paradigm. All the considerations on the electric charge of stars were traditionally directed, following the classical work by Shvartsman (1970)\cite{s70} all the way to the fundamental book by Punsly,\cite{punsly_book} to the presence of a net charge on the star surface in a steady state condition. The star can be endowed of rotation and magnetic field and surrounded by plasma, like in the case of Goldreich \& Julian (1969),\cite{gj69} or, in the case of absence of both magnetic field and rotation, the electrostatic processes can be related to the depth of the gravitational well, like in the treatment of Shvartsman (1970).\cite{s70} However, in neither cases it is possible to reach the condition of the overcritical field needed for pair creation nor have the condition of no baryonic contamination discussed in sections~\ref{dyadosphere},~\ref{era1} and essential for the dyadosphere formation. The basic conceptual point is that GRBs are maybe the most violent transient phenomenon occurring in the universe and so the condition for the dyadosphere creation have to be searched in a transient phenomenon. The solution is related to the most transient phenomenon occurring in the life of a star: the process of gravitational collapse.

Having acquired such a fundamental understanding, the next step is to estimate the amount of polarization needed in order to reach the fully relativistic condition
\begin{equation}
\frac{Q}{M\sqrt{G}}=1\; .
\label{gc_eq1}
\end{equation}
Recalling that the charge to mass ratio of a proton is $q_p/\left(m_p\sqrt{G}\right)=1.1\times 10^{18}$, it is enough to have an excess of one quantum of charge every $10^{18}$ nucleons in the core of the collapsing star to obtain an extreme EMBH after the occurrence of the gravitational collapse. Physically this means that we are dealing with a process of charge segregation between the core and the outer part of the star which has the opposite sign of net charge in order to enforce the overall charge neutrality condition. We here emphasize the name ``charge segregation'' instead of the name ``charge separation'' in order to contrast a very mild charge surplus created in different part of the star, keeping the overall charge neutrality, from the much more extreme condition of charge separation in which all the charges of the atomic component of the star are separated. It is indeed reassuring that such a core, endowed with charge segregation, is indeed stable with respect to the Fermi-Chandrasekhar criteria for the stability of self-gravitating stars duly extended from the magnetic to the electric case: the electric energy of such a core is consistently smaller than its gravitational energy.\cite{bor02}

Such a condition of charge segregation between the core and the oppositely charged star surface layer can be reached under a very large number of physical conditions. We consider, for simplicity, one of the oldest example: the one of a star endowed with both a magnetic field and rotation. It is proved that a typical magnetic field expected for the ISM is $B_\circ\sim 10^{-5}\, G$.\cite{fe01} We further assume, consistently with the data which we have acquired and verified in the present article (see sections~\ref{era4},~\ref{power-law}), that also in the galaxy where GRB~991216 occurred the ISM has an average density of $n_{ism}=1\, proton/cm^3$. From this value of density we have that an ISM cloud with mass $M\sim 10M_\odot$ occupies a sphere of radius $R_\circ\sim 1.4\times 10^{19}\, cm$. If this sphere collapse to a star with radius $R=R_\odot$, from the flux conservation we obtain that it is enough for this star to rotate with the most reasonable angular speed
\begin{equation}
\Omega\sim\frac{\xi Mc\sqrt{G}}{R_\odot R_\circ^2 B_\circ}
\label{omega}
\end{equation}
to conclude that the progenitor star core is endowed of a charge to mass ratio equal to $\xi$. In the extreme case of Eq.(\ref{gc_eq1}) we have $\xi=1$ and so the angular speed is $\Omega\sim 1.1\times 10^{-3}\, rad/s$ --- i.e. one round in $1.5\, hr$ --- and correspondingly we have smaller $\Omega$ values for $\xi < 1$.\cite{bor02} Clearly the overall neutrality is guaranteed by the oppositely charged baryonic matter which is the one measured by the $B$ parameter in the EMBH model (see sections~\ref{era2}--\ref{era3}). The smallness of the $B$ value clearly points to the absence of an extended envelope of the progenitor star.

The formation process of such an electromagnetised progenitor star will be clearly affected by the presence of differential rotation, the consequent amplification of the magnetic field and a variety of magnetohydrodynamical problems which will affect somewhat the simplicity of the heuristic Eq.(\ref{omega}). Similarly the process of gravitational collapse of such a progenitor star endowed with rotation will lead to complex phenomena of ``gravitationally induced electromagnetic radiation''\cite{ja73} and of ``electromagnetically induced gravitational radiation''\cite{ja74} which will tend to reduce both the eccentricity and the angular velocity of the collapsing core. The general outcome of gravitational collapse will be a Kerr-Newmann spacetime. It is interesting that such a general case will break the degeneracy in $\left(\mu,\xi\right)$ described in section~\ref{fp}.\cite{rbcfx02e_paperII} In this article we have addressed the much simpler case of a solution in which $\left(cL\right)/\left(GM^2\right) \ll 1$ and the treatment can be well approximated by a collapse described by a Reissner-Nordstr\"{o}m geometry.

In addition to this scenario, based on the role of magnetic field and rotation, we are as well pursuing the possible generation of the charge segregation by quantum effects at the surface of the almost Fermi degenerate core. This most straightforward analysis also leads to a Reissner-Nordstr\"{o}m geometry.

In both these cases the Reissner-Nordstr\"{o}m geometry appears indeed to be the relevant model for GRB~991216 as discussed in the previous sections. We shall return to non spherical configuration in forthcoming publications and/or when requested by observational evidence.\cite{rbcfx02e_paperII}

Turning now from this general scenario to a more detailed analysis of a Reissner-Nordstr\"{o}m geometry, some preliminary necessary steps have to be accomplished. In Cherubini, Ruffini \& Vitagliano (2002)\cite{crv02} we have considered the gravitational collapse of a charged spherical shell with selected boundary conditions: either starting from infinite distance with a null or non null kinetic energy, or imploding from a finite distance initially at rest. A new analytic solution has been obtained for such a boundary condition, corresponding both to a collapse into an already formed EMBH or to a collapse in Minkowsky space. In both cases we have followed the process of gravitational collapse all the way to the self closure of the shell by the formation of an horizon.

Using this analytic solution it has been possible to clarify the independent physical components, contributing to the formation of the EMBH irreducible mass.\cite{rv02a} Surprisingly, the irreducible mass does not directly depend on the electromagnetic energy of the imploding shell: it is uniquely a function of the initial baryonic mass, of its gravitational energy and of the kinetic energy of the implosion. The electromagnetic energy is stored around the EMBH and can be extracted by two very different process as a function of the electromagnetic field strength. {\bf a)} When the electric field on the collapsing shell is smaller than ${\cal E}_c$, the process of energy extraction occurs in the effective EMBH ergosphere\cite{dr73,dhr74} by a sequence of discrete high energy events, with energy up to $10^{21}$--$10^{27}$ eV. Such sources can be of relevance for the explanation of the ultra high energy cosmic rays.\cite{bcrx02} {\bf b)} When the electric field on the collapsing shell is larger than ${\cal E}_c$, the conditions relevant to the present article are fulfilled. The energy extraction process occurs in the dyadosphere and a much larger number of electron and positron pairs are created with typical energies of the order of $10$ MeV which are relevant for the process considered in the present paper.

It is interesting that the clarification obtained in Ruffini \& Vitagliano (2002a)\cite{rv02a} has allowed a deeper understanding of the essential role of the gravitational and kinetic implosion energies and the storage of the electromagnetic energy in the entire region surrounding the EMBH horizon. It has been shown in Ruffini \& Vitagliano (2002b)\cite{rv02b} that the central point can simply be summarized: the Coulomb repulsion of the collapsing matter reduces the kinetic energy of implosion leading to a smaller value of the irreducible mass and consequently to a larger value of the extractable energy.

\begin{figure}[htbp]
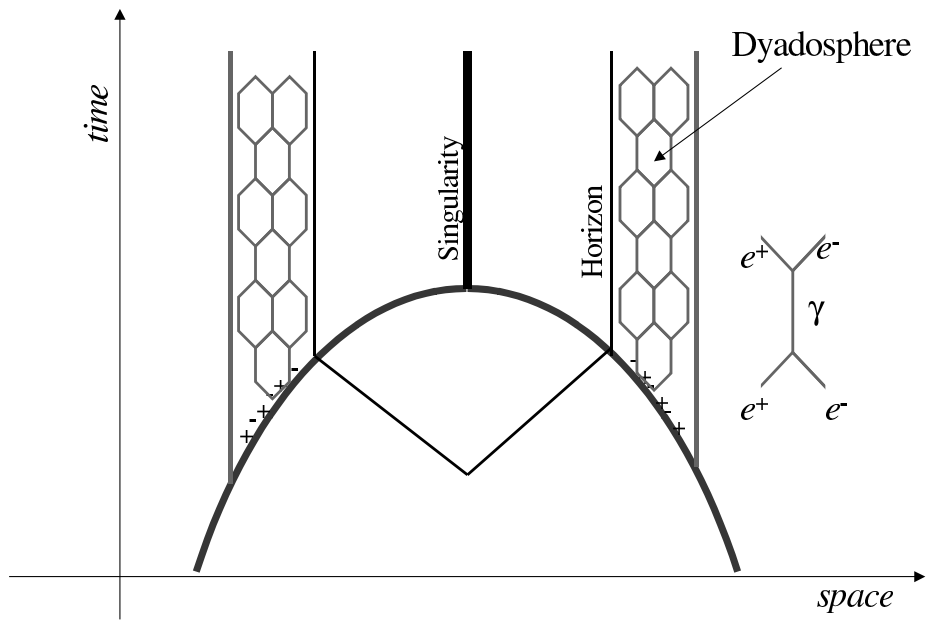

\PSFIG{dyaform}{\hsize}{0}
\caption{Space-time diagram of the collapse process leading to the formation of the dyadosphere. As the collapsing core crosses the dyadosphere radius the pair creation process starts, and the pairs thermalize in a neutral plasma configuration. Then also the horizon is crossed and the singularity is formed.}
\label{dyaform}
\end{figure}

Having so established and clarified the basic conceptual processes of the energetic of the EMBH, we are now ready to approach, using the new analytic solution obtained, the dynamical process of vacuum polarization occurring during the formation of an EMBH as qualitatively represented in Fig.~\ref{dyaform}. The study of the dyadosphere dynamical formation as well as of the electron-positron plasma dynamical evolution will lead to the first possibility of directly observing the general relativistic effects approaching the EMBH horizon.

Before closing we would like to emphasize once more a basic point: all the considerations presented in the description of the preceding eras are based on the approximations in the description of the dyadosphere presented in section \ref{dyadosphere}. This treatment is very appropriate in estimating the general dependence of the energy of the P-GRB, the kinetic energy of the ABM pulse and consequently the intensity of the afterglow, as well as the overall time structure of the GRB and especially the time of the release of the P-GRB in respect to the moment of gravitational collapse and its relative intensity with respect to the afterglow. If, however, is addressed the issue of the detailed temporal structure of the P-GRB and its detailed spectral distribution, the above dynamical considerations on the dyadosphere formation are needed.\cite{rvx02} In turn, this detailed analysis is needed if the general relativistic effects close to the horizon formation have to be followed. As expressed already in section.~\ref{new}, all general relativistic quantum field theory effects are encoded in the fine structure of the P-GRB. As emphasized in section~\ref{fp}, the only way to differentiate between solutions with same $E_{dya}$ but different EMBH mass and charge is to observe the P-GRBs in the limit $B\to0$, namely, to observe the short GRBs.

\section{Conclusions}\label{conclusions}

This paper is a consequence of a process of revision, reanalysis and further expansion of all the results presented in the previous articles\cite{lett1,lett2,lett3,prx98,rswx99,rswx00}. In this process all the numerical estimates have been reanalysed for consistency. Most noticeable we have found a missing factor of $2\pi$ in the values of the experimental data on GRB~991216 we had used in our previous works which led to an underestimate of the total energy of the source.

This process of revision, far from being just a detailed computational verification, has given us the opportunity to rethink the entire GRB process in an unitary description starting from the moment of gravitational collapse all the way up to the latest phases of the afterglow and in identifying the three fundamental actors of the GRB phenomenon:
\begin{enumerate}
\item $E_{dya}$. Having reanalysed in section~\ref{dyadosphere} the physics of the dyadosphere we have pointed out in Fig.~\ref{muxi} that the same value of $E_{dya}$ can be obtained from an entire family of $\left(\mu,\xi\right)$ parameters (i.e. $E_{dya}$ is degenerate in $\left(\mu,\xi\right)$). We have then shown in the reexamination of all the GRB eras that all the results depend only on the value of $E_{dya}$ and not on the particular value of $\left(\mu,\xi\right)$ (see sections~\ref{era2},\ref{era3},\ref{era4},\ref{era5}). The only exception to this occurs in the era I (see section~\ref{era1}) which is the only one relevant for short GRBs.
\item $B$. The crucial role played by the baryonic remnant of the progenitor star in determining the relative intensity ratio and the time delay between the P-GRB and the E-APE has been summarized already in the two Figs.~\ref{crossen}--\ref{dtab} in the introduction.
\item ISM. The density $n_{ism}$ of the interstellar medium and its inhomogeneities appears to have a fundamental role in the intensity and the temporal substructures of the E-APE and the afterglow. In order to identify such a crucial role, however, the correct relativistic space-time relations expressed by the RSTT paradigm are needed as amply exemplified in sections~\ref{approximation}--\ref{substructures}.
\end{enumerate}

The observational data agree with the predictions of the model on:\\
1) the intensity ratio, $1.58\times 10^{-2}$,  between the P-GRB and the E-APE, which strongly depends on the parameter $B$,\\
2) the absolute intensities for both the P-GRB and the E-APE, respectively $7.54\times 10^{51}$ erg  and  $4.75\times 10^{53}$,\\
3) the arrival time of the P-GRB and the peak of the E-APE, respectively $8.41 \times 10^{-2}$ s and $19.87$ s.

These results can certainly be considered the greatest success of the EMBH theory.

Before closing, we like to draw some specific conclusions based on the first fundamental parameter of the EMBH theory: $E_{dya}$. It is clear that $E_{dya}$ is the fundamental parameter which determine the general energetic requirements of the GRB~991216. This energetics strongly depends on the possible existence or absence of beaming in the radiation process. In turn, as expressed in sections~\ref{approximation}--\ref{power-law}, the presence of beaming is led back to the power-law index $n$ of the afterglow. The general conclusions reached on $E_{dya}$ can be summarized as follow:
\paragraph*{\bf 1.1}
The value of $n$ is a function of the transformation between $t \rightarrow t_a \rightarrow t_a^d$ (see section~\ref{approximation}). These transformations are a function of the entire relativistic regime of the world line of the source (see section~\ref{arrival_time}). By systematically neglecting this information the current works in the GRBs literature have obtained incorrect $n$ values (see Tab.~\ref{tab2} and section~\ref{power-law}).\\
\paragraph*{\bf 1.2}
The value $n_{theo} = -1.6$, which we have obtained within the EMBH theory, in the region of interest for the observations, based on the assumptions of spherical symmetry, fully radiative condition in the emission process and constant density of ISM, is in agreement with observed $n_{obs} = -1.616 \pm 0.067$ (see sections~\ref{approximation},~\ref{power-law}). No evidence of beaming is therefore found is GRB~991216 (see section~\ref{power-law}).\\
\paragraph*{\bf 1.3}
For GRB~991216 $E_{dya}=4.83\times 10^{53}$ erg is found in the EMBH theory. This value is systematically larger than the ones quoted in the current literature by Panaitescu \& Kumar (2001)\cite{pk01} and by Halpern et al. (2000)\cite{ha00} due to the fact that they respectively consider beaming angles of $3^\circ-4^\circ$ and $6^\circ$. These considerations have been shown to be untenable in section~\ref{power-law}. There is still a difference of $\sim 28\%$ between the total energy implied by the EMBH theory ($4.83\times 10^{53}$ erg) and the value quoted by Halpern ($E_{dya}=6.7\times 10^{53}$ erg) in the case of spherical emission. We trust that this is a consequence of the underlying assumption of the spectral distribution of the radiation assumed by Halpern et al. (2000)\cite{ha00} (see e.g. Frail et al., 2001\cite{f01}), which should be reassessed on the ground of our theoretical results. See also paper II.\cite{rbcfx02e_paperII}

We turn now to the second fundamental parameter of the EMBH theory: $B$. It is essential in explaining the difference between the so called ``long bursts'' and ``short bursts'' (see sections~\ref{new},~\ref{shortlongburst}). The general conclusions reached on $B$ can be summarized as follows:
\paragraph*{\bf 2.1}
The most general GRB contains three different components: the P-GRB, the E-APE and the rest of the afterglow. The ratio between the P-GRB and the E-APE intensity and their temporal separation is a function of the $B$ parameter (see Figs.~\ref{crossen}--\ref{dtab}). The best fit is obtained for $B=3.0\times 10^{-3}$ (see section~\ref{bf}). We recall that in the present case for $B < 2.5\times10^{-5}$ the energy of the P-GRB would be larger than the one of the E-APE and the energy of the dyadosphere would be mainly emitted in what have been called the ``short bursts'', while for $B > 2.5\times10^{-5} $ the energy of the E-APE would predominate and the energy of the dyadosphere would be mainly carried by the ABM pulse and emitted in the afterglow.
\paragraph*{\bf 2.2}
The difficulties encountered by {\em all} theoretical models, through the years, in order to explain the so called ``long bursts'' are resolved in a drastic way (see section~\ref{shortlongburst}). The so called ``long bursts'' are {\em not} bursts at all. They represent just the E-APE which was interpreted as a burst only due to the noise threshold in the BATSE observations (see Fig.~\ref{fit_1}). The E-APE is emitted at distances from the EMBH in the range $1.0\times 10^{16}\sim 1.0\times 10^{17}$ cm, see Tab.~\ref{tab1}, namely well outside the size of the progenitor star and already deep in interstellar space. The fact that the crossing of such distance, which is a typical dimension of an interstellar cloud, appears to occur in arrival time in only $\sim 100$ seconds is perfectly explained by the relativistic transformations encoded in the RSTT paradigm corresponding to a gamma factor between $100$ and $300$ (see section~\ref{arrival_time} and Tab.~\ref{tab1}). This effect would be interpreted within a classical and incorrect astronomical picture by a ``superluminal'' behaviour propagating at $\sim 3.6\times 10^4c$ (see Tab.~\ref{tab1}).
\paragraph*{\bf 2.3}
In the limit $B \to 0$ the entire dyadosphere energy is emitted in the P-GRB. These events represents the ``short bursts'' class, for which the afterglow intensity is smaller than the P-GRB emission and below the actual observational limits (see section~\ref{new}). It is interesting that the proposed differentiation between the ``short bursts'' and ``long bursts'' within the EMBH theory is merely due to the amount of baryonic matter in the remnant, described by the $B$ parameter, and totally independent from the process of gravitational collapse which is clearly identical in both cases. This explains at once the recently found conclusion that the distribution of short and long GRBs have essentially the same characteristic peak luminosity.\cite{s01} Also the result expressed in Fig.~\ref{energypeak} that the average temperature corresponding to the P-GRB emission does increase for decreasing values of the $B$ parameter can explain the observed fact that the ``short bursts'', which are obtained in the limit $B\to 0$, are systematically harder than ``long bursts''.\cite{ka93}

Finally, the EMBH theory offers an unprecedented tool in order to map with great accuracy all the matter distribution around the newly formed EMBH from the horizon all the way to the ISM. This concept was pioneered by Dermer \& Mitman (1999)\cite{dm99} who proposed to use GRB sources as ``tomographic images of the density distributions of the medium surrounding the sources of GRBs''. It is important to emphasize that the very precise reading of the matter distribution encoded in the data of the P-GRB, the E-APE and the afterglow in GRB~991216 is in marked disagreement with the matter distribution postulated by the ``collapsar'' scenario.\cite{p98,w93,mw99} This conclusion is evidenced not only by the absence of beaming already mentioned above, but also for the paucity of the baryonic matter encountered by the PEM pulse in its way out from the EMBH. There is no evidence for the presence either of a baryonic disk component nor of a conspicuous baryonic remnant. We actually have $B=3.0\times 10^{-3}$. The general conclusions reached on this topic can then be summarized as follow:
\paragraph*{\bf 3.1}
Starting from the inside out we have, from the electrodynamics of the dyadosphere, that the baryonic contamination in that region has to be much smaller than $10^8 {\rm g/cm^3}$ (see section~\ref{era1}). This condition can be achieved in the formation of an EMBH. The same electrodynamical process would be hampered in the formation of a neutron star due to the high density and baryonic contamination. Among all process of discharge of the overcritical electromagnetic field the pair creation is the most effective one due to the very short time scale of the order of $\frac{\hbar}{m_ec^c\alpha}\sim 10^{-19}$ seconds (see section~\ref{dyadosphere}).
\paragraph*{\bf 3.2}
Unlike the case of formation of a neutron star, the mass of the remnant of the progenitor star is very small indeed. This mass, determined by $B$, is very accurately inferable from the relative intensity and temporal distance between the P-GRB and the E-APE (see above). In the present case we have $M_B \sim 8.1\times 10^{-4} M_{\odot}$. The presence of the remnant is also important for guaranteeing the overall charge neutrality of the system formed by the oppositely charged collapsing core and the remnant. It has been pointed out in section~\ref{gc} that this condition of charge separation between the collapsing core and the remnant occurs only during the relevant part of the gravitational collapse process which, we recall, for a $10M_{\odot}$ is of the order of 30 seconds.
\paragraph*{\bf 3.3}
The structure of the E-APE and the afterglow gives as well an unprecedented tool in order to estimate the average density and filamentary structure in the ISM: the structures down to a fraction of seconds observed in the E-APEs, the so called ``long bursts'' of the current literature (see Fig.~\ref{grb991216}), can be used in order to map the filamentary structure as well as the size of interstellar clouds surrounding the EMBH (see section~\ref{substructures}). When all the geometrical and relativistic effects are duly taken into account the intensity and the average profile of the E-APE and of the afterglow point to an average value of the ISM density $n_{ism} \sim 1 proton/cm^3$ in very good agreement with a large variety of independent estimates. The very late phases of the afterglow gives information of the induced supernova collapse (see section~\ref{gsts} and Ruffini et al., 2001c\cite{lett3}) which will be addressed in a forthcoming publication. Since now we can assert that the correct space-time sequence based on the RSTT paradigm is in contrast with the ``supranova'' scenario\cite{vs98,vs99,va99} which was based on a nonrelativistic consideration in ultrarelativistic regimes (see section~\ref{gsts}).

This concludes the presentation of the basic model which is now ready to be applied to additional sources. If we look to the future we can see three main topics to be addressed with special attention:
\begin{enumerate}
\item We have performed a more detailed description of beaming, of the angular spreading and of the spectral properties which is going to be the subject of paper II.\cite{rbcfx02e_paperII} Since now, we can assert that this more detailed treatment supports all general conclusions obtained in the present paper.
\item If one is interested in the detailed effects of general relativity and relativistic field theory, all the attention should be directed to the structure of the short bursts. This needs the development of detailed theoretical works on the approach of the horizon of the black hole and the associated electrodynamical process. From the description presented in this paper of an already formed and averaged dyadosphere (see Fig.~\ref{dens}) we have to move to the treatment of its dynamical formation (see Fig.~\ref{dyaform}). Such an analysis describing the approach to the formation of the horizon of the EMBH, within the EMBH theory, is in advanced phase of development\cite{crv02,rv02a,rv02b,rvx02} see Fig.~\ref{dyaform}. Some preliminary results have appeared in Bianco, Ruffini \& Xue (2001).\cite{brx00}
\item From the observational point of view, the detailed observations of the yet unexplored region in the range up to $10^2$ seconds in Fig.~\ref{final} and the corresponding observations of the ``short bursts'' by a new class of space missions with higher sensitivity than the BATSE instrument appear to be of great importance. Such observations should allow to directly observe for the first time the general relativistic and extreme quantum field theory effects connected to the process of formation of the EMBH. It can be of some interest to explore the possibility of observing in these regimes the ``gravitationally induced electromagnetic radiation''\cite{ja73} and the ``electromagnetically induced gravitational radiation''\cite{ja74} phenomena.
\end{enumerate}

\end{document}